\documentclass[fleqn,usenatbib]{mnras}

\usepackage{newtxtext,newtxmath}

\usepackage[T1]{fontenc}

\DeclareRobustCommand{\VAN}[3]{#2}
\let\VANthebibliography\thebibliography
\def\thebibliography{\DeclareRobustCommand{\VAN}[3]{##3}\VANthebibliography}

\usepackage{graphicx}	% Including figure files
\usepackage{amsmath}	% Advanced maths commands
\usepackage{amssymb}	% Extra maths symbols
\usepackage{makecell}
\usepackage{multicol}   % http://ctan.org/pkg/multicols
\title[Dust-obscured star formation at $z \gtrsim 5 $]{Characterizing the contribution of dust-obscured star formation at $z \gtrsim$ 5 using 
18 serendipitously identified [CII] emitters}

\author[I. F. van Leeuwen et al.]{
I. F. van Leeuwen,$^{1}$\thanks{E-mail: vleeuwen@strw.leidenuniv.nl}
R. J. Bouwens$^{1}$, P. P. van der Werf$^{1}$, J. A. Hodge$^{1}$,  S. Schouws$^{1}$,  M. Stefanon$^{2,3}$, \newauthor H. S. B. Algera$^{4,5}$, M. Aravena$^{6}$,  L. A. Boogaard$^{7}$, R. A .A. Bowler$^8$, E. da Cunha$^{9}$, P. Dayal$^{10}$,   R. Decarli$^{11}$, \newauthor V. Gonzalez$^{12}$, H. Inami$^{4}$,  I. de Looze$^{13}$,  L. Sommovigo$^{14}$, B. P. Venemans$^{1}$, F. Walter$^{7,15}$,  L. Barrufet$^{16}$, \newauthor  A. Ferrara$^{17}$, L. Graziani$^{14,18,23}$, A. P. S. Hygate$^{1}$,  P. Oesch$^{19, 20, 21}$, M. Palla$^{13}$, L. Rowland$^{1}$ and \newauthor R. Schneider$^{14,18, 22, 23}$
\\
$^{1}$Leiden Observatory, Leiden University, NL-2300 RA Leiden, the Netherlands \\
$^2$Departament d’Astronomia i Astrofìsica, Universitat de València, C. Dr. Moliner 50, E-46100 Burjassot, València, Spain \\
$^{3}$Unidad Asociada CSIC ``Grupo de Astrofísica Extragaláctica y Cosmología" (Instituto de Física de Cantabria—Universitat de València), Spain \\
$^{4}$Hiroshima Astrophysical Science Center, Hiroshima University, 1-3-1 Kagamiyama, Higashi-Hiroshima, Hiroshima 739-8526, Japan \\
$^5$National Astronomical Observatory of Japan, 2-21-1, Osawa, Mitaka, Tokyo, Japan\\
$^{6}$Instituto de Estudios Astrofísicos, Facultad de Ingeniería y Ciencias, Universidad Diego Portales, Av. Ejército 441, Santiago 8370191, Chile \\
$^7$Max Planck Institute for Astronomy, K\"{o}nigstuhl 17, 69117 Heidelberg, Germany\\
$^8$Jodrell Bank Centre for Astrophysics, Department of Physics and Astronomy, School of Natural Sciences, The University of Manchester, Manchester, M13 9PL, UK \\ 
$^9$ International Centre for Radio Astronomy Research, University of Western Australia, 35 Stirling Hwy, Crawley, WA 6009, Australia \\
$^{10}$Kapteyn Astronomical Institute, University of Groningen, P.O. Box 800, 9700 AV Groningen, The Netherlands \\
$^{11}$INAF - Osservatorio di Astrofisica e Scienza dello Spazio di Bologna, via Gobetti 93/3, Bologna, 40129, Italy \\
$^{12}$Departmento de Astronomia,Universidad de Chile, Casilla 36-D, Santiago 7591245, Chile\\
$^{13}$Sterrenkundig Observatorium, Ghent University, Krijgslaan 281-S9, B-9000 Gent, Belgium\\
$^{14}$Dipartimento di Fisica, Sapienza, Universita di Roma,Piazza le Aldo Moro 5, I-00185 Roma, Italy\\
$^{15}$National Radio Astronomy Observatory, Pete V. Domenici Array Science Center, P.O. Box O, Socorro, NM 87801, USA\\
$^{16}$Institute for Astronomy, School of Physics \& Astronomy, University of Edinburgh, Royal Observatory, Edinburgh, EH9 3HJ, UK\\
$^{17}$Scuola Normale Superiore, Piazza dei Cavalieri 7, 56126, Pisa, Italy\\
$^{18}$INAF/Osservatorio Astronomico di Roma, via Frascati 33, I 00078 Monte Porzio Catone, Roma, Italy\\
$^{19}$Cosmic Dawn Center (DAWN), Denmark\\
$^{20}$Niels Bohr Institute, University of Copenhagen, Jagtvej 128, DK- 2200, Copenhagen N, Denmark \\
$^{21}$Department of Astronomy, University of Geneva, Chemin Pegasi 51, 1290 Versoix, Switzerland\\
$^{22}$Sapienza School for Advanced Studies, Sapienza Università di Roma, P.le Aldo Moro 2, 00185 Roma, Italy\\
$^{23}$INFN, Sezione di Roma 1, P.le Aldo Moro 2, 00185 Roma, Italy \\
}

\date{Accepted XXX. Received YYY; in original form ZZZ}

\pubyear{2024}

\begin{document}
\label{firstpage}
\pagerange{\pageref{firstpage}--\pageref{lastpage}}
\maketitle

% Abstract of the paper
\begin{abstract}
%This is a simple template for authors to write new MNRAS papers.
%The abstract should briefly describe the aims, methods, and main results of the paper.
%It should be a single paragraph not more than 250 words (200 words for Letters).
%No references should appear in the abstract.

\noindent We present a new method to determine the star formation rate (SFR) density of the Universe at $z \gtrsim 5$  that includes the contribution of dust-obscured star formation. For this purpose, we use a [CII] (158 $\mu$m) selected sample of galaxies serendipitously identified in the fields of known $z\gtrsim 4.5$ objects to characterize the fraction of obscured SFR. The advantage of a [CII] selection is that our sample is SFR-selected, in contrast to a UV-selection that would be biased towards unobscured star formation. We obtain a sample of 23 [CII] emitters near star-forming (SF) galaxies and QSOs -- three of which we identify for the first time -- using previous literature and archival ALMA data. 18 of these serendipitously identified galaxies have sufficiently deep rest-UV data and are used to characterize the obscured fraction of the star formation in galaxies with SFRs $\gtrsim 30\ \text{M}_{\odot} \  \text{yr}^{-1}$. We find that [CII] emitters identified around SF galaxies have $\approx$63\% of their SFR obscured, while [CII] emitters around QSOs have $\approx$93\% of their SFR obscured. By forward modeling existing wide-area UV luminosity function (LF) determinations, we derive the intrinsic UV LF using our characterization of the obscured SFR. Integrating the intrinsic LF to $M_{UV}$ = $-$20 we find that the obscured SFRD contributes to $>3\%$ and $>10\%$ of the total SFRD at $z \sim 5$ and $z \sim 6$ based on our sample of companions galaxies near SFGs and QSOs, respectively. Our results suggest that dust obscuration is not negligible at $z\gtrsim 5$, further underlining the importance of far-IR observations of the $z\gtrsim 5$ Universe.

\end{abstract}

% Select between one and six entries from the list of approved keywords.

%%%%%%%%%%%%%%%%%%%%%%%%%%%%%%%%%%%%%%%%%%%%%%%%%%

%%%%%%%%%%%%%%%%% BODY OF PAPER %%%%%%%%%%%%%%%%%%

\section{Introduction}

One of the most exciting frontiers in extragalactic astronomy is understanding the build-up of stars in galaxies during the epoch of reionization.  A common way to characterize this early build-up is through the volume-averaged star formation rate density (SFRD) of the Universe.  Thanks to a wealth of multi-wavelength observations from both space-based and ground-based observatories, the SFRD at $z \leq 
3$ has already been characterized in detail \citep[e.g.,][]{Madau2014}. The general consensus is that the star formation rate increases with cosmic time until it reaches a peak at $z\sim 2$, the so-called `Cosmic Noon', and then decreases again to the present epoch. 

Reaching further back in time to $z > 3$, observations that are available to constrain the SFRD have been mostly limited to the rest-UV \citep[e.g.,][]{Madau2014}. However, far-IR probes are necessary due to the impact of dust attenuation in substantially obscuring UV light from bright star-forming (SF) galaxies in the early Universe.  Surveying the Universe in the far-IR has been challenging due to the limited spatial resolution of earlier wide-area probes resulting in substantial source confusion from lower-redshift sources (e.g. \citealt{nguyen2010, everett2020}). Moreover, higher spatial resolution probes such as are now available with the Atacama Large Millimeter/submillimeter Array (ALMA) are limited by their relative small field-of-view (FOV) \citep[e.g.,][]{Hodge2020}.

While some of the debate in the literature regarding the evolution of the SFRD has focused on the evolution at $z>8$ \citep[e.g.][]{McLeod2016, Oesch2018}, substantial uncertainties also exist at $z\sim5$-8 from the contribution of dust-obscured galaxies.  For example, several recent results  \citep[e.g.][]{CaseySurvey, gruppioni2020,Fudamoto_2020,Zavala2021, Khusanova_21, Loiacona_alpine, talia2021} have suggested that the obscured contribution of the SFRD at high redshift could be as large  as $\sim$20-60\% of the total SFRD. Moreover, recent work by \citet{algera} has demonstrated that the dust-obscured star formation at $z \sim$ 7 could contribute as much as $\sim$30-50\% of the SFRD, significantly more than earlier estimated based on UV data alone. There are many examples of galaxies at $z \sim$ 7 that are $\gtrsim 90\%$ obscured both in observations (e.g. \citealt{marrone2018,hygate2023}) and in theoretical models (e.g. \citealt{ferrara22}).

As previously discussed, progress in quantifying the fraction of obscured star formation in the early Universe has been challenging due to the limited spatial resolution of earlier probes of far-IR light and the limited areas surveyed by ALMA.  Recently, there have been efforts to overcome ALMA's small field-of-view and survey much wider areas.  Perhaps the widest area examples of such surveys are MORA \citep[2 mm Mapping Obscuration to Reionization with ALMA:][]{Zavala2021,Casey2021_MORA} and ex-MORA (Long et al., in prep.), covering 184 arcmin$^2$ and 0.2 deg$^2$, respectively, as well as the substantial A$^3$COSMOS effort \citep{Liu2019_A3COSMOS} where a 280 arcmin$^2$ survey area is constructed by combining the significant amount of ALMA observations over the COSMOS field.\footnote{https://sites.google.com/view/a3cosmos}  Wide-area surveys such as these can provide us with a useful unbiased probe of the prevalence of dusty star-forming galaxies in the distant Universe.  Nevertheless, an important disadvantage to this approach is also the large observational cost requiring not only wide area surveys in the far-IR, but also a pursuit of line scans to determine redshifts for the identified dusty sources.

Here we present a new method to characterize the obscured star formation in the $z > 4$ Universe, using sensitive ALMA observations targeting bright ISM cooling lines to search for star-forming galaxies in the neighborhoods of bright QSOs and massive galaxies.  Bright star-forming galaxies are readily identifiable via line emission in bright ISM cooling lines such as [CII] (157.74 $\mathrm{\mu \text{m}}$).  Given the clustering of galaxies, line emission from dusty star-forming galaxies tends to be found close to the targeted sources not only in frequency, but also in angular position on the sky. We will therefore also refer to these serendipitously identified galaxies as `companion' galaxies. Previous studies by e.g. \citet{Decarli_2017} and \citet{venemans_kilo} have shown that quasars are especially good places to look for bright star-forming companions, with $\sim$50\% of the quasars they examined showing at least one companion galaxy (see also \citealt{Meyer22}). Massive galaxies ($M_{*} \gtrsim 10^{8.5}$ $\mathrm{M_{\odot}}$) at high redshifts are also excellent targets to use for efficient searches for companion galaxies \citep{Loiacona_alpine,fudamoto_nature}. To date about 40 companion galaxies have been discovered at $z \gtrsim$ 4 with ALMA observations \cite[e.g.,][]{Decarli_2017, trakhtenbrot17, miller2020, venemans_kilo, Loiacona_alpine, fudamoto_nature}.

An important advantage to identifying star-forming galaxies using far-IR line emission is the insensitivity to dust obscuration or the brightness of sources in the rest-frame UV.  [CII] is an especially valuable line to use in this regard due to it being known to trace the total star formation rate in `normal' star-forming galaxies (e.g. \citealt{delooze2014}) and also thanks to its particular brightness.  As a result, [CII]-selected sources can thus be selected entirely on the basis of their SFRs, and sensitive rest-UV observations of such sources can be used to determine the extent to which star formation in the high-redshift Universe is obscured or not.  [CII] line emission is known to have multiple origins but is believed to arise mostly from photodissociation regions (PDRs) in high redshift galaxies (e.g. \citealt{hollenbach,wolfire22}). As [CII] has a low ionization potential (11.3 eV) and is therefore easy to excite, it traces a large fraction of the gas reservoir of a galaxy \citep{wolfire03}. 

In demonstrating the power of our new methodology, we make use of a significant sample of 18 [CII]-selected companion galaxies to quantify the fraction of star formation in $z\sim4$-7 galaxies that is obscured by dust compared to their total SFR. These [CII]-emitting companion galaxies are identified in the neighborhoods of 161 
[CII]-emitting galaxies and QSOs at $z>4$ and have deep enough rest-UV observations to determine if star formation in a source is obscured or not.  These 18 galaxies have been found as companions to galaxies drawn from 75 [CII]-emitting galaxies in ALPINE \citep{alpine, bethermin2020}, 25 [CII]-emitting galaxies in REBELS (\citealt{rebels}: see also Schouws et al. 2024, in prep.), 27 [CII]-emitting QSOs studied by \citet{venemans_kilo}, and 34 [CII]-emitting galaxies and QSOs from the ALMA archive. Based on the fraction of obscured star formation (from comparing rest-UV and far-IR observations) in our [CII]-selected samples, we forward model wide-area UV luminosity function (LF) results to recover the intrinsic luminosity function. The intrinsic UV luminosity function differs from the observed LF in that it accounts for the impact of dust obscuration on both the brightness of galaxies in the rest-UV and their selectability. Finally, we compute a dust-corrected SFRD at $z \gtrsim 5$. 

The organization of the paper is as follows. In Section~\ref{sec:methods}, we present both the archival ALMA data we use in this analysis as well as the process by which we both reduce and calibrate that data.  Additionally, we present our procedure for finding [CII]-emitting companion galaxies and how we estimate the star formation rates in the rest-UV and IR from the observations.  In Section~\ref{results}, we present the results, including the intrinsic luminosity function and the implied SFRD.  In Section~\ref{discussion}, we discuss both our results and the uncertainties and finally conclude with a summary of our findings in Section~\ref{conclusions}.  In this work all magnitudes are given in the AB system \citep{Oke1983}.  Moreover, we assume a \citet{Chabrier2003} initial mass function (IMF) (across a mass range of 0.1-100 M$_{\odot}$) and adopt a flat $\Lambda$CDM cosmology with h = 0.7, $\Omega_{\text{M}}$ = 0.3, and $\Omega_{\Lambda}$ = 0.7. For reference, 1" corresponds to 5.7 kpc at $z = 6$  using this set of cosmological parameters. 

\section{Methods}
\label{sec:methods}

\subsection{Observational data}
\label{sec:obs_data}
In this study we pursue a systematic search for [CII]-emitting companion galaxies at similar redshifts as the bright [CII]-emitting targets.  We make use of the [CII]-emitting companion galaxies identified around 75 [CII]-emitting galaxies from the ALPINE large program \citep{alpine,bethermin2020} at $z=4$-6, 25 [CII]-emitting galaxies from the REBELS large program \citep{rebels} at $z=6.5$-7.7, and 27 [CII]-emitting QSOs from \citet{venemans_kilo} at $z=6.0$-7.6. 

\citet{Loiacona_alpine} report 12 [CII]-emitter candidates around UV-selected $z=4$-6 galaxies in the ALPINE survey, and \citet{fudamoto_nature} report 2 [CII]-emitting galaxies around UV-selected sources in REBELS.  Around the 27 [CII]-emitting QSOs studied by \citet{venemans_kilo}, 27 line-emitting galaxies are identified, but only 20 of these appear to lie at frequencies $<$2000 km/s from the target [CII]-emitting source and thus are likely at $z\sim6$-7.   In addition to the dust-obscured companions of \citet{fudamoto_nature}, another less obscured companion galaxy is found in the REBELS sample (Schouws et al. 2024, in prep.) that we present for the first time in this work and will also be included in our analysis.

In addition to making use of the [CII]-emitting companion galaxies already identified by searches over ALPINE, REBELS, and $z>6$ QSO samples, we have also executed a search for additional companion galaxies using archival ALMA data.  For this search, we identify ALMA projects from the ALMA Science Archive\footnote{\url{https://almascience.nrao.edu/aq/}} that have targeted the [CII] line in massive galaxies or quasars at  $z \geq$ 6, which lie in ALMA band 6 (211-275 GHz) from $z\sim8$ to $z\sim6$.  We focused on ALMA observations acquired at low or modest spatial resolution ($\gtrsim0.2$ arcsec) to avoid over resolving the [CII] emission, which can frequently be quite spatially extended \citep{fujimoto_19, Fujimoto_20, fudamoto_22}. The examined observations have angular resolutions from $\sim$0.2 to $\sim$1.2 arcsec. A more extensive discussion on the resolutions of the observations can be found in Sec.~\ref{neighbours}. A complete list of examined projects and sources and their corresponding angular resolution can be found in Table~\ref{tab:all_projects} from Appendix~\ref{sec:archive}. In total we examine the ALMA data cubes for 55 targets where [CII] is targeted.  Of those 55, we find clear ($\gtrsim$5$\sigma$) [CII] detections in 34 targets.  We search for companion galaxies in the archival observations of the 34 targets where [CII] is detected. In total 161 targets are examined for companion galaxies when we include the searches done in previous literature as described above.

We obtained the calibrated measurement sets produced by the ALMA observatory pipeline from the European ALMA Regional Centre. Before imaging the measurements sets, the data are time-averaged over a period of 30 seconds to reduce the total data volume (e.g. Schouws et al. 2024, in prep.).  The flux loss after time-averaging the visibilities over 30 seconds is less than 1 percent.  
\footnote{The average fraction of intensity that is maintained after time-averaging over a time interval $\tau_a$ can be approximated by  $R_{\tau} = 1 - 1.08\times 10^{-9} \ \text{s}^{-2} (\frac{\theta}{\theta_{\text{HPBW}}})^2 \tau_\text{a}^2$, with $\mathrm{\frac{\theta}{\theta_{HPBW}}}$ the distance to the phase center with respect to the half-power beamwidth. For details see \citet{Bridle1999}.}

To image the measurement sets we use the Common Astronomy Software Applications (CASA) software for ALMA data \citep{CASA} (version 5.7.0). We use the TCLEAN task with natural weighting, to maximize our sensitivity to companion galaxies. We clean the data to a depth of 2$\sigma$ using `auto-multithresh' \citep{multithresh}. The cell sizes of the images are set such that there are approximately 5 pixels per beam. Moreover, we use the \textit{uvcontsub} task to subtract the continuum emission in the \textit{uv}-space using the default zeroth order polynomial excluding the channels within $3 \times$ the FWHM of the [CII] line from the central source. 

\begin{figure}
	\includegraphics[width=\columnwidth]{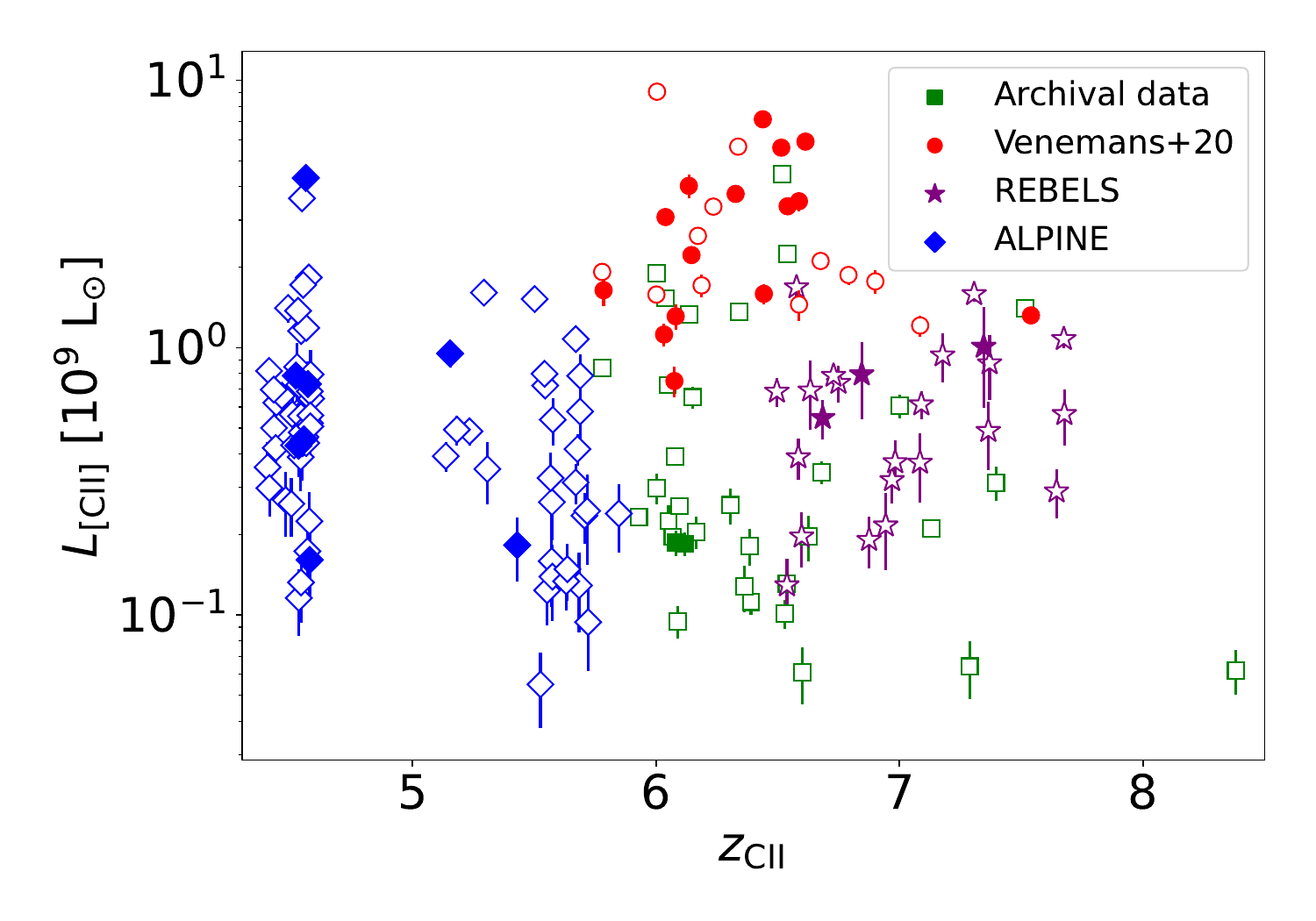}
    \caption{The redshifts and [CII] luminosities of the 161 ALMA targets we use to search for a sample of [CII]-emitting companion galaxies.  The [CII] luminosity of the archival sources is derived based on the peak flux and the redshift is from the peak frequency of [CII]. The red circles correspond to the 27 quasars with [CII] detections where \citet{venemans_kilo} has already described a companion search, while the purple stars and blue diamonds correspond to the 25 [CII]-detected galaxies from REBELS \citep{rebels} and 75 [CII]-detected galaxies from ALPINE \citep{alpine}, where companion searches have been done (\citealt{fudamoto_nature,Loiacona_alpine}; see also Schouws et al. 2024, in prep.). 
    Green squares indicate the 34 [CII]-detected ALMA targets where no searches for [CII]-emitting companion galaxies have thus far been reported in the literature, and we provide for the first time here.
    Filled markers indicate targeted galaxies with serendipitously detected companions, while open markers are targeted galaxies without detected companions.\label{fig:parameter_space}}
\end{figure}

\begin{figure}
    \centering
    \includegraphics[width=\columnwidth]{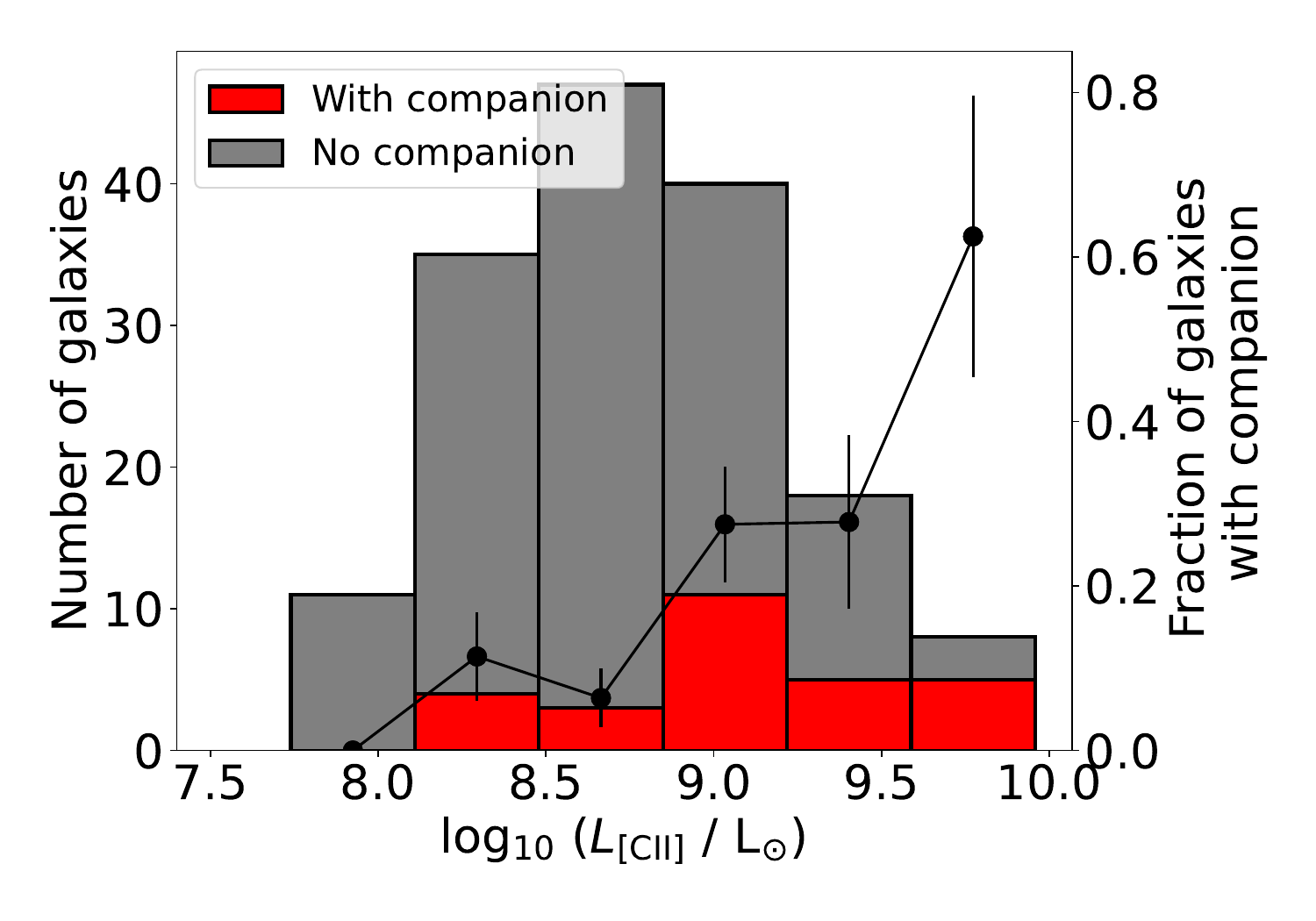}
    \caption{The number of ALMA targets where a companion source is found (\textit{red}) and not found (\textit{grey}) vs. the [CII] luminosity of the primary ALMA target. The fraction of galaxies with at least one [CII]-emitting companion galaxy (\textit{black line}) increases as a function of the [CII] luminosity of the target. The error on the fraction is given by the standard deviation of a binomial distribution.}
    \label{fig:compfrac}
\end{figure}

Figure~\ref{fig:parameter_space} provides a convenient summary of the redshifts and [CII] luminosities of ALMA targets used for the construction of our [CII]-selected sample of companion galaxies. In Figure~\ref{fig:compfrac} the number of targets with companion galaxies as a function of their [CII] luminosity is shown. We see that fraction of galaxies with at least one companion galaxy increases with increasing luminosity.

\subsection{Line-search algorithm}
\label{MF3D}

In the ALMA archival fields where we search for companion sources around [CII]-emitting galaxies and QSOs (green squares in Figure~\ref{fig:parameter_space}), we make use of the Matched Filtering in 3D (MF3D) code \citep{Pavesi_MF3D}.  Given that matched filtering works well for data cubes in which the emission is not very extended and thus well described by 2D Gaussians, MF3D is very effective in searching for line emission from high-redshift galaxies.  To detect line emission MF3D uses 1D Gaussian frequency templates and 2D circular Gaussian spatial templates that are convolved with the ALMA data cubes. 

The advantage of MF3D over normal matched filtering is that MF3D takes into account the spatial correlation present in noise in interferometric datacubes.  The noise in the different channels of the data cubes are nevertheless uncorrelated and therefore the spectral width of the best fitting template matches approximately with the line width of the emission line. A useful reference for more details on the MF3D is \citet{Pavesi_MF3D}.

Detailed analyses have already been done on line identification with MF3D by Schouws et al. (in prep.) for the REBELS survey and it was found that lines with a signal-to-noise ratio (SNR) $\geq$ 6.2 showed a purity in excess of 95 percent. Purity (or fidelity) is defined as $\mathrm{1 - \frac{N_{neg}}{N_{pos}}}$, so at SNR $\geq$  6.2 for each positive peak with that SNR there are only 0.05 negative peaks. In order to decrease the possibility of selecting noise peaks, we set a constraint of SNR $\geq$  6.2. In addition to its identification of positive peaks, MF3D also identifies negative peaks arising from noise. To maximize the robustness of any candidate [CII] line we identify, we require the SNR of a potential galaxy candidate to be larger than the SNR of any negative peaks found in a data set. With SNR $\geq$ 6.2 we expect $<10 \%$ of sources to be affected by flux boosting in excess of $3 \sigma$ and expect a median flux boosting of $<21 \%$ (see \citealt{decarli20}).

The width of the frequency templates used in our search ranges from 100 km s$^{-1}$ to 800 km s$^{-1}$, and the width of the spatial templates ranges from 0 to 3 times the beam size. We define neighbours as serendipitously detected sources with similar redshifts as the targeted (quasar host) galaxy.  We calculate the redshift of the candidates with the central frequency of the [CII] peak.
We consider candidates that have a velocity difference with the target galaxy that is smaller than 2000 km s$^{-1}$ (e.g. \citealt{venemans_kilo}). For our final sample (Sec~\ref{obscuredSFR}), we require sources to show $\leq$ 500 km s$^{-1}$ velocity difference to minimise the impact of contamination from lower-redshift line-emitting sources on our scientific results. The velocity difference is defined by: 

\begin{equation}
    \mathrm{\Delta v} = \frac{z_{\mathrm{comp}} - z_{\mathrm{target}}}{1 + z_{\mathrm{target}}} c 
    \label{eq:veldiff}
\end{equation}

\noindent where $z_{\mathrm{comp}}$ and $z_{\mathrm{target}}$ are the redshifts of the (potential) companion and target galaxy, respectively. At redshift $z\sim6$, a velocity difference of 2000 km s$^{-1}$  (500 km s$^{-1}$) corresponds to $\Delta z \sim 0.05$  ($\Delta z \sim 0.01$). We also apply this criteria to the companion galaxies originally identified by \citet{Loiacona_alpine}, \citet{venemans_kilo}, and \citet{fudamoto_nature} to construct our sample of [CII]-emitting companion galaxies. 

To further confirm that the source is at high redshift and the detected lines do not correspond to a CO transition at lower redshifts, we examine any available imaging data bluewards of the Lyman-alpha transition and look for possible detections. Due to hydrogen between the observer and the galaxy, photons with a rest-frame wavelength bluewards of 1216 \AA \ will be absorbed. The Lyman-alpha transition shifts to bluer wavelengths for sources at lower redshifts. Detections at bluer wavelengths than the Lyman-alpha transition at $z > 5$ indicate that we are observing a low redshift source instead, meaning that the galaxy candidate is a foreground galaxy. We did not find detections in bluer bands for the candidate sources where available. It is worth remarking that this would have only small impact in removing dusty, low-redshift galaxies from our selection due to the faintness of these sources at bluer wavelengths, and our requirement that companion galaxies show a small velocity relative to the primary targets would be the main way of minimising contamination from this type of source (see Sec~\ref{obscuredSFR} and Appendix~\ref{contamination}).

To summarize, a peak found by MF3D is considered to be a potential neighbour when:
\begin{itemize}
    \item The SNR is equal to or larger than 6.2. 
    \item There are no negative peaks with an absolute SNR value greater than or equal to the SNR of the peak.
    \item The velocity difference between the central source and the peak is smaller than or equal to 2000 km s$^{-1}$ (500 km s$^{-1}$ for the final sample).
    \item The galaxy is not detected in filters bluewards of the Lyman-alpha transition.
\end{itemize}

\subsection{Moment-0 and continuum maps}
\label{mom0}

For line candidates found by MF3D that meet the requirements of a neighbour galaxy the moment-0 and continuum maps are made in an iterative way with CASA. An initial moment-0 map is created with the task \textit{immoments}, where we include the frequencies within 2 $\times$ the full width at half maximum (FWHM) of the line. As a starting point we use the FWHM and central frequency of the peak as found with by MF3D. For each source a signal-to-noise weighted spectrum is created only including the pixels that have values higher than 3$\sigma$ in the initial moment-0 map. A Gaussian (or double Gaussian if there is a double peak) is fitted to the spectrum and the FWHM and central frequency are extracted. A new moment-0 map is created with the updated values for the FWHM and the central frequency. This process is repeated and new spectra are made until the values for the FWHM and the central frequency converge. With these final values the definite moment-0 map is created. We also use these values to create the continuum maps. In this case we use TCLEAN on the measurements sets with \textit{specmode=`mfs'}. For the continuum map only the channels outside 2 $\times$ FWHM of the line are used. The [CII] and continuum fluxes are calculated from the moment-0 map and the continuum map, respectively. We fit the sources iteratively with the \textit{imfit} task of CASA until the flux converges. Flux values quoted in this work are integrated fluxes. For companion sources that are not in the phase centre of the ALMA observation, and for which \textit{uvcontsub} will not work properly, we make sure to subtract any significant (3$\sigma$) continuum from the [CII] flux we obtain.

\subsection{Rest-UV photometry}
\label{Rest-uv}

We use rest-UV observations from the Hubble Space Telescope (HST), the UltraVISTA survey (UVISTA), the Dark Energy Camera Legacy Survey (DECaLS), the Canada-France-Hawaii Telescope Legacy Survey (CFHTLS) and Hyper Suprime-Cam (HSC) on the Subaru Telescope.  We focus on rest-UV observations at a rest-frame wavelength of  1500 {\AA} at $z \sim 4$-8.  The approximate sensitivity of the relevant observations is summarized in Table~\ref{tab:filters}.

A large fraction of the companion galaxies found around ALPINE galaxies are located over the COSMOS field, and therefore we can make use of the deep multi-wavelength observations available there.  We use the photometry code MOPHONGO \citep{labbe2006, labbe2010b,labbe2010a, labbe2013,labbe2015} to perform this photometry.  MOPHONGO uses high-resolution data to perform aperture photometry on low-resolution, deep optical data. Moreover, MOPHONGO models both flux from the sources themselves and neighboring objects in the same field such that it can subtract the light originating from nearby neighbors. Therefore, the aperture used with MOPHONGO can be larger and we adopt an aperture diameter of 2 arcsec. We have used the same observational data for the sources in the COSMOS field as used in \citet{stefanon2019}.

For the companion galaxies of REBELS we use the UV measurements in the J-filter of UltraVISTA in COSMOS provided by \citet{fudamoto_nature} and VISTA VIDEO in XMM-LSS (Stefanon et al, in prep.).

The companion sources we utilize from \citet{venemans_kilo} are exclusively observed with HST, some in multiple IR (rest-UV) filters. For these sources the HST observations are stacked in order to increase the signal to noise ratio and to increase the number of detections. This is done by performing an inverse variance weighted stack of the images, where the variance is based on the uncorrelated RMS noise in the image (typically the noise starts behaving as uncorrelated when the image is rebinned by a factor of 10). 

To extract the UV fluxes of the sources with exclusively HST data we cannot use MOPHONGO and use Source Extractor (SExtractor) \citep{sourceextractor}  instead. SExtractor can detect and deblend objects in astronomical images and measure the corresponding fluxes. The algorithm starts by estimating the background signal. This allows the program to compute the relative flux and magnitude of objects. SExtractor has a double imaging mode where an image with the positions of the objects can be supplied. SExtractor then provides the relative magnitudes and fluxes of the source within an aperture diameter of 0.8 arcsec. We choose this aperture as a compromise between encapsulating all the flux of a source and excluding the flux of neighbouring sources. We make sure that we do the necessary aperture corrections for the HST IR filters as the encircled energy in an aperture of 0.8" diameter is $\approx 0.84$\footnote{https://www.stsci.edu/hst/instrumentation/wfc3/data-analysis/photometric-calibration/ir-encircled-energy}.

Moreover, for the galaxies found within this work that do not have HST imaging and are not in the COSMOS field we use the data from DECaLS and perform aperture photometry with SExtractor as explained for the \citet{venemans_kilo} sources.

\begin{figure*}
    \centering
    \includegraphics[width = \textwidth, trim={0 7cm 0 5cm},clip]{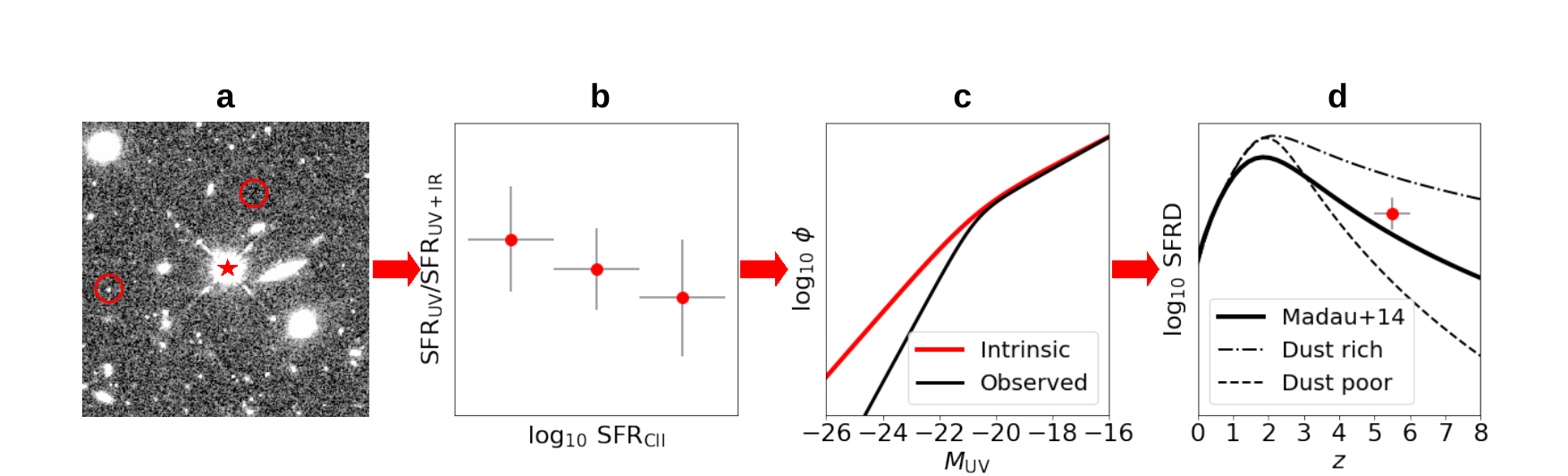}
    \caption{A schematic illustration of our procedure for estimating the contribution that obscured star formation makes to the SFR density at  $z \gtrsim $ 5.  \textbf{a)} Use ALMA observations around bright galaxies and QSOs to select a sample of companion galaxies selected purely through a [CII] emission line (\S\ref{neighbours}).  
\textbf{b)} Leverage observations of these [CII]-emitting companion galaxies in the rest-UV to quantify the unobscured fraction of the SFR (\S\ref{obscuredSFR}: red solid circles).
\textbf{c)} Use a forward modeling procedure to infer the intrinsic LF (red line) that reproduces the observed UV LF (black line) when we apply the obscured fraction of SFR that we find (\S\ref{correctedLF}).
\textbf{d)} Calculate the SFRD at $z \gtrsim $ 5 by integrating the intrinsic luminosity function (\S\ref{correctedSFRD}). We will be able to see if [CII]-selected galaxies imply dust-obscuration fractions consistent with a more dust-poor (dashed line) or dust-rich (dash-dotted line) Universe \citep{casey2018brightest}. For context, the right-most panel includes the \citet{Madau2014} SFR density as a thick solid line.}
    \label{fig:method}
\end{figure*} 

We opted to exclude companions (5 in total) where the radial distance to the QSO target is $<$15 kpc ($\sim 2.6$") as QSOs are particularly bright at rest-UV wavelengths. Our motivation for excluding these sources is due to potential blending between flux from the companion source and the QSO target. For one companion (P167-13C1), we use the measurements from \citet{mazzuchelli} that do PSF modeling and therefore can reliably extract the flux from the companion that is close ($<$15 kpc) to the quasar host galaxy.

\begin{table}
    \centering
    \begin{tabular}{lcc}
    \hline
         Filter name & 5$\sigma$
         depth \\ \hline
         \textbf{Ground-based} \\ \hline
         HSC $z$ &  $25.9^{1}$ \\
         CFHTLS $z$ &  $25.2^{1}$\\
         SSC $z^+$ & $25.0^{1}$ \\
         UVISTA $J$ & $25.4^{1}$\\
         DECaLS $z$  &  $22.5^{2}$ \\
          \hline
         \textbf{HST} \\ \hline
         
         F125W & 27.8$^{3}$  \\
         F140W & 28.0$^{3}$  \\
         F160W & 27.8$^{3}$ \\ \hline
    \end{tabular}
    \caption{The depths of the filters of rest-UV imaging data used in this work.
    \newline $^{1}$ Depths from \citet{stefanon2019}.
    \newline $^{2}$ For an emission line galaxy with half-light radius of 0.45 arcsec.
    \newline $^{3}$ For a point source at $z = 6$ with 30 minutes exposure time. Calculated with the HST ETC.}
    
    \label{tab:filters}
\end{table}

\subsection{Star formation rates}
\label{SFR}
The [CII] luminosity is calculated directly from the [CII] flux using: 

\begin{equation}
   L_{\text{[CII]}} / L_{\odot} = 1.04 \times 10^{-3} \ \nu_{\text{[CII], obs}} \ F_{\text{[CII]}}\  D_{L}^2
   \label{eq:lcii}
\end{equation}

\noindent where $F_{\text{[CII]}}$ is the [CII] flux in $\text{Jy km s}^{-1}$, $\nu_{\text{[CII], obs}}$ the central frequency of the [CII] emission in GHz and $D_L$ the luminosity distance in Mpc \citep{solomon}. 
\citet{delooze2014} derived a relation for the total SFR of $z\sim 0$ galaxies based on their [CII] luminosity. The relation we use in this work is based on the starburst sample from \citet{delooze2014} as this has been found to agree best with high redshift sources (e.g. \citealt{kohandel2019, pallottini2022}): 
\begin{equation}
    \text{log}_{10}\ (\text{SFR} / \text{M}_{\odot}\ \text{yr}^{-1}) = -7.06 + 1.00 \times \text{log}_{10}(L_{\text{[CII]}})
\end{equation}
The $L_{\text{[CII]}}-$SFR relation does not appear to evolve significantly from $z\sim4$-7 to $z\sim0$ \citep{Schaerer_20, Carniani18, carniani20}, and so we will use this relation to calculate the total SFR of sources in our analysis.

With the relative UV magnitudes and redshifts of the sources we calculate the absolute magnitudes and the UV luminosity:
\begin{equation}
    L_{\text{UV}}/L_{\odot} = 10^{\frac{M_{\text{UV}} - M_{\mathrm{sun}}(\lambda)}{-2.5}}
\end{equation}

\noindent where $M_{\text{sun}} (\lambda)$ is the absolute magnitude of the Sun as measured with a filter with wavelength $\mathrm{\lambda \approx 1500} $ \AA \ under the assumption that $\nu L_{\nu} = \text{constant}$.
To convert the UV luminosities to star formation rates we use the following relation from \citet{Madau2014}:
\begin{equation}
    \text{SFR}_{\text{UV}} / \text{M}_{\odot}\ \text{yr}^{-1} = 0.63\times K_{\text{UV}} \times L_{\text{UV}} 
\end{equation}
$K_{\text{UV}}\ (= 1.15 \times 10^{-28}\ \text{M}_\odot\ \text{yr}^{-1}\ \text{erg}^{-1}\ \text{s}\ \text{Hz})$ is the conversion factor from UV luminosity to SFR, while the factor 0.63 is needed to convert from a Salpeter to a Chabrier IMF.

The continuum flux of the sources allows us to constrain the IR luminosity. To estimate the IR luminosity we assume a SED of the dust and integrate the emission between the rest-frame wavelengths of $8$ and $1000 \ \mu \text{m} $. We assume that the dust is heated by star formation only. The dust SED can then be described by a modified blackbody with emissivity index $\beta_{\text{em}} =  2.03$ and $T_{\text{dust}} = 46$ Kelvin following the approach of \citet{Inami2022} \footnote{Assuming Milky Way-like dust and using the median dust temperature for the REBELS galaxies that is found with the approach of \citet{sommovigo2022}. \citet{sommovigo2022} furthermore find the median temperature of the ALPINE galaxies to be 48 K, and therefore our assumption of the dust temperature is also consistent with the ALPINE sample.}. As flux is always measured relative to the CMB we correct the IR luminosity for this.
At higher redshifts, the CMB temperature is higher and contributes more significantly to the flux that is measured. The CMB correction factor is given by: 

\begin{equation}
    f_{\text{CMB}} = \frac{F_{\text{int}} + F_{\text{CMB}}}{F_{\text{int}}} = 1 - \frac{B_\nu(T_{\text{CMB}})}{B_\nu(T_{\text{dust}})}
\end{equation}
where $F_{\text{int}}$ corresponds to the intrinsic flux of the source, $F_{\text{CMB}}$ the flux measured originating from the CMB, $B_\nu$ the emission of a blackbody as a function of temperature and $T_{\text{CMB}} = (2.73\ \textrm{K}) \times (1+z)$ \citep{dacunha2013}. 
The IR luminosity is then calculated by the following equation:
\begin{equation}
    L_{\text{IR}}/L_{\odot} = 7.70\times 10^3 \ \frac{S_{\text{cont,}158\mu m} }{f_{\text{CMB}}} \frac{D_L^2}{1 + z_c}
\end{equation}

\noindent with $S_{\text{cont}, 158\mu m}$ is the continuum flux in mJy measured at a rest frame frequency of $\sim 158\ \mathrm{\mu m}$, $z_c$ the redshift corresponding to the central frequency of the [CII] emission line and $D_L$ the luminosity distance  (analogous to \citet{venemans_kilo}). There is a systematic uncertainty of a factor of $\approx$2-3 on this luminosity as the shape of the SED is unknown \citep{venemans2018}.\footnote{We adopt a systematic uncertainty of a factor of 3, as we observe this systematic scatter in the IR luminosity when we vary the dust temperature by 10K and $\beta_{\mathrm{em}}$ between 1.5-2.5. }
The IR SFR is calculated similarly to the UV SFR using the following equation:
\begin{equation}
    \text{SFR}_{\text{IR}} / \text{M}_{\odot}\ \text{yr}^{-1} = 0.94 \times K_{\text{IR}}  \times L_{\text{IR}} 
\end{equation}

\noindent 
$K_{\text{IR}} = 3.88\times 10^{-44}\  \text{M}_\odot\ \text{yr}^{-1}\ \text{erg}^{-1}\ \text{s}$ and $L_{\text{IR}}$ corresponds to the integrated luminosity from $\mathrm{8}$ to $\mathrm{1000\ \mu m}$ \citep{Murphy11}. The factor 0.94 is used to convert from a Kroupa to a Chabrier IMF.

\begin{figure*}
    \centering
    \includegraphics[width=\textwidth,trim={0cm 6.5cm 0cm 0},clip]{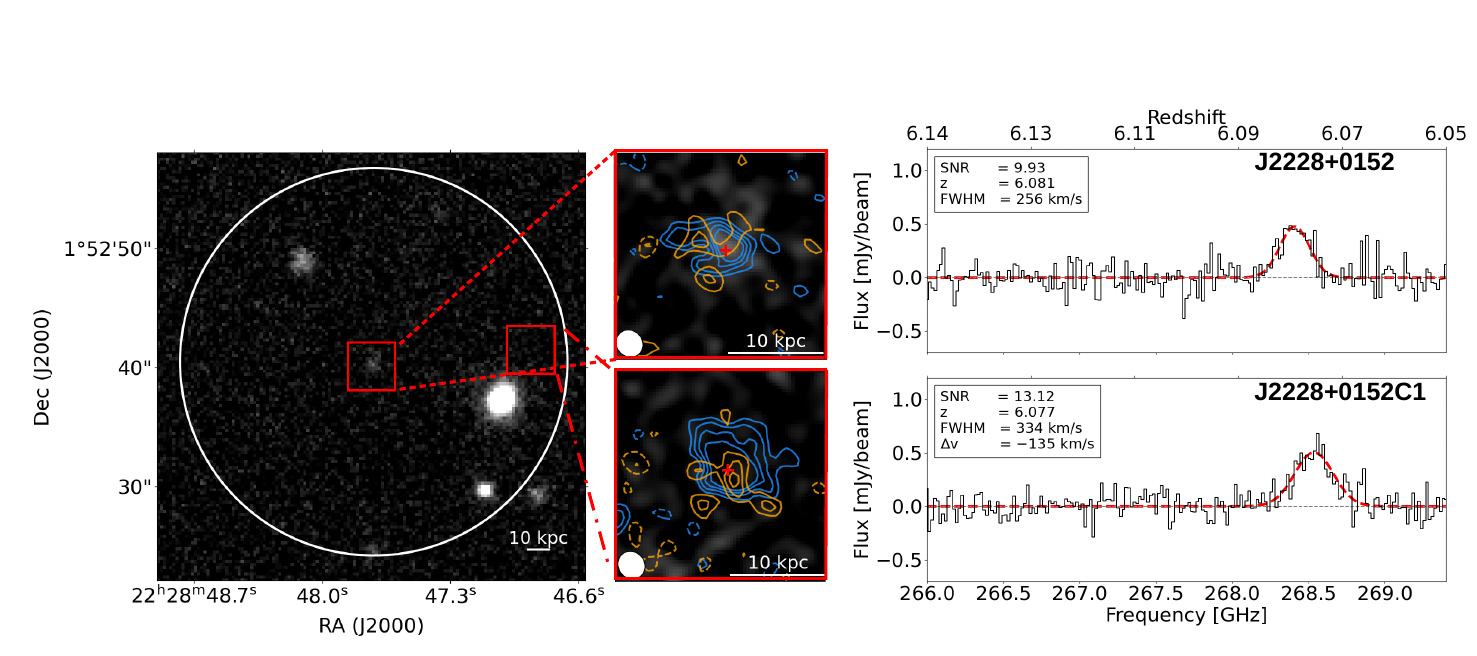}
    \caption{An illustration of a newly discovered [CII]-emitting companion galaxy (\textit{lower panels}) around ALMA QSO target J2228+0152 (\textit{upper panels}).  The leftmost panel shows an image in the rest-UV ($z$-filter from the DECaLS survey) of both the QSO and the [CII]-emitting companion galaxy, along with the FOV of ALMA (\textit{white circle}).  The upper and lower left panels present zoomed-in images of the QSO and companion galaxy.  The solid blue and orange contours indicate those regions detected at $2\sigma$, $3\sigma$, $4\sigma$, $5 \sigma$ in the moment-0 and continuum map, respectively (negative contours are shown with the dashed lines).  The red crosses indicate the peak of the [CII] emission found with MF3D and the ellipses in the left lower corners indicate the ALMA beam size. The upper and lower rightmost panels show extracted spectra (binned at 20 km s$^{-1}$) for the QSO and companion.  The dashed red lines show Gaussian fits to the detected [CII] lines. The SNR reported here is the SNR as found by MF3D.
    \label{fig:J22}}
\end{figure*}

\section{Results}
\label{results}

In this section, we make use of a substantial sample of 18 [CII]-selected companion galaxies (\S\ref{neighbours}) with sensitive rest-UV imaging to characterize the fraction of SFR in $z\sim4$-8 star-forming galaxies that is obscured (\S\ref{obscuredSFR}).  We then make use of the obscuration fraction to derive the intrinsic LF at $z\sim 5$-6 that includes the obscured UV light (\S\ref{correctedLF}).  Finally, we estimate the contribution dust-obscured galaxies make to the SFR density at $z\gtrsim5$ (\S\ref{correctedSFRD}).   Figure~\ref{fig:method} provides a schematic representation of the methodology we use to infer the dust-corrected SFR density at $z\geq 4$ from the present analysis.

\subsection{Newly identified candidate companion galaxies}
\label{neighbours}

The primary objective of this work is to better characterize the obscured fraction of star formation in the early Universe using a [CII]-selected sample of companion galaxies at $z>4$. To do so, we not only make use of [CII]-selected companion sources already identified in various large data sets, e.g. ALPINE and REBELS, but we also conduct searches for [CII]-emitting companion galaxies in the archival ALMA data around $z>6$ targets showing robust [CII] lines.  In this work, we analyze the ALMA cubes for 34 separate sources showing [CII] emission.

With the line-searching algorithm MF3D we found 2 candidate neighbour galaxies in the archival data and a third previously unreported candidate from REBELS. The first two sources were identified as part of ALMA project 2017.1.00541.S with UV imaging observations available from the DECaLS survey.  We summarize the properties of the three neighbour galaxies in Table~\ref{tab:values_nb}.

One of the companion galaxies was identified around quasar host galaxy J2228+0152, at $z_{\mathrm{ly\alpha}} = 6.08$. We find the quasar host galaxy to reside at $z_{\mathrm{[CII]}} = 6.081$ consistent with \citet{matsuoka2018subaru}. For this quasar a neighbour candidate is found at $z_{\mathrm{[CII]}} = 6.077$, which we will call J2228+0152C1. 
The left panel of Fig. ~\ref{fig:J22} shows the position of the neighbour galaxy to the central quasar host galaxy and a zoomed-in image on the location of peaks found by MF3D (red crosses). 
On the right, a section of the corresponding SNR-weighted spectrum of the peaks is shown with a Gaussian fit to the [CII] line. J2228+0152C1 has a more significant [CII] line than the central galaxy but is not detected in any of the filters of the DECaLS survey. The 3$\sigma$ detection this companion galaxy shows in the dust-continuum suggests this source is highly dust-obscured. In Table~\ref{tab:sextractor2} the aperture fluxes are shown as measured with SExtractor along with their corresponding magnitudes. 

The second companion galaxy identified is around the quasar host galaxy J1208-0200 found with the Subaru Telescope at a redshift $z_{\mathrm{ly\alpha}} = 6.2$ \citep{matsuoka2018subaru}. Using the [CII]-line we found the quasar host galaxy to reside at $z = 6.117$, consistent with \citet{Izumi2019}. The neighbour candidate, which we will call J1208-0200C1, is found at $z_{\mathrm{[CII]}} = 6.153$. We show the quasar host galaxy and the neighbour with their corresponding spectrum in Fig.~\ref{fig:J12}. While both galaxies are detected in [CII], only the quasar host galaxy is detected in the rest-UV. J1208-0200C1 is not visible in the rest-UV observations and therefore is likely highly dust-obscured.

In addition to the previously presented obscured companion galaxies by \citet{fudamoto_nature}, another less obscured companion source is found in the REBELS sample (Schouws et al. 2024, in prep.). In contrast to the sources from \citet{fudamoto_nature} that are not detected in the rest-UV, the companion to REBELS-39 has a 2$\sigma$ detection in the J-band (0.23 $\pm$ 0.11 $\mathrm{\mu Jy}$). In Fig.~\ref{fig:R39} we show this serendipitously identified source along with its target galaxy in a consistent manner as we have done for the sources from archival data.

\begin{table*}
    \centering
    \caption{The properties of [CII]-emitting companion galaxies identified for the first time here.}  
    \label{tab:values_nb}
    \begin{tabular}{l c c c c c c c}
        \hline
        Source & $z\mathrm{_{[CII]}}$ & RA (ICRS) & Dec (ICRS) & $\nu_\text{c} $ (GHz) & FWHM (km s$^{-1}$) & $\mathrm{F_{\text{[CII]}} }$ (Jy km s$^{-1}$) & S$\mathrm{_{\text{cont}}}$ (mJy) \\
        \hline
         J1208-0200C1 & $6.1530 \pm 0.0003$ & 12:08:58.46 & $-$02:00:32.85 & $265.70 \pm 0.015$ & $256 \pm 40$ & $0.312 \pm 0.082 $ &  $ 0.072 \pm 0.035 $ \\
         J2228+0152C1 & $6.0774 \pm 0.0002$ & 22:28:46.83  & +01:52:41.55  & $268.54 \pm 0.01$ & $334 \pm 30$ & $0.352\pm 0.058$ & $0.188 \pm 0.029$\\ 
         REBELS-39C1 & 6.8383 $\pm$ 0.0004 & 10:03:05.31 & +02:18:44.62 & 242.47 $\pm$ 0.01 & 260 $\pm$ 40 & 0.458 $\pm$ 0.053 &  0.021 $\pm$ 0.016 \\
        \hline
    \end{tabular}
\end{table*}

\begin{table*}
\caption{The properties of the sample of neighbouring galaxies included in our analysis.$^{\dagger}$}
\label{all_properties}
\begin{tabular}{ccccccccccc}
\hline
Source & S/N$^*$ & $z_{\mathrm{[CII]}}$ & $L_{\mathrm{IR}}$ & $\mathrm{SFR_{IR}}$  & $L_{\mathrm{UV}}$ & $\mathrm{SFR_{UV}}$ & $L_{\mathrm{[CII]}}$ & $\mathrm{SFR_{[CII]}}$  & $\mathrm{\Delta v}$ & Used?  \\ 
 &  & & $\mathrm{(10^{11}\ L_{\odot})}$& $\mathrm{(M_\odot\ yr^{-1})}$  & $\mathrm{(10^9\ L_{\odot})}$ & $\mathrm{(M_\odot\ yr^{-1})}$ & $\mathrm{(10^{9}\ L_{\odot})}$ & $\mathrm{(M_\odot\ yr^{-1})}$  & $\mathrm{(km\ s^{-1})}$ &  \\ \hline
\multicolumn{10}{c}{\textbf{This analysis}} \\ \hline
J1208-0200C1$^a$ & 6.5 &  6.15 & $ 2.9 ^{+ 6.0 }_{- 2.3 } $ & $ 41 ^{+ 83 }_{- 33 } $ & $ 61 \pm 30 $ & $ 7.3 \pm 3.4 $ & $ 0.305 \pm 0.080 $ & $ 24.9 \pm 13.2 $ & $1546$ \\
J2228+0152C1 & 13.1 & 6.08 & $ 7.5 ^{+ 15.0}_{- 5.1 } $ & $ 104 ^{+ 209 }_{- 71 } $ & $ 0.4 \pm 29.2 $ & $ 0.1 \pm 3.5 $ & $0.338 \pm 0.056 $ & $ 27.6 \pm 13.5 $ & -135 & \checkmark \\
REBELS-39C1 &  9.4 &  6.84 & $ 1.0 ^{+ 2.2 }_{- 1.0 } $ & $ 14 ^{+ 31 }_{- 14 } $ & $ 410 \pm 196 $ & $ 61 \pm 29 $ & $ 0.521 \pm 0.060 $ & $ 42.7 \pm 20.3 $ & $-252$  & \checkmark \\ \hline
\multicolumn{10}{c}{\textbf{\citet{venemans_kilo}$^d$}}  \\ \hline
J0100+2802C1& 5.9 & 6.32 & $ 1.1 ^{+ 3.3 }_{- 2.6 } $ & $ 15 ^{+ 47 }_{- 37 } $ & $ 69.6 \pm 73.2 $ & $ 13 \pm 13 $ & $ 0.813 \pm 0.132 $ & $ 66.6 \pm 32.5 $ & $-112$ & \checkmark \\
J0305-3150C1 & 6.0 &6.61 & $ 30 ^{+ 61 }_{- 21 } $ & $ 424 ^{+ 851 }_{- 293 } $ & $ 5.9 \pm 9.5 $ & $ 0.9 \pm 1.5 $ & $ 1.23 \pm 0.31 $ & $ 101 \pm 53 $ & $-245$ & \checkmark \\
J0842+1218C1 & 18.9 &6.07 & $ 6.2 ^{+ 12.6 }_{- 4.9 } $ & $ 86 ^{+ 176 }_{- 69 } $ & $ 5.3 \pm 7.0 $ & $ 1.0 \pm 1.2 $ & $ 1.73 \pm 0.12 $ & $ 142 \pm 66 $ & $-431$ & \checkmark \\
J0842+1218C2 & 7.5 &6.06 & $ 4.7 ^{+ 9.6 }_{- 3.8 } $ & $ 66 ^{+ 135 }_{- 53 } $ & $ 34.2 \pm 7.3 $ & $ 6.1 \pm 1.3 $ & $ 0.411 \pm 0.086 $ & $ 33.6 \pm 17.0 $ & $-476$ & \checkmark \\
J1319+0950C2 & 5.6 &6.14 & $ 7.5 ^{+ 15.2 }_{- 5.4 } $ & $ 105 ^{+ 212 }_{- 75 } $ & $ 9.6 \pm 11.9 $ & $ 1.6 \pm 2 $ & $ 0.341 \pm 0.078 $ & $ 27.9 \pm 14.4 $ & 407 & \checkmark \\
P231-20C1 & 6.0 &6.55 & $ 1.4 ^{+ 3.1 }_{- 1.8 } $ & $ 19 ^{+ 44 }_{- 25 } $ & $ -0.1 \pm 6.7 $ & $ 0.0 \pm 1.1 $ & $ 0.203 \pm 0.043 $ & $ 16.6 \pm 8.4 $ & $-1615$ \\
J2054-0005C1 & 5.8 &6.04 & $ 16 ^{+ 33 }_{- 12 } $ & $ 226 ^{+ 459 }_{- 170 } $ & $ 1.8 \pm 4.5 $ & $ 0.3 \pm 0.8 $ & $ 0.522 \pm 0.142 $ & $ 42.8 \pm 22.9 $ & $-26$  & \checkmark \\
J2100-1715C1 & 15.4 &6.08 & $ 145 ^{+ 291 }_{- 98 } $ & $ 2030 ^{+ 4060 }_{- 1370 } $ & $ -6.7 \pm 12.3 $ & $ -1.2 \pm 2.2 $ & $ 7.76 \pm 1.03 $ & $ 636 \pm 305 $ & $-8$ & \checkmark \\
P167-13C1$^c$ & 18.8 &6.51 & $ 7.6 ^{+ 15.3 }_{- 5.3 } $ & $ 106^{+ 213 }_{- 74 } $ & $ 62.4 \pm 25.0 $ & $ 11 \pm 4 $ & $ 1.34 \pm 0.09 $ & $ 109 \pm 51 $ & $-118$ & \checkmark \\
\hline
\multicolumn{10}{c}{\textbf{\citet{Loiacona_alpine}}} \\ \hline
D\textunderscore C\textunderscore 842313$^a$ & 35.5 & 4.54 & $ 208 ^{+ 417 }_{- 139 } $ & $ 2910 ^{+ 5820 }_{- 1940} $ & $ 94.2 \pm 15.7 $ & $ 14 \pm 2 $ & $ 4.56 \pm 0.14 $ & $ 373 \pm 172 $ & $-829$ \\
D\textunderscore C\textunderscore 665626& 12.5 &4.58 & $ 8.7 ^{+ 17.6 }_{- 6.1 } $ & $ 122 ^{+ 245 }_{- 85 } $ & $ 11.7 \pm 12.3 $ & $ 1.7 \pm 1.8 $ & $ 0.716 \pm 0.069 $ & $ 58.6 \pm 27.6$ & $-23$ & \checkmark \\
v\textunderscore c\textunderscore5101209780 & 13.0 & 4.57 & $ 1.6 ^{+ 3.5 }_{- 1.9 } $ & $ 22 ^{+ 50 }_{- 26 } $ & $ 301 \pm 16.3 $ & $ 45 \pm 2 $ & $ 1.56 \pm 0.18 $ & $ 128 \pm 61 $ & $-56$& \checkmark \\
D\textunderscore C\textunderscore 818760& 10.3 &4.57 & $ 8.2 ^{+ 16.4 }_{- 5.6 } $ & $ 114 ^{+ 229 }_{- 78 } $ & $ 9.1 \pm 16.0 $ & $ 1.4 \pm 2.4 $ & $ 0.390 \pm 0.038 $ & $ 31.9 \pm 15.0 $ & 241 & \checkmark \\
D\textunderscore C\textunderscore 787780 & 15.5& 4.51 & $ 8.8 ^{+ 17.7 }_{- 6.0} $ & $ 123 ^{+ 248 }_{- 84 } $ & $ 472 \pm 13.9 $ & $ 71 \pm 2 $ & $ 1.43 \pm 0.09 $ & $ 117 \pm 54$ & $-107$ & \checkmark \\
D\textunderscore C\textunderscore 873321 & 8.5 & 5.16 & $ -0.2 ^{+ 3.4 }_{- 3.3 } $ & $ -3 ^{+ 47 }_{- 47 } $ & $ 72.2 \pm 16.8 $ & $ 9.7 \pm 2.3 $ & $ 1.86 \pm 0.23 $ & $ 152 \pm 73 $ & 56 & \checkmark \\
D\textunderscore C\textunderscore 378903 & 8.1 &5.42 & $ 6.3 ^{+ 12.9 }_{- 4.8 } $ & $ 89 ^{+ 180 }_{- 67 } $ & $ 49.3 \pm 14.2 $ & $ 6.4 \pm 1.8 $ & $ 0.637 \pm 0.049 $ & $ 52.2 \pm 24.4 $ & $-294$ & \checkmark \\
v\textunderscore c\textunderscore5100822662 & 6.6 &4.52 & $ 3.7 ^{+ 7.6 }_{- 3.0 } $ & $ 52 ^{+ 107 }_{- 42 } $ & $ 26.5 \pm 5.9 $ & $ 4 \pm 1 $ & $ 0.363 \pm 0.006 $ & $ 29.7 \pm 13.7 $ & $-8$ & \checkmark \\
D\textunderscore C\textunderscore 859732 & 6.2 & 4.54 & $ -4.8 ^{+ 11.3 }_{- 6.9 } $ & $ -67 ^{+ 158 }_{- 97 } $ & $ -1.1 \pm 9.9 $ & $ -0.2 \pm 1.5 $ & $ 0.59 \pm 0.13 $ & $ 48.3 \pm 24.7 $ & 369 & \checkmark \\ \hline
\multicolumn{10}{c}{\textbf{\citet{fudamoto_nature}}} \\ \hline
REBELS-12C1$^b$  &  8.2 & 7.35 & $ 3.6 ^{+ 7.2 }_{- 2.6 } $ & $ 50 ^{+ 101 }_{- 36 } $ & $ 169 \pm 253 $ & $ 24 \pm 35 $ & $0.645 \pm 0.072 $ & $ 52.8 \pm 25.0 $ & 118 \\
REBELS-29C1 &  10.3 &6.68 & $ 8.2 ^{+ 16.5 }_{- 5.6 } $ & $ 115 ^{+ 230 }_{- 77 } $ & $ -50.7 \pm 67.7 $ & $ -7.7 \pm 10.3 $ & $ 0.778 \pm 0.102 $ & $ 63.7 \pm 30.5 $ & $-118$ & \\
\hline
\end{tabular}
 \begin{flushleft}
    $^{\dagger}$   A galaxy is used to infer the obscured star formation rate only if the UV observations are deep enough (1$\sigma $ depth would detect the UV if the galaxy is maximally 85 percent obscured) and it is likely to be at high redshift ($\mathrm{\Delta v \leq 500\ km\ s^{-1}}$). Only galaxies with a $\mathrm{\checkmark}$ in the rightmost column are used to compute the unobscured fraction of the SFR as function of the total SFR.\\     
    $^{*}$ S/N of the [CII] line as reported by MF3D. For the \citet{venemans_kilo} sources this is the S/N found by their independent line-search method.\\  
        $^a$ Excluded from the analysis because of a large velocity offset $\mathrm{|\Delta v|>500}$ km s$^{-1}$ relative to the primary target.\\
        $^b$ Excluded from the analysis because the source lacks sufficiently deep imaging data to probe rest-UV SFRs. \\
        $^c$ We have used the rest-UV fluxes from \citet{mazzuchelli} as explained in \ref{Rest-uv}.\\
        $^d$ The only sources included in this list are those with sensitive rest-UV observations and which lie at least 15 kpc away from the central QSO target (\S\ref{Rest-uv}).  We have adopted the IR and [CII] fluxes as found in \citet{venemans_kilo} to compute the luminosities and SFRs.\\
   \end{flushleft}
\end{table*}

\subsection{[CII]-selected sample}
\label{obscuredSFR}

With the present compilation of companion galaxies -- including the newly identified candidates from \S\ref{neighbours} -- our goal is to determine the fraction of SFR in galaxies that is obscured vs. the total SFR in galaxies.  We do this as a function of the SFR given the well known dependence of dust obscuration on the stellar mass \citep[e.g.][]{Reddy2006,Pannella2009,Pannella2015,whitaker2018,Bouwens2020}.  This can be done by looking at the ratio between the unobscured (UV) star formation and the total star formation rate of the neighbouring galaxies found in [CII].  Our use of [CII]-selected galaxies ensures that we are not biased to unobscured galaxies. 

Table~\ref{all_properties} summarizes the properties of the 23 neighbour candidates we examined.  The full sample consists of the three new companion galaxy candidates at $z>6$ found in archival data and around REBELS-39 and the already known companion galaxies from \citet{Loiacona_alpine}, \citet{venemans_kilo}, \citet{fudamoto_nature}.  We examined the companion galaxies that have rest-UV observations and have a velocity difference of $\mathrm{\Delta v < 2000\ km\ s^{-1}}$ with their central galaxies.  In some cases galaxies are not detected in the rest-UV or IR continuum at 3$\sigma$, as can be seen by the errors on their properties. Out of the 23 neighbour candidates, 11 (17) sources have  3$\sigma$ (2$\sigma$) continuum detections. Moreover, out of the 23 sources, 7 (10) are detected with 3$\sigma$ (2$\sigma$) in the rest-UV.

To constrain the obscured fraction, we only consider sources which feature UV observations sensitive enough to detect sources at $1\sigma$ even if 85 percent of the UV light is obscured. Furthermore, we decide to restrict our analysis to those companion sources that have $\mathrm{\Delta v \leq 500\ km\ s^{-1}}$. Using the full companion sample from \citet{venemans_kilo} we quantify the probability of a companion galaxy being a lower redshift interloper as function of  $\mathrm{\Delta v}$ in Appendix~\ref{contamination}. We assume that sources with $\mathrm{\Delta v > 2000\ km\ s^{-1}}$ are contaminants from lower redshift (e.g. CO line emitters). We quantify that for a velocity difference of  $\leq$ 500 km s$^{-1}$ the probability of the neighbour candidate being at low redshift is low ($\sim$13\%).
Applying all of the above criteria, we arrive at a sample of 18 galaxies. 

We did not include the known systems of QSOs host galaxies and massive galaxies with companions such as PJ308-21 \citep{Decarli_2017, decarli2019} and BRI 1202-0725 \citep{carilli13}. They have been excluded from this study as our original focus was on companions at $z > 6$, using the \citet{venemans_kilo} sample. Since we focus on companion galaxies selected by their [CII] emission, newly detected companions using JWST (e.g. \citealt{kashino23, Wang23}) are likewise not included.

\begin{figure}
    \centering
    \includegraphics[width=\columnwidth]{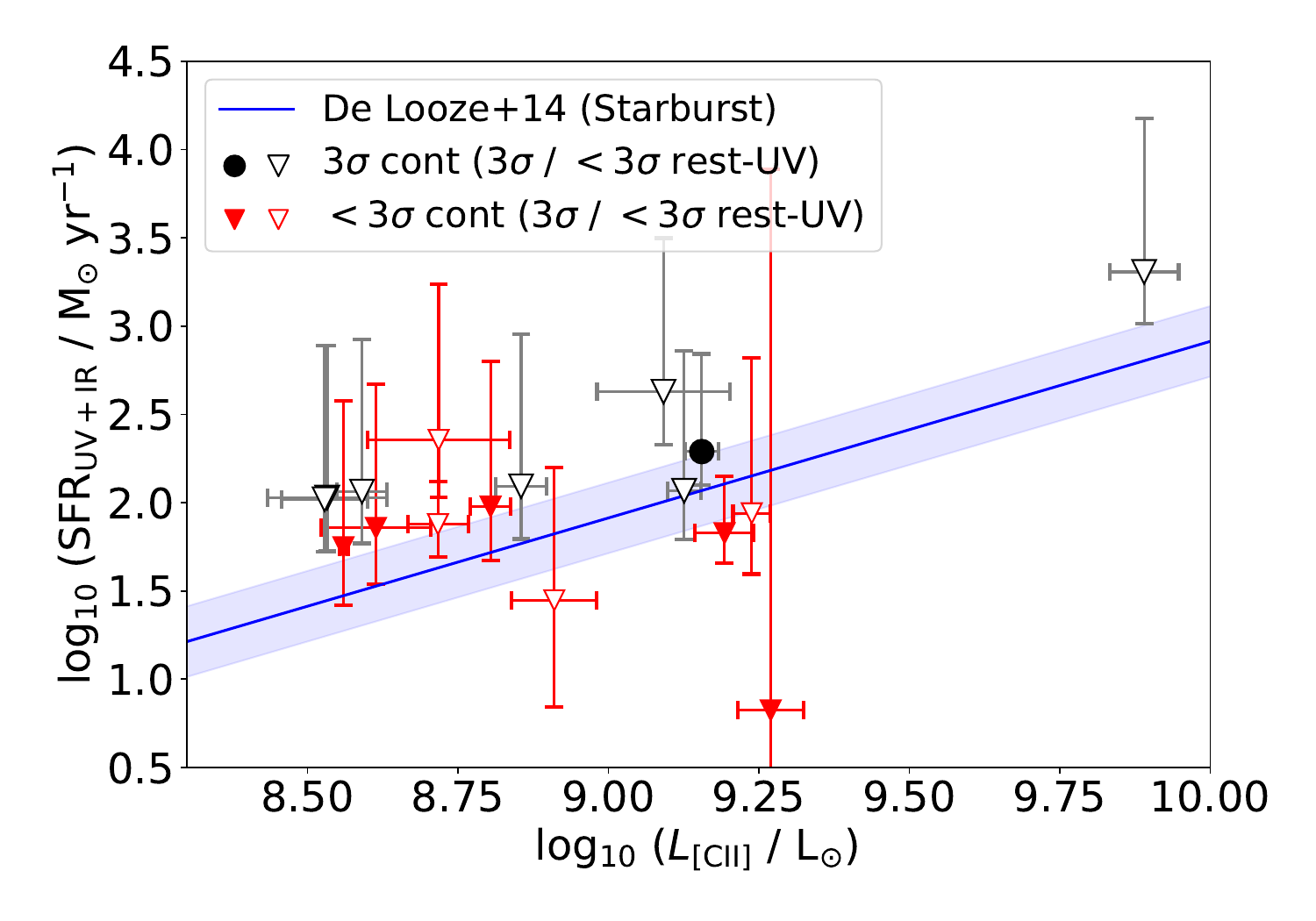}
    \caption{The total SFR $\mathrm{\left(SFR_{UV +IR} \right)}$ as function of the [CII] luminosity of the 18 neighbour galaxies used in this work. In blue the starburst galaxy relation derived by \citet{delooze2014} and the 1$\mathrm{\sigma}$ dispersion is shown. The black filled and non-filled symbols indicate sources for which the dust continuum is detected (3$\sigma$), the red symbols show the continuum non-detections. The filled data points are the galaxies that have a 3$\sigma$ rest-UV detection, empty symbols indicate galaxies undetected in the rest-UV. We include the uncertainty (factor of 3) on the IR luminosity due to the uncertain shape of the dust SED. }
    \label{fig:looze}
\end{figure}

As a check on our analysis (also given the possibility of a small fraction of foreground interlopers which could enter our selection: Appendix~\ref{contamination}), we compare the SFRs we compute based on flux measurements in the rest-UV and IR continuum and the [CII] luminosity based on the detected [CII] lines.  We present the results of this comparison in Figure~\ref{fig:looze}. In blue the starburst relation of \citet{delooze2014} and the 1$\mathrm{\sigma}$ dispersion is shown. The black circles indicate sources in our sample for which the far-IR continuum is detected (3$\sigma$), the red squares correspond to sources with no such continuum detections. Finally, the filled data points correspond to galaxies detected in the rest-UV (3$\sigma$), while the empty symbols indicate galaxies lacking a rest-UV detection. For most sources the error on the total SFR is dominated by the uncertainty introduced in the IR luminosity by the uncertain shape of the dust SED. We see from this figure that our sample agrees with the relation from \citet{delooze2014}, although for some data points we unfortunately have large uncertainties.

\begin{figure}
    \centering
    \includegraphics[width =\columnwidth,trim={0cm 0.1cm 0cm 0cm},clip]{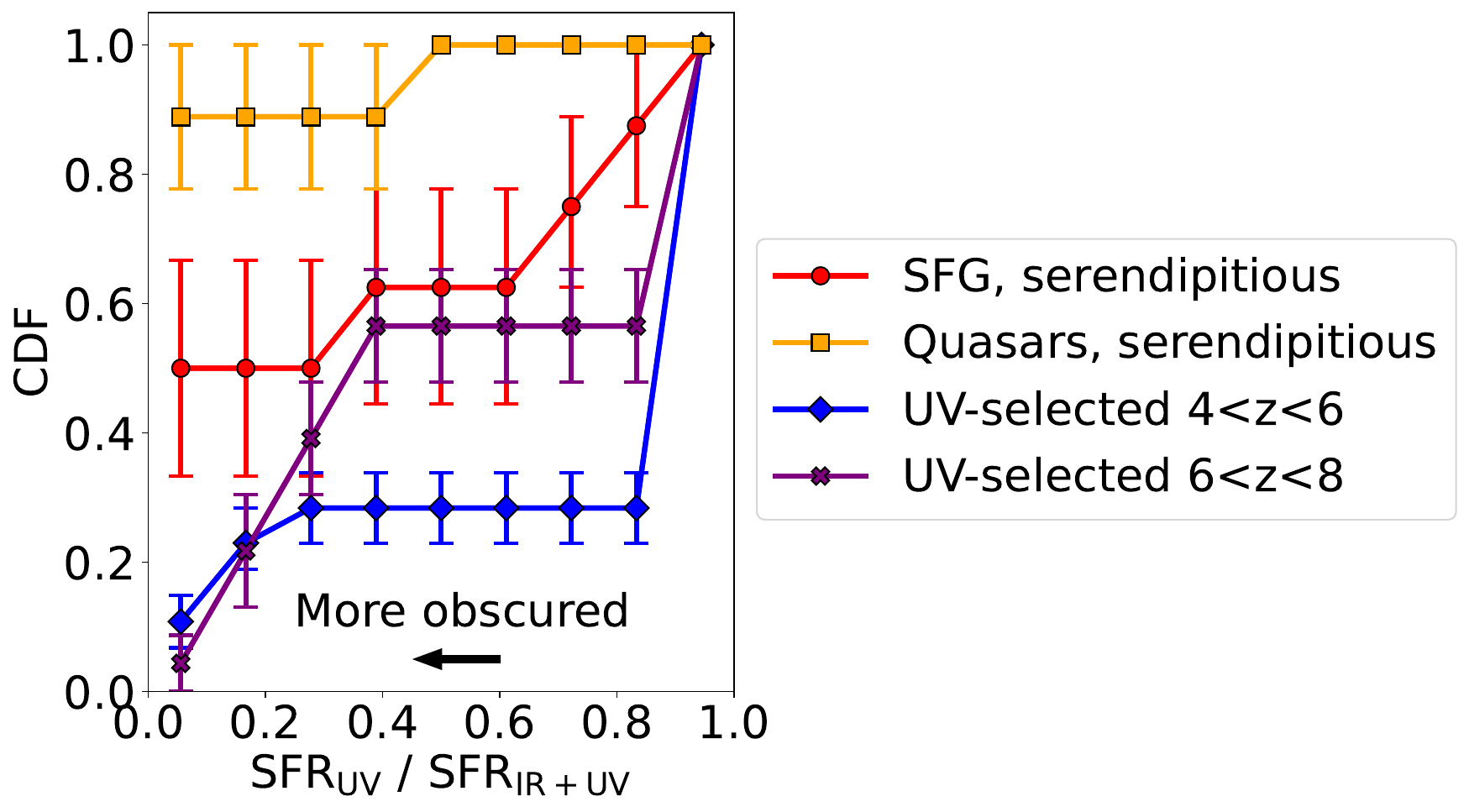}
    \caption{The cumulative distribution function (CDF) of the fraction of unobscured star formation rate for companion galaxies around SF galaxies (red) and QSOs (orange). The CDFs for the UV-selected samples of ALPINE \citep{alpine} and REBELS \citep{rebels} are shown in blue and purple, respectively. The errors are obtained with bootstrapping. The companions around QSOs have lower SFRs in the rest-UV, suggesting that galaxies found in the neighborhoods around QSOs are dustier than around bright star-forming galaxies (but we stress that the current statistics are still small).}
    \label{fig:difference}
\end{figure}

Since our sample of [CII]-emitting galaxies are selected from two different environments, around QSOs and SF galaxies, we examine whether the environment appears to have an impact on the obscured fraction of SFR. In Fig.~\ref{fig:difference} the cumulative distribution function is shown for the fraction of unobscured SFR. Fractions of unobscured SFR that have negative values are treated as having a value of 0 while fractions that are larger than 1 are treated as having a value of 1. We observe that the companion galaxies around QSOs in our sample appear to be significantly more obscured. We therefore decide to create sub samples of the companion galaxies, separated by their environment. The SF galaxy companion sample consists of companions found around ALPINE and REBELS galaxies and has a mean redshift of $z_{\text{SF}} = 5.14$. The sample of companions around QSOs consists of the two companions shown in \S\ref{neighbours} and also companions from \citet{venemans_kilo} and has a mean redshift of $z_{\text{QSO}} = 6.21$. 

\subsection{Obscured star formation}
\label{sec:results_obs}

\begin{figure*}
    \centering
    \includegraphics[width =\textwidth]{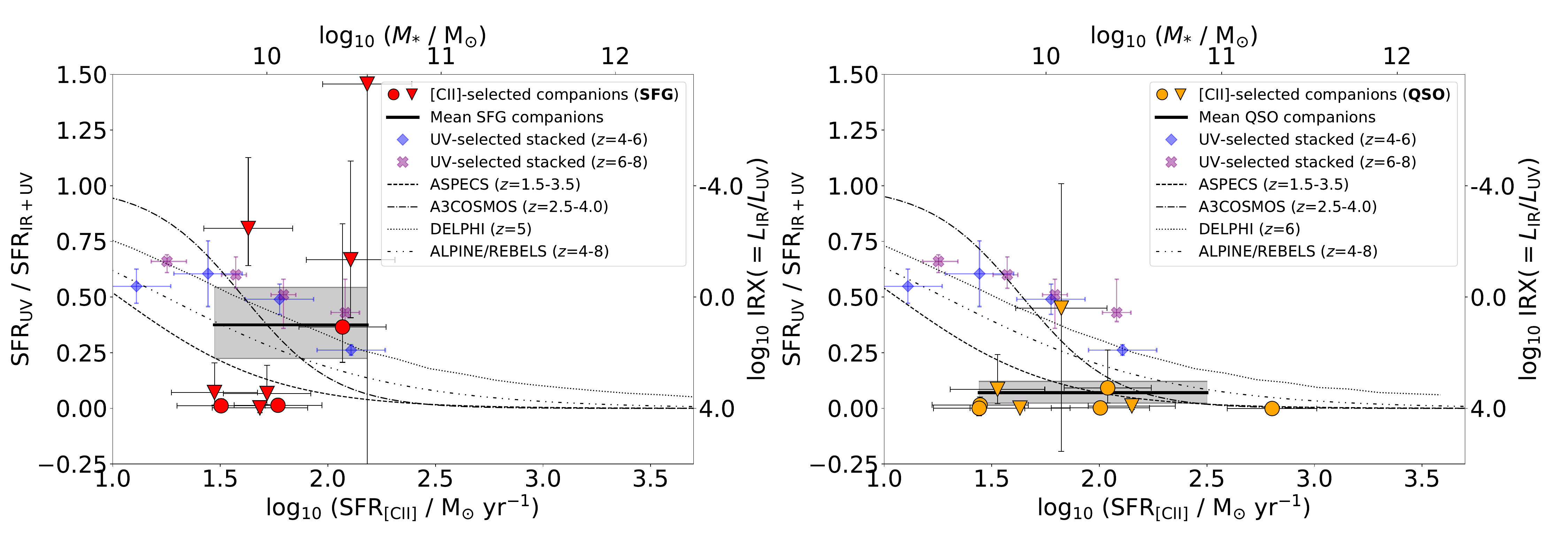}
    \caption{The unobscured fraction of the SFR as a function of the total SFR as measured from the [CII] emission from companions galaxies around SF galaxies (\textit{left}) and companions galaxies around QSOs (\textit{right}). The [CII]-selected galaxies are shown in red and orange for these samples, respectively, and the mean unobscured fraction by the black solid line. The width of the shaded grey region indicates the 16th and 84th percentiles of the mean fraction. Galaxies that are not detected in the far-IR are indicated by an upside down triangle. The blue diamonds show the relation derived for the stacked UV-selected ALPINE galaxies \citep{bethermin2020}, while the purple crosses show the relation based on the stacked REBELS sample \citep{algera23}. The dotted lines show the $z \sim 5$ and $z \sim 6$ relations from \citet{mauerhofer} from the DELPHI semi-analytic model calibrated with ALPINE and REBELS galaxies. Also shown are the IRX-$M_{*}$  relations from ASPECS \citep{Bouwens2020},  A$^3$COSMOS \citep{fudamoto20} and ALPINE/REBELS \citep{bowler23} assuming the main sequence relation from \citet{speagle14} for $z = 5$ and $z = 6$. The top axis shows the stellar masses assuming the same relation from \citet{speagle14} at $z = 5$ and $z = 6$.}
    \label{fig:obs}
\end{figure*}

We have chosen to characterize the fraction of obscured star formation in galaxies as a function of their total SFRs as measured by [CII] -- given our ability to readily quantify this SFR and not e.g. the stellar mass. In Figure~\ref{fig:obs}, we show individual constraints on the obscured fraction as a function of the [CII]-derived SFRs for both the companions near QSO and SF galaxies. If a galaxy is not detected (not a $3\sigma$ detection) in the continuum and/or rest-UV we use the flux as measured (no upper limits)\footnote{We note that the median SNR of the dust continuum observations for our sample is $\sim 3$.}. 

Previous work by e.g. \citet{Bouwens2020, fudamoto20, bowler23} on UV-selected galaxies show that the infrared excess (IRX)  increases for larger stellar masses ($M_{*}$). As the relation between the fraction of obscured or unobscured star formation and total SFR is unknown, we use the IRX-$M_{*}$ relation as a starting point.  We can convert IRX to the fraction of  unobscured star formation rate using:
\begin{equation}
\label{eq:obs}
    \begin{aligned}
        \mathrm{\frac{SFR_{UV}}{SFR_{UV+IR}}} &= \mathrm{\left(1 + \frac{SFR_{IR}}{SFR_{UV}}\right)^{-1}}  = \left(1 + C_{1} \times \frac{L_{\text{IR}}}{L_{\text{UV}}}\right)^{-1} \\ 
        & = \left(1 + C_{1}\times \mathrm{IRX} \right)^{-1} \\
    \end{aligned}
\end{equation}

\noindent with $C_{1} = K_{\text{IR}} / K_{\text{UV}}$ and IRX = $L_{\text{IR}}/L_{\text{UV}}$.

We then use the dependence of IRX on $M_{*}$ ($\mathrm{IRX} \propto M_{*}^{B}$) and assume the best-fit main sequence relation ($M_{*} \mathrm{\propto  SFR_{tot}}$) of \citet{speagle14}  (eq. 29) to have an unobscured star formation rate fraction as a function of total SFR:  
\begin{equation}
\label{eq:obscured}
   \frac{\text{SFR}_{\text{UV}}}{\text{SFR}_{\text{UV+IR}}} =   (1 +  A \times \text{SFR}_{[\text{CII]}}^{B})^{-1}
\end{equation}
where $A$ and $B$ are free parameters. 

This relation is fitted to both the SF galaxy and QSO companion [CII]-selected samples. We find $B$ to be consistent with zero as there is no clear dependence of the obscuration fraction on the total SFR for the sources in our sample. Therefore for simplicity we fix $B$ to zero. Our fit can then be simplified by taking the mean of the (un)obscured fraction of the samples. The median (and 16th/84th percentiles) we find for the mean unobscured fraction of the SF galaxy and QSO sample when we generate $10^6$ bootstrap samples are $\mathrm{f_{unobscured, SFG}} = 0.37^{+0.17}_{-0.15}$ and $\mathrm{f_{unobscured, QSO}} = 0.07^{+0.05}_{-0.05}$. Adopting the functional form given in Eq.~\ref{eq:obscured}, when fixing $B = 0$, we find $A_{\mathrm{SFG}} = 1.6^{+1.7}_{-0.8}$ and $A_{\mathrm{QSO}} =11.6^{+15.6}_{-5.4} $ (Table~\ref{tab:summary}). The unobscured fractions for the individual galaxies together with the mean unobscured fractions are shown in Fig~\ref{fig:obs}.
Errors shown on the individual data points result from the formal errors on the star formation rates, the scatter on the IR luminosity introduced by the uncertain dust temperature and dust emissivity and the scatter on the [CII] SFR. The top axis shows stellar masses assuming the main sequence relation of \citet{speagle14} at $z = 5$ and $z = 6$. As the galaxies in our sample do not all have a large amount of multi-wavelength observations, we could not securely determine their stellar masses.

For comparison, we also show the UV-selected galaxies of ALPINE with blue diamonds and the UV-selected galaxies of REBELS with purple crosses. Also shown is the relation found by \citet{mauerhofer} at $z \sim 5 -6$ with the DELPHI semi-analytic model baselined against ALPINE and REBELS observations. Moreover we plot the IRX-$M_{*}$ relations from \citet{Bouwens2020}, \citet{fudamoto20} and \citet{bowler23} using Eq.~\ref{eq:obs}. The SF galaxy companion sample suggests an obscuration fraction ($f_\mathrm{{obs, median}}= 63 \%$) similar to those expected from UV-selected samples. The [CII]-selected galaxies suggest a relation for galaxies around QSOs that implies that more SFR is obscured ($f_\mathrm{{obs, median}}= 93 \% $) than what is inferred from UV-selected galaxies (as indicated by the relations from ASPECS, A3COSMOS, ALPINE and REBELS).  The high obscuration fraction of the companion galaxies near QSOs is remarkable. Regardless of the origin of this increased dust attenuation, we clearly observe a difference between QSO and SF galaxy environments and hence we treat the samples separately in the rest of this paper.

In Appendix~\ref{contamination} we have used the \citet{venemans_kilo} sample to estimate the fraction of low redshift contaminants we would expect in our sample. According to these calculations 13$\%$ of our sample of companion galaxies could correspond to lower redshift sources which happened to show CO line emission at approximately the same frequency as [CII] for the main target. We thus look at the effect this would have on the derived mean obscured fraction as shown in Fig.~\ref{fig:obs}. Using the ASPECS sample at $z \sim 1.3$-2.5 from \citet{Bouwens2020} as a reference low redshift sample, we would expect most low redshift emitters to have an IRX $ >4$, which according to Eq.~\ref{eq:obs} would correspond to $\gtrsim 80 \%$ of the star formation to be obscured. We have repeated our fitting procedure while we randomly remove one source from the SF and QSO companion sample that is $>80 \%$ obscured to simulate the effect of excluding a potential contaminant from our sample. We find that the newly calculated unobscured fractions are within $20\%$ of the unobscured fraction described above for the full sample and therefore within the errors on the mean unobscured fraction.

\begin{figure*}
    \centering
    \includegraphics[width=\textwidth,trim=0cm 13.5cm 0cm 0cm,clip]{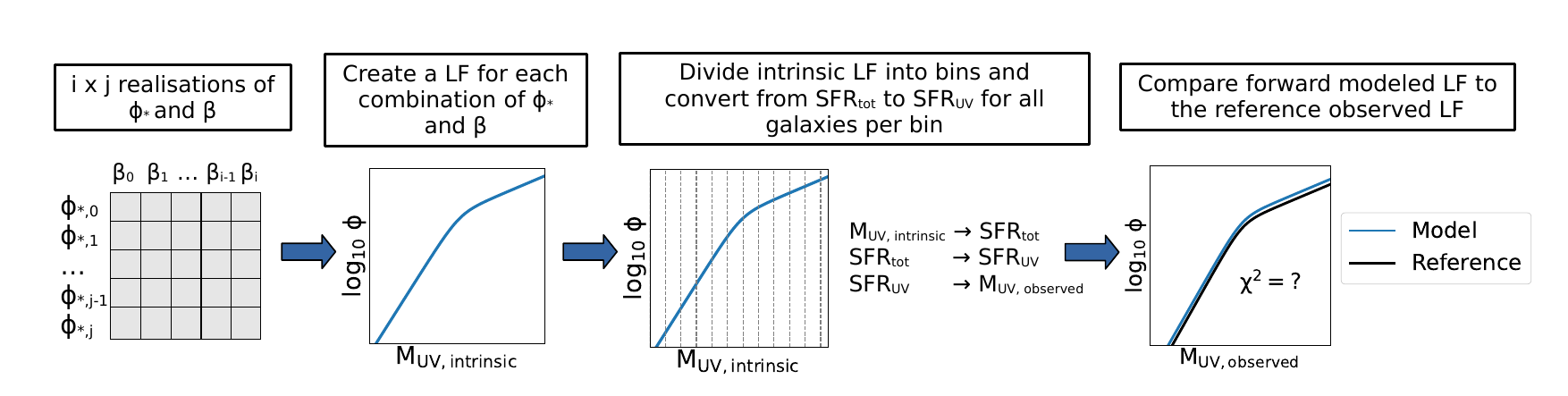}
    \caption{A schematic overview of our forward modeling procedure that is used to recover the intrinsic UV LF. 1) We create i $\times$ j realisations of $\phi_{*}$ and $\beta_{\mathrm{DP}}$, which serve as the input parameters of the intrinsic LF. 2) For every combination of  $\phi_{*}$ and $\beta_{\mathrm{DP}}$ we compute a model LF. 3) We segregate the model intrinsic LF across various bins in magnitudes and for all the galaxies in a bin we convert the total SFR to the UV SFR from the distribution we find in Fig.~\ref{fig:obs}. 4) Finally, we compare the model observed LF to the reference observed LF from \citet{Harikane2022} and compute $\chi^2$.}
    \label{fig:method_LF}
\end{figure*}

\subsection{The Intrinsic UV Luminosity Function}
\label{correctedLF}

\begin{figure*}
    \centering
    \includegraphics[width=\textwidth,trim=5cm 0.2cm 5cm 0cm,clip]{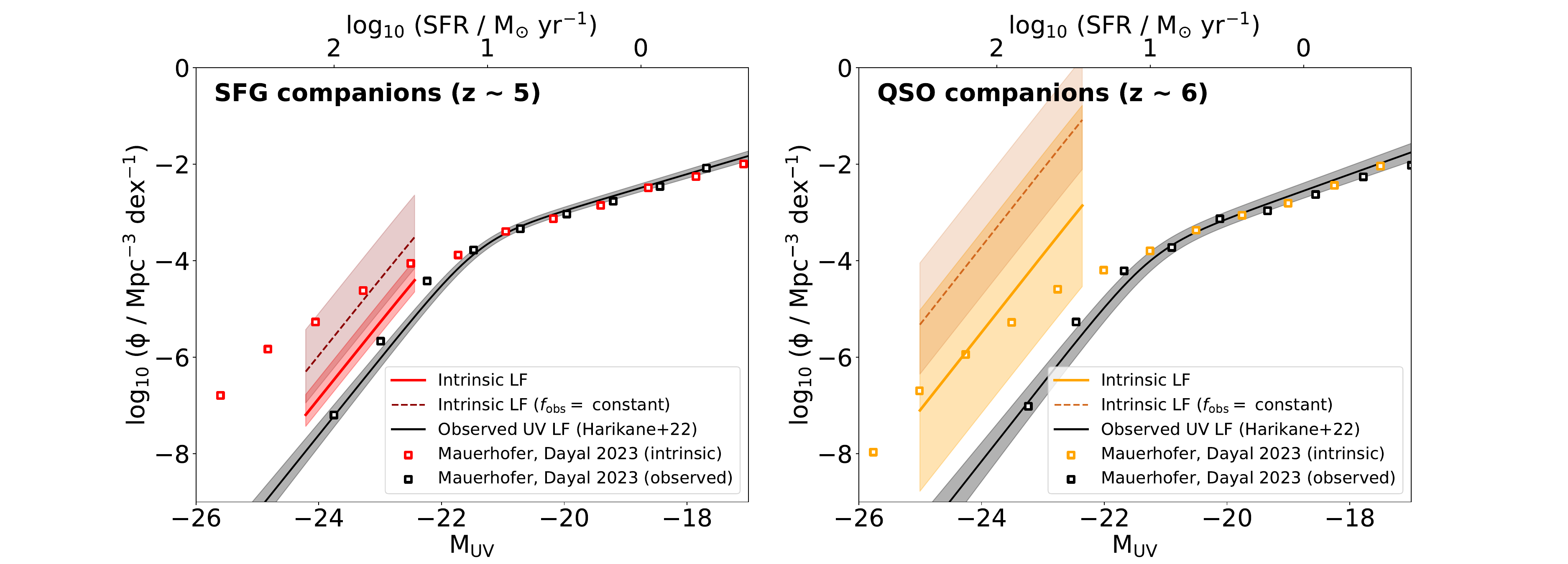}
    \caption{The observed UV luminosity function at $z\sim 5$ (left panel) and $z
    \sim6$ (right panel) from \citet{Harikane2022} ({\it black line}) and the intrinsic UV luminosity function we obtain with the dust obscuration correction procedure we describe in this paper based on the characteristics of companion galaxies around SF galaxies and QSOs ({\it red and orange line, respectively}).  The width of the shaded regions ({\it shown in red and orange}) indicates the 16th and 84th percentiles of the observed and intrinsic LFs. Shown with dashed lines are the intrinsic LFs when instead of using the distribution of obscured fractions we adopt the mean obscuration instead (see Sec.~\ref{correctedLF}). Clearly, the dust corrections we infer here substantially increase the volume density of intrinsically luminous sources at $z \gtrsim 5$. This is also shown by the predictions of the intrinsic and observed UV LF from \citet{mauerhofer} that are calibrated with ALPINE and REBELS data at  $z\sim 5$ and $z\sim6$. We only infer the intrinsic LF over the SFR range sampled by our companion galaxy sample.}
    \label{fig:UV_luminosity_function}
\end{figure*}

Having established the unobscured star formation rate from the [CII]-selected neighbours, we use this characterization to recover the intrinsic UV luminosity function from the observed UV LF. The observed LF differs from the intrinsic LF in that both the luminosity of sources and their inclusion in the LF can be impacted by dust obscuration. For our reference UV LF from the observations, we have selected to use the double power law galaxy UV LF from \citet{Harikane2022} that utilizes the very wide-area ($\sim$300 deg$^2$) Hyper Suprime-Cam survey. 

A double power law luminosity function is given by:

\begin{equation}
    \phi(M) = \frac{\phi_{*}}{10^{0.4(M-M_{*})(\alpha_{\mathrm{DP}} + 1)} + 10^{0.4(M-M_{*})(\beta_{\mathrm{DP}} + 1)}}
    \label{eq:DPL}
\end{equation}

with $\phi_*$ the normalization density, $M_*$ the characteristic magnitude, $\alpha_{\mathrm{DP}}$ the faint-end slope and $\beta_{\mathrm{DP}}$ the bright-end slope. To recover the intrinsic UV LF, we use a forward modeling methodology. Fig.~\ref{fig:method_LF} shows a schematic overview of our procedure. The steps of our forward modeling procedure are the following:

\begin{enumerate}
  \item We allow the parameters of the input intrinsic LF ($\phi_{*}$ and $\beta_{\mathrm{DP}}$) to span a sufficient range of parameters such that we are able to reproduce the observed LFs with the distribution of obscured fractions of SFR. $M_*$ and $\alpha_{\mathrm{DP}}$ are fixed to the determinations obtained by \citet{Harikane2022}. 
   We create 1000 x 100 realisations of $\phi_*$ and $\beta_{\mathrm{DP}}$.
  \item For each combination of $\phi_*$ and $\beta_{\mathrm{DP}}$ we create the (intrinsic) LF. Then the intrinsic LF is divided up into various bins of magnitudes. For all galaxies within a bin we have the intrinsic magnitudes and therefore $\text{SFR}_{\text{tot}}$. To convert this to observed magnitudes and $\text{SFR}_{\text{UV}}$ we apply the distribution of $f_{\text{unobscured}}$ we find for our sample and use: $\text{SFR}_{\text{UV}} = \frac{\text{SFR}_{\text{UV}}}{\text{SFR}_{\text{tot}}} \times \text{SFR}_{\text{tot}} = f_{\text{unobscured}} \times \text{SFR}_{\text{tot}}$.
  \item We then sum the contribution from all dust-corrected galaxies in the previous step to arrive at a model UV LF and compare it to the observed UV LF of \citet{Harikane2022}. We do this by obtaining the $\chi^2$ value. To ensure that our derived LF results are fully guided by our observational constraints, we only infer the intrinsic LF for the SFR range that is covered by our sample of companion galaxies.
\end{enumerate}

We repeat this procedure 1000 times while we randomly sample from the distribution of $f_{\text{unobscured}}$ we find in Sec.~\ref{sec:results_obs} allowing  $f_{\text{unobscured}}$ to vary within its uncertainties. We then correct the SFR of each galaxy in the bin of the UV LF with each of the values in the distribution, while we make sure to conserve the number of sources in each bin according to the input intrinsic LF. When we sample from the distribution, we make sure to only correct the intrinsic UV LF with physical values of $f_{\text{unobscured}}$. This means that $f_{\text{unobscured}} < 0 $ is treated as 0 and $f_{\text{unobscured}} > 1$ is treated as 1. Subsequently, this ensures that galaxies with $f_{\text{unobscured}} = 0$ that would be non-detected in rest-UV observations, do not contribute to the UV LF. This method therefore takes into account that the intrinsic UV LF includes the correction on the luminosity of the sources as well as their inclusion in the UV LF according to their detectability. For each trial in these simulations, we save the input LF that has the smallest $\chi^2$ value. Since we use the distribution of obscured fractions directly and therefore the obscuration does not depend on the SFR, we find that we can reproduce the observed LF by only varying $\phi_*$ and keeping $\beta_{\mathrm{DP}}$ constant to the value of \citet{Harikane2022}.
In Fig.~\ref{fig:UV_luminosity_function} we show the intrinsic UV LF obtained by using the distribution of obscured fractions of SFR of the SFG and QSO companion samples with red and orange solid lines, respectively. The solid black lines show the observed UV LF from \citet{Harikane2022} at $z \sim 5$ (left panel) and $z \sim 6$ (right panel) and the shaded grey area shows the 16th and 84th percentiles of $10^5$ realisations of the luminosity function as we vary the input parameters (e.g. faint-end slope) according to the uncertainties specified for the parameters. The intrinsic UV LF is a factor $\sim 6$ and $\sim450$ larger than the observed UV LF based on the distribution of obscured fractions for SFG and QSO companions, respectively.

It is worthwhile comparing these results (utilizing the full distribution of obscured fractions) with what we would obtain if we assumed the same fixed obscured value for every source. We have performed this experiment with the obscured fraction of the SFG companion sample using both our forward-modeling code and analytically where

\begin{equation}
    M_{\text{UV, intrinsic}} = M_{\text{UV, observed}} - 2.5 \times \text{log}_{10}(f_{\text{unobs}})
\end{equation}

\noindent and find consistent results. We show the median intrinsic luminosity function in Fig.~\ref{fig:UV_luminosity_function} with the dark red and dark orange dotted lines based on the mean dust obscuration from the SF galaxy and QSO companion sample, respectively. We see that for both samples this procedure results in significantly more galaxies at the bright end of the luminosity function than our primary procedure of sampling from the distribution of obscured fractions. The large differences are due to the fact that in our primary approach we account for the few high-SFR galaxies in our sample that are only subject to minimal dust obscuration rather supposing that all galaxies are subject to the same obscuration. This makes it possible to reproduce the observed LF with a much smaller overall correction and therefore the simplified approach presented in this paragraph could easily result in a SFRD that is substantially overestimated.

\subsection{The SFRD at $z \gtrsim $ 5}
\label{correctedSFRD}

With the intrinsic luminosity function obtained from forward modeling we can calculate the SFRD at $z \gtrsim $ 5. In this work our SF galaxy and QSO companion sample probe a limited range in SFRs,  $\approx 30 - 150$ and $\approx 30 - 320\ \text{M}_{\odot} \  \text{yr}^{-1}$, respectively. First we obtain the SFRD of these ranges by numerically integrating the LF of both samples that have been obtained using the distribution of $f_{\mathrm{unobscured}}$. We obtain $\text{log}_{10} (\text{SFRD} / \text{M}_{\odot} \  \text{yr}^{-1} \ \text{Mpc}^{-3})$ = $-3.36_{-0.24}^{+0.42}$ for the SF galaxy and $-1.85_{-1.68}^{+2.08}$ for the QSO companion sample. For comparison, if we would integrate the observed UV LF over the same range we obtain $\text{log}_{10} (\text{SFRD} / \text{M}_{\odot} \  \text{yr}^{-1} \ \text{Mpc}^{-3})$ = $-4.11_{-0.18}^{+0.17}$ and $-4.51_{-0.33}^{+0.29}$ at $z \sim 5$ and $z\sim 6$, respectively. If we compare the results from the observed and intrinsic LF solely integrated over the range of SFRs probed in this work, we find a median increase in density of $\approx 6$ and $\approx 
500$. 

We also integrate to $-17$ mag ($0.20\ \text{M}_{\odot} \  \text{yr}^{-1}$), as is more commonly done and will make it easier to compare to literature results. When we integrate over  magnitudes that are not probed by the SFRs in this work we assume the observed UV LF. Essentially we assume that at those magnitudes there is no dust obscuration and therefore this provides us with a lower limit on the SFRD. We obtain $\text{log}_{10} (\text{SFRD} / \text{M}_{\odot} \  \text{yr}^{-1} \ \text{Mpc}^{-3})$ = $-1.81_{-0.08}^{+0.09}$ for the SF galaxy and $-1.55_{-0.40}^{+1.78}$ for the QSO companion sample. If we would integrate the observed UV LF to $-17$ mag we obtain $\text{log}_{10} (\text{SFRD} / \text{M}_{\odot} \  \text{yr}^{-1} \ \text{Mpc}^{-3})$ = $-1.83_{-0.08}^{+0.08}$ and $-1.94_{-0.16}^{+0.16}$ at $z \sim 5$ and $z\sim 6$, respectively. The difference between the observed and intrinsic LF originates from the dust-obscured star formation, therefore the dust-obscured SFRD is $\text{log}_{10} (\text{SFRD} / \text{M}_{\odot} \  \text{yr}^{-1} \ \text{Mpc}^{-3})$ = $-3.44_{-0.33}^{+0.47}$ and $-1.85_{-1.72}^{+2.08}$, which is $2^{+4}_{-1}\%$ and $55^{+44}_{-53}\%$ of the total intrinsic SFRD according to the median values for the SF galaxy and QSO companion samples, respectively. 

As the samples in this work probe the bright-end of the luminosity function, we also take a look at the fractional contribution that dust-obscured star formation provides to the SFRD integrated down to $-$20 mag ($ \sim 3\ \text{M}_{\odot} \  \text{yr}^{-1}$). We obtain $\text{log}_{10} (\text{SFRD} / \text{M}_{\odot} \  \text{yr}^{-1} \ \text{Mpc}^{-3})$ = $-2.25_{-0.09}^{+0.10}$ for the SF galaxy and $-1.77_{-0.74}^{+1.99}$ for the QSO companion sample. The contribution from dust-obscured star formation is $7^{+11}_{-4}\%$ and $84^{+16}_{-74}\%$ (or equivalently $>3\%$ and  $>10\%$) of the total intrinsic SFRD integrated to $-$20 mag leveraging results from the SF galaxy and QSO companion samples, respectively.

We note that the dust-obscured SFRDs are obtained from Monte Carlo sampling both the observed and intrinsic SFRD with their corresponding errors and subtracting the observed SFRD from the intrinsic SFRD to account for the asymmetric errors.

\begin{figure*}
    \centering
    \includegraphics[width=\textwidth]{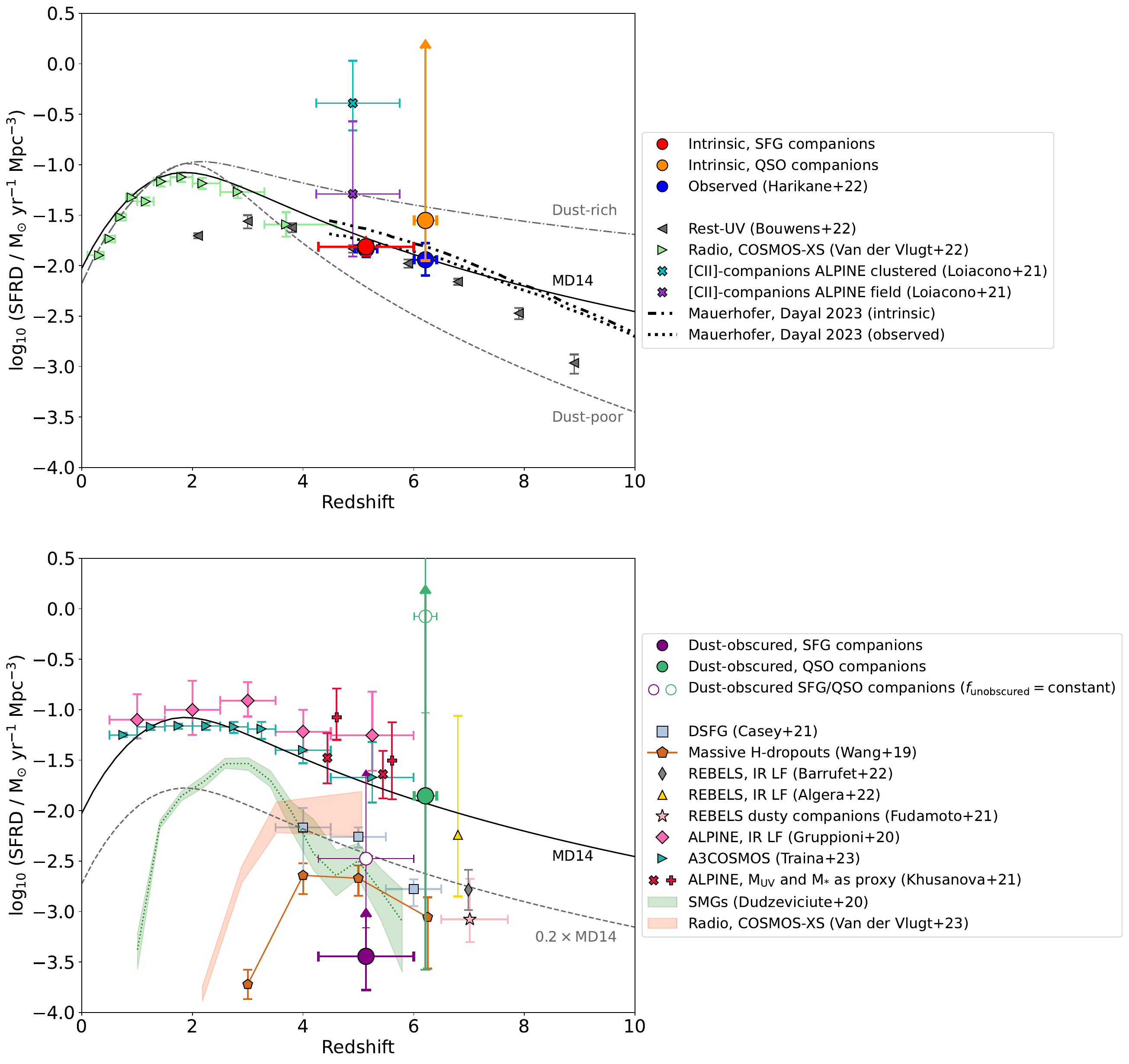}
    \caption{Constraints on the SFR density as a function of redshift. \textit{Top panel:} The red and orange solid circles show the SFR density we estimate from our intrinsic UV luminosity functions based on dust obscuration from companions around SF galaxies and QSOs, respectively, integrated to $-17$ mag. We only infer the intrinsic LF and dust corrections over the SFR range directly sampled by our work and therefore the dust corrected SFRDs we derive are lower limits. The blue solid circles show the SFR density when we integrate the observed UV LF from \citet{Harikane2022} to $-17$ mag. The solid black line indicates the fiducial SFR density evolution from \citet{Madau2014}.  Meanwhile, the dashed/dotted lines correspond to the dust-rich and dust-poor models from \citet{casey2018brightest} and intrinsic and dust-attenuated models from \citet{mauerhofer}. Furthermore, we show the constraints on the rest-UV SFRD from \citet{bouwens2022} (\textit{grey triangles}), the SFRD from radio-observations from \citet{Vlugt22} (\textit{green triangles}) and SFRDs from the [CII]-LF from the ALPINE companions from \citet{Loiacona_alpine} (\textit{aqua and purple crosses}). 
    \textit{Lower panel:} The purple and green solid circles show the contribution of dust-obscured sources to the SFR density, based on the dust obscuration from SF galaxy and QSO companions, respectively. The empty purple and green circles show the dust-obscured SFRD when we assume a constant mean obscuration instead of using the distribution of obscured fractions. Other estimates of the contribution of obscured sources to the SFRD are shown, including those of \citet{Casey2021_MORA} (\textit{light blue triangles}), \citet{Wang2019} (\textit{orange pentagons}), \citet{Barrufet} (\textit{gray thin diamond}), \citet{algera} (\textit{yellow triangle}), \citet{fudamoto_nature} (\textit{pink star}),  \citet{gruppioni2020} (\textit{pink diamonds}),  \citet{Traina} (\textit{aqua squares}), \citet{Khusanova_21} (\textit{red cross and plus sign}), \citet{dud+20} (\textit{green shaded region}) and \citet{vlugt23} (\textit{orange shaded region}).}
    \label{fig:SFRD}
\end{figure*}

We present our results in Table~\ref{tab:summary} and Fig.~\ref{fig:SFRD} where they are also compared with the SFRD values from the literature. The solid black line indicates the SFRD evolution from \citet{Madau2014} (MD14) and the dashed/dotted lines indicate the dust-rich and dust-poor models from \citet{casey2018brightest}. Both the intrinsic and observed SFRD inferred from the companions near SF galaxies agree with MD14. The companions near QSOs imply a dustier Universe with the intrinsic SFRD while the observed SFRD agrees with MD14.

We have shown in Sec.~\ref{correctedLF} that we obtain significantly larger intrinsic UV LF if we use the mean obscured fraction instead of the distribution. The corresponding obscured SFRD to these UV LFs are: $\text{log}_{10} (\text{SFRD} / \text{M}_{\odot} \  \text{yr}^{-1} \ \text{Mpc}^{-3})$ = $-2.48_{-0.68}^{+0.88}$ and $-0.08_{-0.96}^{+1.41}$ for the mean obscured fraction from the SFG and QSO companion sample, respectively. These results are presented in the lower panel of Fig.~\ref{fig:SFRD} with the empty purple and green markers.

As mentioned before, we find no relation between the obscured fraction and the total SFR. However, from the IRX-$M_{*}$ relation, one might reasonably expect a trend between the obscuration and the total SFR of a source. We therefore look into how a potential relation between obscuration and SFR would change our results. For this exercise we adopt the mass dependence found by \citet{bowler23} in the IRX-$M_{*}$ relation they derive based on ALPINE and REBELS galaxies. In Fig.~\ref{fig:obs} we can see that the relation from \citet{bowler17} indeed shows a similar slope between obscuration and SFR as the stacked results from ALPINE \citep{bethermin2020} and REBELS \citep{algera23}.
To illustrate the impact such a dependence can have on our results, we introduce a  small change in our procedure of correcting for the impact of dust obscuration by
forward modeling the observed distribution of obscured fraction of the SFG companions. The change we make is to scale up or down the fractions according to the SFR of sources in the intrinsic LF using the same slope as Bowler et al. (2022) find for the IRX-M* relation. In the case of the obscuration being dependent on SFR we not only vary $\phi_*$, but make sure to also vary $\beta_{\mathrm{DP}}$. Adopting the \citet{bowler23} dependence on SFR, we obtain a $\text{log}_{10} (\text{SFRD} / \text{M}_{\odot} \  \text{yr}^{-1} \ \text{Mpc}^{-3})$ that is 0.37 dex smaller than our fiducial result assuming no dependence.

\section{Discussion}
\label{discussion}

\subsection{Comparisons with earlier selections of [CII]-emitting Companion Galaxies}

In this work we search for [CII]-emitting sources in the neighborhoods of luminous galaxies and quasars.  Neighbours are identified through the detection of SNR $\geq$ 6.2 [CII] lines in the same ALMA data cubes as the target sources and with the requirement that the velocity difference with the primary target is $\mathrm{\Delta v \leq 2000\ km\ s^{-1}}$.  Our SNR $\geq$ 6.2 is based on a purity of $>$ 95 percent (Schouws et al. 2024, in prep.) for serendipitous detections with this SNR. Therefore, the probability that selected peaks originate from noise is very low, i.e., $<$5\%.  We have examined the ALMA data cubes corresponding to 34 [CII]-detected target sources and found 2 neighbour candidates: J1208-0200C1 and J2228+0152C1. We expanded our sample with the neighbour galaxies from \citet{Loiacona_alpine}, \citet{venemans_kilo}, \citet{fudamoto_nature} and REBELS-39 (Schouws et al. in prep.). We adopt the selection criteria of $\mathrm{\Delta v \leq \ 500\ km\ s^{-1}}$ to create a sample of galaxies that are most likely to be at $z \gtrsim 5$. However, we cannot exclude the possibility that some neighbours are not [CII] emitters, but e.g. mid-J CO lines at a lower redshift. While we have verified that for the sources with available multi-wavelength observations there is no emission bluewards of the Lyman-alpha transition, fully excluding the low redshift possibility would require the detection of additional emission lines or spectral features in sources with spectroscopy. We have estimated in Sec.~\ref{sec:results_obs} the impact of excluding a possible lower redshift contaminant on the mean obscured fraction and found the result to still be within the errors on the mean obscured fraction. Multi-wavelength observations are also needed to robustly constrain the stellar masses of the sample, in order to more closely tie our results to the stellar mass of sources and not just their SFRs through [CII].

One important property of our sample that can indicate that we are observing [CII] is the equivalent width of the candidate [CII] emission lines. Our sample (with dust detections) has a high median EW of $\approx 1.1 \mu \text{m} \approx 2000\ \text{km}\ \text{s}^{-1}$. High EW ($> 1000\ \text{km}\ \text{s}^{-1}$) are also found for [CII]-emission from e.g. Lyman break galaxies \citep{capak15} at $z \sim 5-6$ and QSO host galaxies at $z > 6.6$ \citep{Venemans2016} and local starbursts \citep{sargsyan2014}. In contrast to this, the median EW of CO at $z \sim 1-2.5$ is typically lower \citep{decarli2016}, e.g., $\approx 1000\ \mathrm{km\ s}^{-1}$.  

As previously stated we find two neighbour galaxies from our own examination of 34 archival ALMA observations. Out of the 34 observations analysed in this work, 18 of these are observations of quasars. The 2 neighbour galaxies are found to accompany quasars, therefore 11 percent of the examined quasars have a neighbour galaxy. In comparison, \citet{venemans_kilo} found 13 quasar host galaxies with companions that would qualify as a neighbour in 27 quasar observations, almost 50 percent. In Fig.~\ref{fig:parameter_space} we see that the targets of \citet{venemans_kilo} have higher [CII] luminosities compared to our sample of archival observations. 
Likely one of the most important factors in the lower prevalence of companion around the bright galaxy samples we consider vs. that seen in the \citet{venemans_kilo} are the halo masses of the targets considered.  Given their brightness and rarity, the bright QSOs studied by \citet{venemans_kilo} almost certainly have much higher halo masses and therefore much more significant clustering \citep{MoWhite1996} than most of the sources we consider.  While halo mass is likely the largest factor, it is also possible that feedback from the QSO could contribute to the prevalence of nearby companions by boosting their SFRs, as \citet{Zana2022} and \citet{ferrara2023} discuss. Moreover, the \citet{venemans_kilo} data is relatively deep, with quasars that have on source observation times larger than an hour, likely resulting in more significant serendipitous detections. 

Another possibility for the low amount of companions that are found could be due to the resolution of the observations. We restricted our search to archival observations with an angular resolution of $ \gtrsim 0.2$ arcsec. We have found the [CII]-emitting companions J1208-0200C1 and J2228+0152C1 that have observations with resolutions of $\sim 0.4$ arcsec. The majority (21) of the 34 observations used to search for companion galaxies have resolutions of $\gtrsim 0.4$ arcsec. It is however possible that we over resolve the [CII]-emission in those observations we consider that have higher spatial resolution and thus we miss companions there. As we do not obtain the intrinsic UV LF from the number of companion galaxies that are found, but rather based on the obscured fraction of the companions, we do not expect this to be an issue for our analysis. However, it could be an issue if there is a trend between the obscuration of a galaxy and source size, for example if some extended sources that are missed are generally less dust-obscured. If that would be the case we might overestimate the obscuration if we have missed some more extended companion galaxies.

One last contributing factor to why we do not find as many companions as in previous work around QSOs might be due to us only considering archival observations of sources without published companions in our search. Given the lesser frequency with which researchers report null results, we might be focusing on sources where companion searches have already been done, but the results are not deemed interesting enough to publish. 

\subsection{Obscured star formation derived from [CII]-Selected Samples}

We estimate the obscured star formation based on the FIR emission we infer from the (dust) continuum of the galaxies in our sample. When we convert continuum fluxes to IR luminosities it is necessary to rely on assumptions on the dust temperature and emissivity as we cannot constrain these parameters of the dust SED. While there is literature that suggests dust might be hotter at higher redshift, there is also a large scatter in dust temperatures observed (e.g. \citealt{schreiber18, viero2022, sommovigo22temp}). As IR luminosity scales with the dust temperature and dust emissivity ($L_{\text{IR}} \propto M_{\text{dust}} T_{\text{dust}}^{\beta_{\mathrm{em}} + 4}$) this translates into a larger uncertainty on the dust-obscured star formation. We include a systemic uncertainty of a factor of 3, to account for the dust temperature varying by 10K and $\beta_{\mathrm{em}}$ between 1.5-2.5. If the dust in galaxies in our sample were systematically colder than our fiducial assumptions, the IR luminosities and obscured SFR would decrease. Conversely, if the dust were, systematically hotter, this would increase the obscured SF. 

An important challenge for the present inferences is the very small numbers of galaxies in our samples, particularly at lower SFRs.  Clearly, larger samples of companion sources are required to extend the range of SFRs where the characteristics of companion galaxies can be well sampled and to improve the statistical confidence on these results.  Yet, even with the present small selection of companion sources, we can see that dust-obscured star formation contributes quite substantially to the SFR in individual star-forming galaxies at $z >$ 4.

As we have seen in Sec.~\ref{correctedSFRD}, whether the obscured fraction is characterized by the distribution of obscured fractions or a mean obscured fraction can have a significant impact on the resulting SFRD. The use of the mean obscured fraction is a simplified approach that would very likely lead to an overestimate of the dust-obscured SFRD, due to the omission of the impact of the modest fraction of sources where the dust extinction is substantially less. Therefore, we prefer to characterize the obscured fraction with the distribution of obscured fractions and it is even more important to obtain larger samples in order to better sample the distribution of obscured fractions.

\subsection{The use of [CII] as a SFR tracer}

Having constructed a sample of [CII]-selected companion galaxies, we aimed to use this selection to characterize the relation between the [CII] SFR of a galaxy and the fraction of the SFR that is directly visible in the rest-UV.  The advantage of performing this characterization on a [CII]-selected sample is that we would not expect the results to suffer from the same selection biases that would impact the same characterization performed on galaxies identified with a UV selection, as is the case for the ALPINE and REBELS galaxies themselves.

We find the total SFR as obtained from the UV and IR of our galaxies and the $L_{\text{[CII]}}$ of the galaxies to generally follow the relation from \citet{delooze2014}. This is consistent with e.g. results from the high-resolution SERRA simulations that finds $z \sim 8 $ galaxies to lie on the standard [CII]-SFR relation \citep{pallottini2022}. However, there is also literature questioning the use of [CII] as a tracer of SFR at high redshift (e.g. \citealt{Liang2023}). \citet{Liang2023} find with the use of the FIRE simulations that early Epoch of Reionization galaxies could have a [CII] deficit due to their low metallicities. If such a [CII]-deficit is present in our galaxy sample, this would imply less obscured star formation rates at a certain [CII] SFR. It may furthermore suggest that we may overestimate the obscured SFR if our sample is significantly deficient in [CII]. However, [CII] deficits are mostly observed in very bright and compact sources such as ULIRGs ($L_{\text{IR}} \geq 10^{12}\ \mathrm{L_{\odot}}$), SMGs and QSOs. 3 out of the 18 galaxies in our sample have IR luminosities corresponding to ULIRGs, however these galaxies do not show a greater [CII] deficit than the other galaxies in the sample. As we do not have estimates on the gas properties (e.g. metallicities) of the galaxies in our sample we cannot use the $L_{\text{[CII]}}$-SFR relation as described in \citet{Liang2023}.

One further caveat to the above argument concerns the origin of the [CII] emission. [CII] emission is believed to originate mainly from photodissociation regions at high redshift \citep{hollenbach, wolfire22}. However, it is unclear what fraction of [CII] emission originates from the neutral gas. \citet{croxall2017origins} find that in local star forming galaxies 60$\%$ to 80$\%$ of the [CII] emission is from neutral gas. This neutral gas is heated by the photo-electric effect on small dust grains and cooled via [CII] emission. Therefore, since an unknown fraction of the [CII]-emission in high redshift galaxies could be due to photoelectric heating, it could subsequently bias us to slightly dustier galaxies. We note that this method is still preferred over UV-selection since the [CII] line is not affected by dust attenuation and we find our results to be in agreement with other selection methods seen in Sec.~\ref{sec:dust_obs}. 

The use of [CII] as a SFR tracer could also be questioned if the [CII] and IR or UV show significant offsets, as the emission would probe different environments. We do not observe such large offsets, but the resolution of the [CII] observations used in this work is limited and therefore we cannot conclude whether these offsets are present. Lastly, some of the [CII] emission might correspond to the halo emission of a galaxy. With high-resolution modeling \citet{Schimek23} find that up to $10\%$ of the emission of [CII] is from the circumgalactic medium (CGM). If a significant percentage of [CII] emission does indeed result from the CGM in contrast to the disks of galaxies, this would decrease the reliability of [CII] as a SFR tracer.

\subsection{Characterizing the obscured fraction of SFR from [CII] emitters found in overdense regions}

One important uncertainty is connected to our overall methodology employed here. Due to the use of companion galaxies we are biased to overdense regions. We emphasize that we do not use the companion galaxies to derive the UV LF results directly as we base our results on the LF function from \citet{Harikane2022}. We only use the properties of the companion galaxies to infer the dust obscuration at $z \gtrsim$ 5 to circumvent the bias that one introduces by UV-selection. For our method we assume that the [CII]-selected companion galaxies used to compute the dust obscuration and the UV-selected galaxies used to compute the LF are drawn from the same distribution of star-forming galaxies. Moreover, we assume that this population of star-forming galaxies dominate the UV LF (especially at the bright end) and is the main contributor to the SFRD at high redshift. Although we are not biasing our LF to overdense regions, we might ask ourselves if  companion galaxies identified in overdense regions around massive galaxies or QSOs  are more likely to be dust-obscured at a given SFR than galaxies in lower density environments. 
While posing this question, it is important to keep in mind that the richest environments contribute an increasingly large fraction of the SFR density at high redshift, with $\sim$30\% of the SFR density at $z \sim6$ expected to originate from only $\sim$5\% of the cosmic volume \citep{chiang17}.

In theory, it is possible that galaxies in overdense regions are more dusty given that outflows and starbursts can enrich the IGM with metals. This has indeed been shown to be the case in studies of simulated galaxies (e.g. \citealt{graziani20, dicesare23}). The Astraeus framework \citep{astraeus} that fully couples galaxy formation and reionization has been used to show that radiative feedback preferentially quenches star formation more in low-mass galaxies ($\mathrm{M_{halo} < 10^{9.5} M_{\odot}}$) in overdense environments, affecting their star formation rates \citep{legrand23}. However, the metal and dust contents of early galaxies in the framework are primarily determined by a combination of their stellar masses and star formation rates \citep{Ucci23}. For example, low-mass halos with high star formation rates show extremely low metallicities due to outflows removing their dust and metal contents. Moreover, the FLARES simulations find that both SFR and metallicity depend primarily on the stellar mass of sources and show no strong dependence on the environment \citep{Flares_wilkins}.  Additionally, \citet{champagne_2018} find no evidence for an overdensity of continuum sources around  $z > 6$ quasars, in contrast to the overdensity of [CII] emitters, but concede that deeper observations may be needed to arrive at more robust conclusions. 

Interestingly, we observe the companion galaxies around QSOs to be more dust-obscured than companion galaxies with similar SFRs around SF galaxies. The inferred mean obscured fractions imply that companion galaxies around QSOs have SFRs that are on average 93$\%$ obscured, whereas companion galaxies around SF galaxies have less obscured SFR, at $63 \%$. The suggestion that QSO environments might be more dusty is not new, as this has been used to explain the lack of overdensities around QSOs in the rest-UV (e.g. \citealt{garcia2019}). Generally, the QSO observations have higher spatial resolution than the observations of star-forming galaxies. This could mean that we are biased to more compact systems and as mentioned before we could overestimate the obscuration if there is a trend between the obscuration of a galaxy and the source size. To understand if QSO environments are statistically more dusty and what the underlying physical scenarios are, we need larger and deeper samples. Moreover, we would benefit from even higher resolution data to understand whether an interaction between QSO and the companions is evident and could be the cause of increased dust obscuration. Larger samples could further help us to correct for a potential bias due to overdensity, e.g. if we could quantify obscuration as function of overdensity factor.

The SF-galaxy sample shows some low-attenuation outliers as we can clearly see in Fig.~\ref{fig:obs}. The number of galaxies would be too small to conclude whether this is a statistically different population of galaxies. However, regardless of the size of sample we do not observe these objects to be special (e.g. they are not at larger velocity differences than then rest of our sample). As these outliers are non-detected in the continuum we cannot infer whether the IR and rest-UV emission in these objects are spatially offset ([CII] and the rest-UV are spatially coincident) and if this could be responsible for the low-attenuation observed in these objects.

\begin{table*}Considering the importance of the SFR IR in this work, we prefer to use the work of Murphy+11. 
    \centering
    \caption{Summary of our key results for galaxies at $z\sim5$-6.}
    {\renewcommand{\arraystretch}{1.2}
    \begin{tabular}{c c c}
    \hline
     \multicolumn{3}{c}{\textbf{Unobscured SFR}}\\ \hline
     Mean unobscured fraction &  $\mathrm{f_{unobscured, SFG}} = 0.33^{+0.16}_{-0.14}$ &  $\mathrm{f_{unobscured, QSO}} = 0.07^{+0.05}_{-0.05}$ \\
     $\mathrm{\frac{SFR_{UV}}{SFR_{UV} + SFR_{IR}} } =   (1 + A \times \text{SFR}_{\text{[CII]}}^{B})^{-1}$, $ B = 0$ & $A_{\mathrm{SFG}} = 1.6^{+1.7}_{-0.8}$ & $A_{\mathrm{QSO}} =11.6^{+15.6}_{-5.4} $\\ \hline
     \multicolumn{3}{c}{\textbf{Star formation rate densities (lower limits)} }\\ \hline
      & $\mathrm{log_{10} (SFRD_{\text{SFG}}\ /\ M_{\odot}\ yr^{-1}\ Mpc^{-3})}$ & $\mathrm{log_{10} (SFRD_{\text{QSO}}\ /\ M_{\odot}\ yr^{-1}\ Mpc^{-3})}$ \\ 
     Intrinsic (Including obscured, $M_{\text{UV}}< -17$) &  $-1.81_{-0.08}^{+0.09}$ & $-1.55_{-0.40}^{+1.78}$  \\
     Observed (Based on rest-UV, $M_{\text{UV}} < -17$) & $-1.83_{-0.08}^{+0.08}$ & $-1.94_{-0.16}^{+0.16}$ \\
    Dust-obscured ($M_{\text{UV}}< -17$) & $-3.44_{-0.33}^{+0.47}$ & $-1.85_{-1.72}^{+2.08}$ \\ 
    \hline
    \end{tabular}}
    \label{tab:summary}
\end{table*}

\subsection{Evolution of the SFRD}
In Figure~\ref{fig:SFRD}, we show the SFRD we derive by integrating our intrinsic UV LF to $-$17 mag  (\textit{red and orange circles}), as well as the contribution from dust-obscured star formation alone (\textit{purple and green circles}) and the SFRD without any dust correction (\textit{blue circles}). When integrating to $-17$ mag, we make use of the dust-uncorrected UV LF from \citet{Harikane2022} at the fainter magnitudes we do not probe in this work. Therefore, we show our SFRDs as lower limits, as we have assumed zero attenuation for galaxies with lower SFRs. Besides the relation found by \citet{Madau2014} plotted with the solid black line, we also show the dust-poor and dust-rich models from \citet{casey2018brightest}. The dust-poor model indicates the SFRD where dusty star-forming galaxies (DSFGs) contribute $< 10 \%$ to the SFRD at $z > 4$, whereas the dust-rich model shows the SFRDs where DSFGs contribute $\sim 90 \%$ of the SFRD at z > 4.

\subsubsection{Comparison of total SFRD}

The top panel of the figure shows that the SFRD from the observed UV LF from \citet{Harikane2022} also agrees with the rest-UV SFRD (integrated to $-17$ mag) from \citet{bouwens2022}. The intrinsic SFRD from the dust-obscured SF galaxy companions agrees with the total SFRD inferred from radio (3 GHz) observations \citep{Vlugt22}, which is not surprising given the fact that dust obscuration would also have no impact on radio probes. The high SFRD inferred on basis of QSO companions is suggestive of a potentially much higher contribution from dust-obscured star formation at $z > 5$, but there might be environmental biases which impact this result and cause it to be higher than in other probes.

As we have made use of the sample of companion galaxies identified in ALPINE from \citet{Loiacona_alpine} we want to see if our results agree with the SFRD based on the [CII] luminosity function from that work. The [CII] LF is constructed from the companion galaxies in the ALPINE observations. Therefore the LF is biased by the use of overdense regions, we can see this clearly as the clustered SFRD is much higher than expected from previous results. We aim to arrive at more representative results by basing our results on the wide-area rest-UV LF results of  \citet{Harikane2022} and then correcting for the impact of dust using our obscured fraction determinations derived using our companion sample. The SFRD computed from the field galaxy found in the ALPINE data by \citet{Loiacona_alpine} is more in line with previous literature and our own SFRD estimates.

Also shown in the top panel of Figure~\ref{fig:SFRD} are the results of the DELPHI semi-analytic model by \citet{mauerhofer} integrated to $-17$ mag. In this model the maximum star-formation efficiency and fraction of supernova energy coupled to gas are baselined to galaxies from ALPINE and REBELS. Their intrinsic UV LF indeed implies more SFRD than the dust-attenuated (or observed) LF.

\subsubsection{Dust-obscured SFRD}
\label{sec:dust_obs}

It is interesting to compare the present estimates of the dust-obscured star formation from the [CII]-selected companions with estimates from previous studies as shown in the lower panel of Figure~\ref{fig:SFRD}, including those by \citet{Casey2021_MORA}, \citet{Wang2019}, \citet{Barrufet}, \citet{algera}, \citet{fudamoto_nature}, \citet{gruppioni2020},\citet{dud+20}, \citet{Traina}, \citet{Khusanova_21} and \citet{vlugt23}. In comparing to previous work in the literature, it is important that we pay close attention to the SFR limits that various studies probe. We compute the SFRD from integrating our obtained intrinsic luminosity function from 0.20  $\text{M}_{\odot} \  \text{yr}^{-1}$ ($-$17 mag) to $\approx 150$ and  $\approx 320$ $\text{M}_{\odot} \  \text{yr}^{-1}$ for the SF galaxy and QSO companion sample, respectively. Important to note is that for the magnitudes fainter than those that are sampled by the SFG and QSO companions, we assume no dust-obscuration, and therefore our SFRD estimates are lower limits.

Starting with the dust-obscured SFRD from the \citet{Casey2021_MORA} 2mm probe down to $\sim 2 \times 10^{12}\ \mathrm{L_{\odot}}$ (200 $\mathrm{M_{\odot}\ yr^{-1}}$), we find a lower and higher dust-obscured SFRD we derive from our SF galaxy and QSO companion sample, respectively. Moving towards slightly deeper probes of obscured star formation in the $z>4$ universe, we consider the search for sub-millimetre galaxies (SMGs) provided by \citet{dud+20} to $S_{870} \geq 1 $ mJy, corresponding to $\mathrm{ \approx 100\ M_{\odot}\ yr^{-1}}$ and the ($M_{*} > 10^{10.3}\ \text{M}_{\odot}$) H-dropouts from \citet{Wang2019}. Again the dust-obscured SFRD from the SFG/QSO companion are smaller/greater than the SFRD from these SMGs. We find a much higher dust-obscured SFRD using the obscured fraction of QSO companions, but this may be due to galaxies near bright QSOs being subject to more dust obscuration. We note however that the uncertainty on the intrinsic and observed SFRD based on the distribution of obscured fractions of the QSO companions are large.

Slightly deeper yet, in \citet{fudamoto_nature}, two companion sources in the REBELS data are described with obscured star formation rates of 40 - 70 $\mathrm{M_{\odot}\ yr^{-1}}$.  Other results for the dust-obscured SFRD from REBELS obtained by \citet{algera} and \citet{Barrufet} arrive at similar SFR density determinations to \citet{fudamoto_nature}. These SFRD estimates are at slightly higher redshift, $z \sim 7$, and they are higher than the obscured SFRD from our SF galaxy companion sample, while they are slightly lower than from our inferences using our QSO companion sample.

Lastly, results from the dust-obscured SFRD from the ALPINE targets by \citet{gruppioni2020} show higher SFRDs than in other studies. This may be due to their SFRD results being integrated down to $\mathrm{ \simeq 10^{-2}\ M_{\odot}\ yr^{-1}}$. Results based on the ALPINE survey by \citet{Khusanova_21} are also shown and indicate a relation between $\mathrm{L_{IR}}$ and both the UV magnitudes and stellar mass. The values shown in Fig.~\ref{fig:SFRD} are the obscured SFRD they infer extrapolating their sample results (obtained to $-$20 mag) to $-$17 mag.
Also shown is the contribution of optically dark radio sources in COSMOS-XS as found by \citet{vlugt23}. At $z \sim 5$ they find radio sources down to SFRs of $\mathrm{ \approx 500\ M_{\odot}\ yr^{-1}}$ and integrate over the whole LF to obtain the dust-obscured SFRD, that agrees with our SFRD based on the QSO galaxy companion sample. Results from \citet{Traina} from the A$^3$COSMOS survey that utilizes archival ALMA data on the COSMOS field to conduct an unbiased probe, are integrated down to the same SFRs as the results from ALPINE and show comparable SFRDs. Our dust-obscured SFRD from the QSO companions agrees with these results, and our results imply that a higher SFRD than observed from UV-selected galaxies with a small dust correction (as shown by the relation from \citealt{Madau2014}) is probable. 

The corrected SFRD estimates from this work demonstrate again that dust-obscured star formation likely contributes meaningfully to the total SFRD of the Universe at $z\gtrsim$ 5. Moreover, we see that SFRD estimates derived based on UV-selected samples may under predict the SFRD due to biases associated with a UV selection. Massive H-dropouts and SMGs contribute significantly to the obscured SFRD and their densities agree with the average dust-obscured SFRD we obtain from the SFG and QSO companion sample together.

For context, we also show the dust-obscured SFRD we obtain if we assume that all sources show the same mean obscured fraction as shown in Fig.~\ref{fig:obs} and do not utilise the full distribution to perform the correction (as in our fiducial calculations). With larger samples, we should be able to better constrain the full distribution of the obscured fraction and its dependence on the total SFR and improve constraints on the dust-obscured SFRD at $z \gtrsim 5$.

\section{Conclusions}
\label{conclusions}

Here we present an entirely new methodology for estimating the contribution from obscured star formation in the high-redshift Universe.  Essential to our methodology is the selection of sources using the [CII] line, which has been shown to be an excellent tracer of the total SFR in galaxies across cosmic time \citep{delooze2014,Schaerer_20,rebels, carniani20} and not to depend sensitively on the extent to which SFR is obscured.  Selection of sources via [CII] line emission can be performed in the same ALMA data sets as used for the characterization of bright galaxies and QSOs using the same line, but key is to construct these samples from sources found at similar redshift as the (UV-selected) target sources. This method of identifying dusty, far-IR bright galaxies is significantly cheaper than the use of blank field surveys as the ALMA FOV is relatively small. Given that the selected galaxies will be identified using their [CII]-derived SFRs alone, we can make use of rest-UV observations (where available) to determine the fraction of their SFR which is obscured.

In constructing [CII]-selected samples of companion galaxies for this purpose, we made use of selections of companion galaxies in the literature from \citet{venemans_kilo}, ALPINE \citep{Loiacona_alpine}, and from REBELS \citep{rebels,fudamoto_nature}, but we also did a search for these sources in ALMA observations of $z >$ 6 galaxies from the archive, considering ALMA observations of 55 separate targets.  Of the 34 sources where a [CII] line was found for the main target, two dust-obscured companion galaxies were found with the line search code MF3D: J1208-0200C1 and J2228+0152C1. Neither neighbour has a detection in the rest-UV, possibly due to obscuration by dust.  A less obscured companion is found around REBELS-39, with a 2$\sigma$ detection in the rest-UV. Figure~\ref{fig:parameter_space} shows the full range of redshifts and [CII] luminosity for targets we considered in constructing our samples.

Using these samples to characterize the fraction of SFR that is obscured, we only consider [CII]-selected companion galaxies where the available rest-UV observations are deep enough to detect sources in the rest-UV even if they are 85\% obscured.  Additionally, to maximize the probability that sources in these [CII]-selected samples are at $z >$ 4, we only consider neighbor candidates whose systemic velocity difference with the target galaxies are $\mathrm{\Delta v \leq 500\ km\ s^{-1}}$.  On the basis of the rest-UV observations, we estimate two fractions of obscured SFR, one for galaxies found in the neighborhoods around SF galaxies and one for galaxies around QSOs. The companion sample near SF galaxies suggest $f_\mathrm{{obs, median}}= 63\%$ similar to those expected from UV-selected samples.  The [CII]-selected galaxies around QSOs indicate $f_\mathrm{{obs, median}}= 93\% $, suggesting that QSO environments are even more dust-enriched.

With the fraction of obscured SFR we derive, we used a forward modeling methodology to determine the intrinsic UV LF from the observed UV LF results by \citet{Harikane2022}. By the intrinsic UV LF, we indicate the UV LF we would expect to obtain prior to the impact of dust on either the rest-UV brightness or selectability of sources. We integrate the intrinsic UV LF to $-17$ mag to find $\text{log}_{10} (\text{SFRD} / \text{M}_{\odot} \  \text{yr}^{-1} \ \text{Mpc}^{-3})$ $\geq$ $-1.81_{-0.08}^{+0.09}$ and $-1.55_{-0.40}^{+1.78}$ applying the distribution of obscured fractions from companion galaxies nearby SF galaxies and QSOs, respectively (Table~\ref{tab:summary}). It is possible the dust obscuration is underestimated due to the lack of constraints we have at low SFRs, although the QSO companion sample suggest that their environments may be dustier than the environment of SF galaxies. Using this new methodology we find that $>2^{+4}_{-1}\%$ and $>55^{+44}_{-53}\%$ of the SFRD is dust-obscured, based on the SF galaxy and QSO companion sample, respectively. If we integrate to $-$20 mag we obtain $\text{log}_{10} (\text{SFRD} / \text{M}_{\odot} \  \text{yr}^{-1} \ \text{Mpc}^{-3})$ = $-2.25_{-0.09}^{+0.10}$ for the SF galaxy and $-1.77_{-0.74}^{+1.99}$ for the QSO companion sample. This is $7^{+11}_{-4}\%$ and $84^{+16}_{-74}\%$ (or equivalently $>3\%$ and $>10\%$) of the total intrinsic SFRD integrated to $-$20 mag for the SF galaxy and QSO companion samples, respectively. This result indicates that the effect of dust-obscuration is substantial at the bright-end of the luminosity function. 

To constrain the obscured SFR relation more robustly, we need more sensitive probes for [CII] emission from companion galaxies and this will also provide us constraints to lower star formation rates. This is particularly important as we are not yet able to detect a dependence of the obscured fraction of SFR, but if the obscured fraction increases with SFR (as expected from most derived IRX-$M_{*}$ relations, e.g., \citealt{bowler23}), this would have a clear impact on the derived SFRD. Moreover, observations in more ALMA bands are needed to probe the full IR SED, to decrease uncertainties on the IR luminosities. Rest-UV observations are also needed for the majority of the companion galaxies that are found in the high redshift Universe, so that these sources can also be used to infer the dust obscuration. Future JWST observations (e.g. from ASPIRE;
program ID 2078; PI: F. Wang) will be beneficial for our understanding of stellar masses of these companion galaxies and confirming the redshifts as found by [CII]. One way of making immediate progress on this front is to make use of the deeper observations being obtained around massive $z>$ 4 galaxies from CRISTAL (2021.1.00280.L, PI:  Herrera-Camus, Rodrigo), an ALMA Large Program approved for execution in cycle 8 targeting 19 massive galaxies in the range $z =4$-6. In addition, we will be making use of the ALMA observations from the $\sim$49-hour Cycle-10 program CISTERN targeting 195 spectroscopically confirmed QSOs and bright galaxies. Among other open questions, the data should allow us to better determine if highly dust-obscured galaxies are substantially more common in QSO environments than they are around star-forming galaxies.

\section*{Acknowledgements}

% The Acknowledgements section is not numbered. Here you can thank helpful
% colleagues, acknowledge funding agencies, telescopes and facilities used etc.
% Try to keep it short.
The authors are very appreciative to our referee, Matthieu Bethermin, for the constructive feedback that significantly improved this paper.
RJB acknowledges support from NWO grants 600.065.140.11.N211 (vrijcompetitie) and TOP grant TOP1.16.057.
JH acknowledges support from the ERC Consolidator Grant 101088676 (VOYAJ) and the VIDI research program with project number 639.042.611, which is (partly) financed by the Netherlands Organization for Scientific Research (NWO). MA acknowledges support from ANID basal grant FB210003. LB acknowledges support by ERC AdG grant 740246 (Cosmic-Gas). RB acknowledges support from an STFC Ernest Rutherford Fellowship [grant number ST/T003596/1]. PD acknowledges support from the NWO grant 016.VIDI.189.162 (``ODIN") and from the European Commission's and University of Groningen's CO-FUND Rosalind Franklin program. HI and HSBA acknowledge support from the NAOJ ALMA Scientific Research Grant Code 2021-19A. IDL acknowledges funding support from ERC starting grant 851622 DustOrigin. AF acknowledges support from the ERC Advanced Grant INTERSTELLAR H2020/740120. LG and RS acknowledge support from the Amaldi Research Center funded by the MIUR program "Dipartimento di Eccellenza" (CUP:B81I18001170001).

This paper makes use of the following ALMA data: ADS/JAO.ALMA\#2017.1.00541.S, ADS/JAO.ALMA\#2017.1.00428.L, ADS/JAO.ALMA\#2019.1.01634.L. ALMA is a partnership of ESO (representing its member states), NSF (USA) and NINS (Japan), together with NRC (Canada), MOST and ASIAA (Taiwan), and KASI (Republic of Korea), in cooperation with the Republic of Chile. The Joint ALMA Observatory is operated by ESO, AUI/NRAO and NAOJ. 

The project leading to this publication has received support from ORP, that is funded by the European Union's Horizon 2020 research and innovation programme under grant agreement No 101004719 [ORP].

This research is based on observations made with the NASA/ESA Hubble Space Telescope obtained from the Space Telescope Science Institute, which is operated by the Association of Universities for Research in Astronomy, Inc., under NASA contract NAS 5–26555. These observations are associated with program(s) 12184, 12974, 13645, 14185, 14876 and 15064.

The Legacy Surveys consist of three individual and complementary projects: the Dark Energy Camera Legacy Survey (DECaLS; Proposal ID \#2014B-0404; PIs: David Schlegel and Arjun Dey), the Beijing-Arizona Sky Survey (BASS; NOAO Prop. ID \#2015A-0801; PIs: Zhou Xu and Xiaohui Fan), and the Mayall z-band Legacy Survey (MzLS; Prop. ID \#2016A-0453; PI: Arjun Dey). DECaLS, BASS and MzLS together include data obtained, respectively, at the Blanco telescope, Cerro Tololo Inter-American Observatory, NSF’s NOIRLab; the Bok telescope, Steward Observatory, University of Arizona; and the Mayall telescope, Kitt Peak National Observatory, NOIRLab. Pipeline processing and analyses of the data were supported by NOIRLab and the Lawrence Berkeley National Laboratory (LBNL). The Legacy Surveys project is honored to be permitted to conduct astronomical research on Iolkam Du’ag (Kitt Peak), a mountain with particular significance to the Tohono O’odham Nation.

NOIRLab is operated by the Association of Universities for Research in Astronomy (AURA) under a cooperative agreement with the National Science Foundation. LBNL is managed by the Regents of the University of California under contract to the U.S. Department of Energy.

This project used data obtained with the Dark Energy Camera (DECam), which was constructed by the Dark Energy Survey (DES) collaboration. Funding for the DES Projects has been provided by the U.S. Department of Energy, the U.S. National Science Foundation, the Ministry of Science and Education of Spain, the Science and Technology Facilities Council of the United Kingdom, the Higher Education Funding Council for England, the National Center for Supercomputing Applications at the University of Illinois at Urbana-Champaign, the Kavli Institute of Cosmological Physics at the University of Chicago, Center for Cosmology and Astro-Particle Physics at the Ohio State University, the Mitchell Institute for Fundamental Physics and Astronomy at Texas A\&M University, Financiadora de Estudos e Projetos, Fundacao Carlos Chagas Filho de Amparo, Financiadora de Estudos e Projetos, Fundacao Carlos Chagas Filho de Amparo a Pesquisa do Estado do Rio de Janeiro, Conselho Nacional de Desenvolvimento Cientifico e Tecnologico and the Ministerio da Ciencia, Tecnologia e Inovacao, the Deutsche Forschungsgemeinschaft and the Collaborating Institutions in the Dark Energy Survey. The Collaborating Institutions are Argonne National Laboratory, the University of California at Santa Cruz, the University of Cambridge, Centro de Investigaciones Energeticas, Medioambientales y Tecnologicas-Madrid, the University of Chicago, University College London, the DES-Brazil Consortium, the University of Edinburgh, the Eidgenossische Technische Hochschule (ETH) Zurich, Fermi National Accelerator Laboratory, the University of Illinois at Urbana-Champaign, the Institut de Ciencies de l’Espai (IEEC/CSIC), the Institut de Fisica d’Altes Energies, Lawrence Berkeley National Laboratory, the Ludwig Maximilians Universitat Munchen and the associated Excellence Cluster Universe, the University of Michigan, NSF’s NOIRLab, the University of Nottingham, the Ohio State University, the University of Pennsylvania, the University of Portsmouth, SLAC National Accelerator Laboratory, Stanford University, the University of Sussex, and Texas A\&M University.

BASS is a key project of the Telescope Access Program (TAP), which has been funded by the National Astronomical Observatories of China, the Chinese Academy of Sciences (the Strategic Priority Research Program “The Emergence of Cosmological Structures” Grant \# XDB09000000), and the Special Fund for Astronomy from the Ministry of Finance. The BASS is also supported by the External Cooperation Program of Chinese Academy of Sciences (Grant \# 114A11KYSB20160057), and Chinese National Natural Science Foundation (Grant \# 12120101003, \# 11433005).

The Legacy Survey team makes use of data products from the Near-Earth Object Wide-field Infrared Survey Explorer (NEOWISE), which is a project of the Jet Propulsion Laboratory/California Institute of Technology. NEOWISE is funded by the National Aeronautics and Space Administration.

The Legacy Surveys imaging of the DESI footprint is supported by the Director, Office of Science, Office of High Energy Physics of the U.S. Department of Energy under Contract No. DE-AC02-05CH1123, by the National Energy Research Scientific Computing Center, a DOE Office of Science User Facility under the same contract; and by the U.S. National Science Foundation, Division of Astronomical Sciences under Contract No. AST-0950945 to NOAO.

%%%%%%%%%%%%%%%%%%%%%%%%%%%%%%%%%%%%%%%%%%%%%%%%%%
\section*{Data Availability}
 
% The inclusion of a Data Availability Statement is a requirement for articles published in MNRAS. Data Availability Statements provide a standardised format for readers to understand the availability of data underlying the research results described in the article. The statement may refer to original data generated in the course of the study or to third-party data analysed in the article. The statement should describe and provide means of access, where possible, by linking to the data or providing the required accession numbers for the relevant databases or DOIs.

All ALMA observations used are publicly available in the ALMA Science Archive (\url{https://almascience.nrao.edu/aq/}). The rest-UV HST observations are obtained with the MAST portal (\url{https://mast.stsci.edu/portal/Mashup/Clients/Mast/Portal.html}). DECaLS observations are acquired from the Legacy Survey website (\url{https://www.legacysurvey.org/}). MF3D is publicly available on github (\url{https://github.com/pavesiriccardo/MF3D}).

%%%%%%%%%%%%%%%%%%%% REFERENCES %%%%%%%%%%%%%%%%%%

% The best way to enter references is to use BibTeX:

\bibliographystyle{mnras}
\bibliography{references}

\begin{thebibliography}{}
\makeatletter
\relax
\def\mn@urlcharsother{\let\do\@makeother \do\$\do\&\do\#\do\^\do\_\do\%\do\~}
\def\mn@doi{\begingroup\mn@urlcharsother \@ifnextchar [ {\mn@doi@} {\mn@doi@[]}}
\def\mn@doi@[#1]#2{\def\@tempa{#1}\ifx\@tempa\@empty \href {http://dx.doi.org/#2} {doi:#2}\else \href {http://dx.doi.org/#2} {#1}\fi \endgroup}
\def\mn@eprint#1#2{\mn@eprint@#1:#2::\@nil}
\def\mn@eprint@arXiv#1{\href {http://arxiv.org/abs/#1} {{\tt arXiv:#1}}}
\def\mn@eprint@dblp#1{\href {http://dblp.uni-trier.de/rec/bibtex/#1.xml} {dblp:#1}}
\def\mn@eprint@#1:#2:#3:#4\@nil{\def\@tempa {#1}\def\@tempb {#2}\def\@tempc {#3}\ifx \@tempc \@empty \let \@tempc \@tempb \let \@tempb \@tempa \fi \ifx \@tempb \@empty \def\@tempb {arXiv}\fi \@ifundefined {mn@eprint@\@tempb}{\@tempb:\@tempc}{\expandafter \expandafter \csname mn@eprint@\@tempb\endcsname \expandafter{\@tempc}}}

\bibitem[\protect\citeauthoryear{{Algera} et~al.,}{{Algera} et~al.}{2023a}]{algera}
{Algera} H. S.~B.,  et~al., 2023a, \mn@doi [\mnras] {10.1093/mnras/stac3195}, \href {https://ui.adsabs.harvard.edu/abs/2023MNRAS.518.6142A} {518, 6142}

\bibitem[\protect\citeauthoryear{{Algera} et~al.,}{{Algera} et~al.}{2023b}]{algera23}
{Algera} H. S.~B.,  et~al., 2023b, \mn@doi [\mnras] {10.1093/mnras/stac3195}, \href {https://ui.adsabs.harvard.edu/abs/2023MNRAS.518.6142A} {518, 6142}

\bibitem[\protect\citeauthoryear{{Barrufet} et~al.,}{{Barrufet} et~al.}{2023}]{Barrufet}
{Barrufet} L.,  et~al., 2023, \mn@doi [arXiv e-prints] {10.48550/arXiv.2303.11321}, \href {https://ui.adsabs.harvard.edu/abs/2023arXiv230311321B} {p. arXiv:2303.11321}

\bibitem[\protect\citeauthoryear{{Bertin} \& {Arnouts}}{{Bertin} \& {Arnouts}}{1996}]{sourceextractor}
{Bertin} E.,  {Arnouts} S.,  1996, \mn@doi [\aaps] {10.1051/aas:1996164}, \href {https://ui.adsabs.harvard.edu/abs/1996A&AS..117..393B} {117, 393}

\bibitem[\protect\citeauthoryear{{B{\'e}thermin} et~al.,}{{B{\'e}thermin} et~al.}{2020}]{bethermin2020}
{B{\'e}thermin} M.,  et~al., 2020, \mn@doi [\aap] {10.1051/0004-6361/202037649}, \href {https://ui.adsabs.harvard.edu/abs/2020A&A...643A...2B} {643, A2}

\bibitem[\protect\citeauthoryear{{Bouwens} et~al.,}{{Bouwens} et~al.}{2020}]{Bouwens2020}
{Bouwens} R.,  et~al., 2020, \mn@doi [\apj] {10.3847/1538-4357/abb830}, \href {https://ui.adsabs.harvard.edu/abs/2020ApJ...902..112B} {902, 112}

\bibitem[\protect\citeauthoryear{{Bouwens}, {Illingworth}, {Ellis}, {Oesch}, {Paulino-Afonso}, {Ribeiro}  \& {Stefanon}}{{Bouwens} et~al.}{2022a}]{bouwens2022}
{Bouwens} R.~J.,  {Illingworth} G.,  {Ellis} R.~S.,  {Oesch} P.,  {Paulino-Afonso} A.,  {Ribeiro} B.,   {Stefanon} M.,  2022a, \mn@doi [\apj] {10.3847/1538-4357/ac618c}, \href {https://ui.adsabs.harvard.edu/abs/2022ApJ...931...81B} {931, 81}

\bibitem[\protect\citeauthoryear{{Bouwens} et~al.,}{{Bouwens} et~al.}{2022b}]{rebels}
{Bouwens} R.~J.,  et~al., 2022b, \mn@doi [\apj] {10.3847/1538-4357/ac5a4a}, \href {https://ui.adsabs.harvard.edu/abs/2022ApJ...931..160B} {931, 160}

\bibitem[\protect\citeauthoryear{Bowler, Dunlop, McLure  \& McLeod}{Bowler et~al.}{2017}]{bowler17}
Bowler R. A.~A.,  Dunlop J.~S.,  McLure R.~J.,   McLeod D.~J.,  2017, \mn@doi [\mnras] {10.1093/mnras/stw3296}, 466, 3612

\bibitem[\protect\citeauthoryear{{Bowler} et~al.,}{{Bowler} et~al.}{2023}]{bowler23}
{Bowler} R.~A.~A.,  et~al., 2023, \mn@doi [arXiv e-prints] {10.48550/arXiv.2309.17386}, \href {https://ui.adsabs.harvard.edu/abs/2023arXiv230917386B} {p. arXiv:2309.17386}

\bibitem[\protect\citeauthoryear{{Bridle} \& {Schwab}}{{Bridle} \& {Schwab}}{1999}]{Bridle1999}
{Bridle} A.~H.,  {Schwab} F.~R.,  1999, in {Taylor} G.~B.,  {Carilli} C.~L.,   {Perley} R.~A.,  eds,  Astronomical Society of the Pacific Conference Series Vol. 180, Synthesis Imaging in Radio Astronomy II. p.~371

\bibitem[\protect\citeauthoryear{{CASA Team} et~al.,}{{CASA Team} et~al.}{2022}]{CASA}
{CASA Team} et~al., 2022, \mn@doi [\pasp] {10.1088/1538-3873/ac9642}, \href {https://ui.adsabs.harvard.edu/abs/2022PASP..134k4501C} {134, 114501}

\bibitem[\protect\citeauthoryear{{Capak} et~al.,}{{Capak} et~al.}{2015}]{capak15}
{Capak} P.~L.,  et~al., 2015, \mn@doi [\nat] {10.1038/nature14500}, \href {https://ui.adsabs.harvard.edu/abs/2015Natur.522..455C} {522, 455}

\bibitem[\protect\citeauthoryear{{Carilli}, {Riechers}, {Walter}, {Maiolino}, {Wagg}, {Lentati}, {McMahon}  \& {Wolfe}}{{Carilli} et~al.}{2013}]{carilli13}
{Carilli} C.~L.,  {Riechers} D.,  {Walter} F.,  {Maiolino} R.,  {Wagg} J.,  {Lentati} L.,  {McMahon} R.,   {Wolfe} A.,  2013, \mn@doi [\apj] {10.1088/0004-637X/763/2/120}, \href {https://ui.adsabs.harvard.edu/abs/2013ApJ...763..120C} {763, 120}

\bibitem[\protect\citeauthoryear{Carniani et~al.,}{Carniani et~al.}{2018}]{Carniani18}
Carniani S.,  et~al., 2018, \mn@doi [\mnras] {10.1093/mnras/sty1088}, 478, 1170

\bibitem[\protect\citeauthoryear{Carniani et~al.,}{Carniani et~al.}{2020}]{carniani20}
Carniani S.,  et~al., 2020, \mn@doi [\mnras] {10.1093/mnras/staa3178}, 499, 5136

\bibitem[\protect\citeauthoryear{{Casey} et~al.,}{{Casey} et~al.}{2018a}]{casey2018brightest}
{Casey} C.~M.,  et~al., 2018a, \mn@doi [\apj] {10.3847/1538-4357/aac82d}, \href {https://ui.adsabs.harvard.edu/abs/2018ApJ...862...77C} {862, 77}

\bibitem[\protect\citeauthoryear{{Casey}, {Hodge}, {Zavala}, {Spilker}, {da Cunha}, {Staguhn}, {Finkelstein}  \& {Drew}}{{Casey} et~al.}{2018b}]{CaseySurvey}
{Casey} C.~M.,  {Hodge} J.,  {Zavala} J.~A.,  {Spilker} J.,  {da Cunha} E.,  {Staguhn} J.,  {Finkelstein} S.~L.,   {Drew} P.,  2018b, \mn@doi [\apj] {10.3847/1538-4357/aacd11}, \href {https://ui.adsabs.harvard.edu/abs/2018ApJ...862...78C} {862, 78}

\bibitem[\protect\citeauthoryear{{Casey} et~al.,}{{Casey} et~al.}{2021}]{Casey2021_MORA}
{Casey} C.~M.,  et~al., 2021, \mn@doi [\apj] {10.3847/1538-4357/ac2eb4}, \href {https://ui.adsabs.harvard.edu/abs/2021ApJ...923..215C} {923, 215}

\bibitem[\protect\citeauthoryear{{Chabrier}}{{Chabrier}}{2003}]{Chabrier2003}
{Chabrier} G.,  2003, \mn@doi [\pasp] {10.1086/376392}, \href {https://ui.adsabs.harvard.edu/abs/2003PASP..115..763C} {115, 763}

\bibitem[\protect\citeauthoryear{{Champagne} et~al.,}{{Champagne} et~al.}{2018}]{champagne_2018}
{Champagne} J.~B.,  et~al., 2018, \mn@doi [\apj] {10.3847/1538-4357/aae396}, \href {https://ui.adsabs.harvard.edu/abs/2018ApJ...867..153C} {867, 153}

\bibitem[\protect\citeauthoryear{{Chiang}, {Overzier}, {Gebhardt}  \& {Henriques}}{{Chiang} et~al.}{2017}]{chiang17}
{Chiang} Y.-K.,  {Overzier} R.~A.,  {Gebhardt} K.,   {Henriques} B.,  2017, \mn@doi [\apjl] {10.3847/2041-8213/aa7e7b}, \href {https://ui.adsabs.harvard.edu/abs/2017ApJ...844L..23C} {844, L23}

\bibitem[\protect\citeauthoryear{{Croxall} et~al.,}{{Croxall} et~al.}{2017}]{croxall2017origins}
{Croxall} K.~V.,  et~al., 2017, \mn@doi [\apj] {10.3847/1538-4357/aa8035}, \href {https://ui.adsabs.harvard.edu/abs/2017ApJ...845...96C} {845, 96}

\bibitem[\protect\citeauthoryear{{De Looze} et~al.}{{De Looze} et~al.}{2014}]{delooze2014}
{De Looze} I.,  et~al., 2014, \mn@doi [\aap] {10.1051/0004-6361/201322489}, \href {https://ui.adsabs.harvard.edu/abs/2014A&A...568A..62D} {568, A62}

\bibitem[\protect\citeauthoryear{{Decarli} et~al.,}{{Decarli} et~al.}{2016}]{decarli2016}
{Decarli} R.,  et~al., 2016, \mn@doi [\apj] {10.3847/1538-4357/833/1/70}, \href {https://ui.adsabs.harvard.edu/abs/2016ApJ...833...70D} {833, 70}

\bibitem[\protect\citeauthoryear{{Decarli} et~al.}{{Decarli} et~al.}{2017}]{Decarli_2017}
{Decarli} R.,  et~al., 2017, \mn@doi [Nature] {10.1038/nature22358}, \href {https://ui.adsabs.harvard.edu/abs/2017Natur.545..457D} {545, 457}

\bibitem[\protect\citeauthoryear{{Decarli} et~al.,}{{Decarli} et~al.}{2019}]{decarli2019}
{Decarli} R.,  et~al., 2019, \mn@doi [\apj] {10.3847/1538-4357/ab297f}, \href {https://ui.adsabs.harvard.edu/abs/2019ApJ...880..157D} {880, 157}

\bibitem[\protect\citeauthoryear{{Decarli} et~al.,}{{Decarli} et~al.}{2020}]{decarli20}
{Decarli} R.,  et~al., 2020, \mn@doi [\apj] {10.3847/1538-4357/abaa3b}, \href {https://ui.adsabs.harvard.edu/abs/2020ApJ...902..110D} {902, 110}

\bibitem[\protect\citeauthoryear{{Di Cesare}, {Graziani}, {Schneider}, {Ginolfi}, {Venditti}, {Santini}  \& {Hunt}}{{Di Cesare} et~al.}{2023}]{dicesare23}
{Di Cesare} C.,  {Graziani} L.,  {Schneider} R.,  {Ginolfi} M.,  {Venditti} A.,  {Santini} P.,   {Hunt} L.~K.,  2023, \mn@doi [\mnras] {10.1093/mnras/stac3702}, \href {https://ui.adsabs.harvard.edu/abs/2023MNRAS.519.4632D} {519, 4632}

\bibitem[\protect\citeauthoryear{{Dudzevi{\v{c}}i{\={u}}t{\.{e}}} et~al.,}{{Dudzevi{\v{c}}i{\={u}}t{\.{e}}} et~al.}{2020}]{dud+20}
{Dudzevi{\v{c}}i{\={u}}t{\.{e}}} U.,  et~al., 2020, \mn@doi [\mnras] {10.1093/mnras/staa769}, \href {https://ui.adsabs.harvard.edu/abs/2020MNRAS.494.3828D} {494, 3828}

\bibitem[\protect\citeauthoryear{{Everett} et~al.,}{{Everett} et~al.}{2020}]{everett2020}
{Everett} W.~B.,  et~al., 2020, \mn@doi [\apj] {10.3847/1538-4357/ab9df7}, \href {https://ui.adsabs.harvard.edu/abs/2020ApJ...900...55E} {900, 55}

\bibitem[\protect\citeauthoryear{{Ferrara} et~al.,}{{Ferrara} et~al.}{2022}]{ferrara22}
{Ferrara} A.,  et~al., 2022, \mn@doi [\mnras] {10.1093/mnras/stac460}, \href {https://ui.adsabs.harvard.edu/abs/2022MNRAS.512...58F} {512, 58}

\bibitem[\protect\citeauthoryear{{Ferrara}, {Zana}, {Gallerani}  \& {Sommovigo}}{{Ferrara} et~al.}{2023}]{ferrara2023}
{Ferrara} A.,  {Zana} T.,  {Gallerani} S.,   {Sommovigo} L.,  2023, \mn@doi [\mnras] {10.1093/mnras/stad299}, \href {https://ui.adsabs.harvard.edu/abs/2023MNRAS.520.3089F} {520, 3089}

\bibitem[\protect\citeauthoryear{{Fudamoto} et~al.,}{{Fudamoto} et~al.}{2020a}]{fudamoto20}
{Fudamoto} Y.,  et~al., 2020a, \mn@doi [\mnras] {10.1093/mnras/stz3248}, \href {https://ui.adsabs.harvard.edu/abs/2020MNRAS.491.4724F} {491, 4724}

\bibitem[\protect\citeauthoryear{{Fudamoto} et~al.,}{{Fudamoto} et~al.}{2020b}]{Fudamoto_2020}
{Fudamoto} Y.,  et~al., 2020b, \mn@doi [\aap] {10.1051/0004-6361/202038163}, \href {https://ui.adsabs.harvard.edu/abs/2020A&A...643A...4F} {643, A4}

\bibitem[\protect\citeauthoryear{{Fudamoto} et~al.,}{{Fudamoto} et~al.}{2021}]{fudamoto_nature}
{Fudamoto} Y.,  et~al., 2021, \mn@doi [\nat] {10.1038/s41586-021-03846-z}, \href {https://ui.adsabs.harvard.edu/abs/2021Natur.597..489F} {597, 489}

\bibitem[\protect\citeauthoryear{{Fudamoto} et~al.,}{{Fudamoto} et~al.}{2022}]{fudamoto_22}
{Fudamoto} Y.,  et~al., 2022, \mn@doi [\apj] {10.3847/1538-4357/ac7a47}, \href {https://ui.adsabs.harvard.edu/abs/2022ApJ...934..144F} {934, 144}

\bibitem[\protect\citeauthoryear{{Fujimoto} et~al.,}{{Fujimoto} et~al.}{2019}]{fujimoto_19}
{Fujimoto} S.,  et~al., 2019, \mn@doi [\apj] {10.3847/1538-4357/ab480f}, \href {https://ui.adsabs.harvard.edu/abs/2019ApJ...887..107F} {887, 107}

\bibitem[\protect\citeauthoryear{{Fujimoto} et~al.,}{{Fujimoto} et~al.}{2020}]{Fujimoto_20}
{Fujimoto} S.,  et~al., 2020, \mn@doi [\apj] {10.3847/1538-4357/ab94b3}, \href {https://ui.adsabs.harvard.edu/abs/2020ApJ...900....1F} {900, 1}

\bibitem[\protect\citeauthoryear{{Garc{\'\i}a-Vergara}, {Hennawi}, {Barrientos}  \& {Arrigoni Battaia}}{{Garc{\'\i}a-Vergara} et~al.}{2019}]{garcia2019}
{Garc{\'\i}a-Vergara} C.,  {Hennawi} J.~F.,  {Barrientos} L.~F.,   {Arrigoni Battaia} F.,  2019, \mn@doi [\apj] {10.3847/1538-4357/ab4d52}, \href {https://ui.adsabs.harvard.edu/abs/2019ApJ...886...79G} {886, 79}

\bibitem[\protect\citeauthoryear{{Gonz{\'a}lez-L{\'o}pez} et~al.,}{{Gonz{\'a}lez-L{\'o}pez} et~al.}{2019}]{GonzalezLopez2019_ASPECS}
{Gonz{\'a}lez-L{\'o}pez} J.,  et~al., 2019, \mn@doi [\apj] {10.3847/1538-4357/ab3105}, \href {https://ui.adsabs.harvard.edu/abs/2019ApJ...882..139G} {882, 139}

\bibitem[\protect\citeauthoryear{{Graziani}, {Schneider}, {Ginolfi}, {Hunt}, {Maio}, {Glatzle}  \& {Ciardi}}{{Graziani} et~al.}{2020}]{graziani20}
{Graziani} L.,  {Schneider} R.,  {Ginolfi} M.,  {Hunt} L.~K.,  {Maio} U.,  {Glatzle} M.,   {Ciardi} B.,  2020, \mn@doi [\mnras] {10.1093/mnras/staa796}, \href {https://ui.adsabs.harvard.edu/abs/2020MNRAS.494.1071G} {494, 1071}

\bibitem[\protect\citeauthoryear{{Gruppioni} et~al.,}{{Gruppioni} et~al.}{2020}]{gruppioni2020}
{Gruppioni} C.,  et~al., 2020, \mn@doi [\aap] {10.1051/0004-6361/202038487}, \href {https://ui.adsabs.harvard.edu/abs/2020A&A...643A...8G} {643, A8}

\bibitem[\protect\citeauthoryear{{Harikane} et~al.,}{{Harikane} et~al.}{2022}]{Harikane2022}
{Harikane} Y.,  et~al., 2022, \mn@doi [\apjs] {10.3847/1538-4365/ac3dfc}, \href {https://ui.adsabs.harvard.edu/abs/2022ApJS..259...20H} {259, 20}

\bibitem[\protect\citeauthoryear{{Hodge} \& {da Cunha}}{{Hodge} \& {da Cunha}}{2020}]{Hodge2020}
{Hodge} J.~A.,  {da Cunha} E.,  2020, \mn@doi [Royal Society Open Science] {10.1098/rsos.200556}, \href {https://ui.adsabs.harvard.edu/abs/2020RSOS....700556H} {7, 200556}

\bibitem[\protect\citeauthoryear{{Hollenbach} \& {Tielens}}{{Hollenbach} \& {Tielens}}{1999}]{hollenbach}
{Hollenbach} D.~J.,  {Tielens} A.~G.~G.~M.,  1999, \mn@doi [Reviews of Modern Physics] {10.1103/RevModPhys.71.173}, \href {https://ui.adsabs.harvard.edu/abs/1999RvMP...71..173H} {71, 173}

\bibitem[\protect\citeauthoryear{{Hutter}, {Dayal}, {Yepes}, {Gottl{\"o}ber}, {Legrand}  \& {Ucci}}{{Hutter} et~al.}{2021}]{astraeus}
{Hutter} A.,  {Dayal} P.,  {Yepes} G.,  {Gottl{\"o}ber} S.,  {Legrand} L.,   {Ucci} G.,  2021, \mn@doi [\mnras] {10.1093/mnras/stab602}, \href {https://ui.adsabs.harvard.edu/abs/2021MNRAS.503.3698H} {503, 3698}

\bibitem[\protect\citeauthoryear{{Hygate} et~al.,}{{Hygate} et~al.}{2023}]{hygate2023}
{Hygate} A.~P.~S.,  et~al., 2023, \mn@doi [\mnras] {10.1093/mnras/stad1212}, \href {https://ui.adsabs.harvard.edu/abs/2023MNRAS.524.1775H} {524, 1775}

\bibitem[\protect\citeauthoryear{{Inami} et~al.,}{{Inami} et~al.}{2022}]{Inami2022}
{Inami} H.,  et~al., 2022, \mn@doi [\mnras] {10.1093/mnras/stac1779}, \href {https://ui.adsabs.harvard.edu/abs/2022MNRAS.515.3126I} {515, 3126}

\bibitem[\protect\citeauthoryear{{Izumi} et~al.}{{Izumi} et~al.}{2019}]{Izumi2019}
{Izumi} T.,  et~al., 2019, \mn@doi [\pasj] {10.1093/pasj/psz096}, \href {https://ui.adsabs.harvard.edu/abs/2019PASJ...71..111I} {71, 111}

\bibitem[\protect\citeauthoryear{{Kashino}, {Lilly}, {Matthee}, {Eilers}, {Mackenzie}, {Bordoloi}  \& {Simcoe}}{{Kashino} et~al.}{2023}]{kashino23}
{Kashino} D.,  {Lilly} S.~J.,  {Matthee} J.,  {Eilers} A.-C.,  {Mackenzie} R.,  {Bordoloi} R.,   {Simcoe} R.~A.,  2023, \mn@doi [\apj] {10.3847/1538-4357/acc588}, \href {https://ui.adsabs.harvard.edu/abs/2023ApJ...950...66K} {950, 66}

\bibitem[\protect\citeauthoryear{{Kepley}, {Tsutsumi}, {Brogan}, {Indebetouw}, {Yoon}, {Mason}  \& {Donovan Meyer}}{{Kepley} et~al.}{2020}]{multithresh}
{Kepley} A.~A.,  {Tsutsumi} T.,  {Brogan} C.~L.,  {Indebetouw} R.,  {Yoon} I.,  {Mason} B.,   {Donovan Meyer} J.,  2020, \mn@doi [\pasp] {10.1088/1538-3873/ab5e14}, \href {https://ui.adsabs.harvard.edu/abs/2020PASP..132b4505K} {132, 024505}

\bibitem[\protect\citeauthoryear{{Khusanova} et~al.,}{{Khusanova} et~al.}{2021}]{Khusanova_21}
{Khusanova} Y.,  et~al., 2021, \mn@doi [\aap] {10.1051/0004-6361/202038944}, \href {https://ui.adsabs.harvard.edu/abs/2021A&A...649A.152K} {649, A152}

\bibitem[\protect\citeauthoryear{{Kohandel} et~al.}{{Kohandel} et~al.}{2019}]{kohandel2019}
{Kohandel} M.,  et~al., 2019, \mn@doi [\mnras] {10.1093/mnras/stz1486}, \href {https://ui.adsabs.harvard.edu/abs/2019MNRAS.487.3007K} {487, 3007}

\bibitem[\protect\citeauthoryear{{Labb{\'e}}, {Bouwens}, {Illingworth}  \& {Franx}}{{Labb{\'e}} et~al.}{2006}]{labbe2006}
{Labb{\'e}} I.,  {Bouwens} R.,  {Illingworth} G.~D.,   {Franx} M.,  2006, \mn@doi [\apjl] {10.1086/508512}, \href {https://ui.adsabs.harvard.edu/abs/2006ApJ...649L..67L} {649, L67}

\bibitem[\protect\citeauthoryear{{Labb{\'e}} et~al.,}{{Labb{\'e}} et~al.}{2010a}]{labbe2010b}
{Labb{\'e}} I.,  et~al., 2010a, \mn@doi [\apjl] {10.1088/2041-8205/708/1/L26}, \href {https://ui.adsabs.harvard.edu/abs/2010ApJ...708L..26L} {708, L26}

\bibitem[\protect\citeauthoryear{{Labb{\'e}} et~al.,}{{Labb{\'e}} et~al.}{2010b}]{labbe2010a}
{Labb{\'e}} I.,  et~al., 2010b, \mn@doi [\apjl] {10.1088/2041-8205/716/2/L103}, \href {https://ui.adsabs.harvard.edu/abs/2010ApJ...716L.103L} {716, L103}

\bibitem[\protect\citeauthoryear{{Labb{\'e}} et~al.,}{{Labb{\'e}} et~al.}{2013}]{labbe2013}
{Labb{\'e}} I.,  et~al., 2013, \mn@doi [\apjl] {10.1088/2041-8205/777/2/L19}, \href {https://ui.adsabs.harvard.edu/abs/2013ApJ...777L..19L} {777, L19}

\bibitem[\protect\citeauthoryear{{Labb{\'e}} et~al.,}{{Labb{\'e}} et~al.}{2015}]{labbe2015}
{Labb{\'e}} I.,  et~al., 2015, \mn@doi [\apjs] {10.1088/0067-0049/221/2/23}, \href {https://ui.adsabs.harvard.edu/abs/2015ApJS..221...23L} {221, 23}

\bibitem[\protect\citeauthoryear{{Le F{\`e}vre} et~al.,}{{Le F{\`e}vre} et~al.}{2020}]{alpine}
{Le F{\`e}vre} O.,  et~al., 2020, \mn@doi [\aap] {10.1051/0004-6361/201936965}, \href {https://ui.adsabs.harvard.edu/abs/2020A&A...643A...1L} {643, A1}

\bibitem[\protect\citeauthoryear{{Legrand}, {Dayal}, {Hutter}, {Gottl{\"o}ber}, {Yepes}  \& {Trebitsch}}{{Legrand} et~al.}{2023}]{legrand23}
{Legrand} L.,  {Dayal} P.,  {Hutter} A.,  {Gottl{\"o}ber} S.,  {Yepes} G.,   {Trebitsch} M.,  2023, \mn@doi [\mnras] {10.1093/mnras/stac3760}, \href {https://ui.adsabs.harvard.edu/abs/2023MNRAS.519.4564L} {519, 4564}

\bibitem[\protect\citeauthoryear{{Liang} et~al.,}{{Liang} et~al.}{2023}]{Liang2023}
{Liang} L.,  et~al., 2023, \mn@doi [arXiv e-prints] {10.48550/arXiv.2301.04149}, \href {https://ui.adsabs.harvard.edu/abs/2023arXiv230104149L} {p. arXiv:2301.04149}

\bibitem[\protect\citeauthoryear{{Liu} et~al.,}{{Liu} et~al.}{2019}]{Liu2019_A3COSMOS}
{Liu} D.,  et~al., 2019, \mn@doi [\apjs] {10.3847/1538-4365/ab42da}, \href {https://ui.adsabs.harvard.edu/abs/2019ApJS..244...40L} {244, 40}

\bibitem[\protect\citeauthoryear{{Loiacono} et~al.}{{Loiacono} et~al.}{2021}]{Loiacona_alpine}
{Loiacono} F.,  et~al., 2021, \mn@doi [\aap] {10.1051/0004-6361/202038607}, \href {https://ui.adsabs.harvard.edu/abs/2021A&A...646A..76L} {646, A76}

\bibitem[\protect\citeauthoryear{{Madau} \& {Dickinson}}{{Madau} \& {Dickinson}}{2014}]{Madau2014}
{Madau} P.,  {Dickinson} M.,  2014, \mn@doi [\araa] {10.1146/annurev-astro-081811-125615}, \href {https://ui.adsabs.harvard.edu/abs/2014ARA&A..52..415M} {52, 415}

\bibitem[\protect\citeauthoryear{{Marrone} et~al.,}{{Marrone} et~al.}{2018}]{marrone2018}
{Marrone} D.~P.,  et~al., 2018, \mn@doi [\nat] {10.1038/nature24629}, \href {https://ui.adsabs.harvard.edu/abs/2018Natur.553...51M} {553, 51}

\bibitem[\protect\citeauthoryear{{Matsuoka} et~al.,}{{Matsuoka} et~al.}{2018}]{matsuoka2018subaru}
{Matsuoka} Y.,  et~al., 2018, \mn@doi [\pasj] {10.1093/pasj/psx046}, \href {https://ui.adsabs.harvard.edu/abs/2018PASJ...70S..35M} {70, S35}

\bibitem[\protect\citeauthoryear{{Mauerhofer} \& {Dayal}}{{Mauerhofer} \& {Dayal}}{2023}]{mauerhofer}
{Mauerhofer} V.,  {Dayal} P.,  2023, \mn@doi [\mnras] {10.1093/mnras/stad2734}, \href {https://ui.adsabs.harvard.edu/abs/2023MNRAS.tmp.2648M} {}

\bibitem[\protect\citeauthoryear{{Mazzucchelli} et~al.,}{{Mazzucchelli} et~al.}{2019}]{mazzuchelli}
{Mazzucchelli} C.,  et~al., 2019, \mn@doi [\apj] {10.3847/1538-4357/ab2f75}, \href {https://ui.adsabs.harvard.edu/abs/2019ApJ...881..163M} {881, 163}

\bibitem[\protect\citeauthoryear{{McLeod}, {McLure}  \& {Dunlop}}{{McLeod} et~al.}{2016}]{McLeod2016}
{McLeod} D.~J.,  {McLure} R.~J.,   {Dunlop} J.~S.,  2016, \mn@doi [\mnras] {10.1093/mnras/stw904}, \href {https://ui.adsabs.harvard.edu/abs/2016MNRAS.459.3812M} {459, 3812}

\bibitem[\protect\citeauthoryear{{Meyer} et~al.,}{{Meyer} et~al.}{2022}]{Meyer22}
{Meyer} R.~A.,  et~al., 2022, \mn@doi [\apj] {10.3847/1538-4357/ac4f67}, \href {https://ui.adsabs.harvard.edu/abs/2022ApJ...927..141M} {927, 141}

\bibitem[\protect\citeauthoryear{{Miller} et~al.}{{Miller} et~al.}{2020}]{miller2020}
{Miller} T.~B.,  et~al., 2020, \mn@doi [\apj] {10.3847/1538-4357/ab63dd}, \href {https://ui.adsabs.harvard.edu/abs/2020ApJ...889...98M} {889, 98}

\bibitem[\protect\citeauthoryear{{Mo} \& {White}}{{Mo} \& {White}}{1996}]{MoWhite1996}
{Mo} H.~J.,  {White} S.~D.~M.,  1996, \mn@doi [\mnras] {10.1093/mnras/282.2.347}, \href {https://ui.adsabs.harvard.edu/abs/1996MNRAS.282..347M} {282, 347}

\bibitem[\protect\citeauthoryear{{Murphy} et~al.,}{{Murphy} et~al.}{2011}]{Murphy11}
{Murphy} E.~J.,  et~al., 2011, \mn@doi [\apj] {10.1088/0004-637X/737/2/67}, \href {https://ui.adsabs.harvard.edu/abs/2011ApJ...737...67M} {737, 67}

\bibitem[\protect\citeauthoryear{{Nguyen} et~al.,}{{Nguyen} et~al.}{2010}]{nguyen2010}
{Nguyen} H.~T.,  et~al., 2010, \mn@doi [\aap] {10.1051/0004-6361/201014680}, \href {https://ui.adsabs.harvard.edu/abs/2010A&A...518L...5N} {518, L5}

\bibitem[\protect\citeauthoryear{{Oesch}, {Bouwens}, {Illingworth}, {Labb{\'e}}  \& {Stefanon}}{{Oesch} et~al.}{2018}]{Oesch2018}
{Oesch} P.~A.,  {Bouwens} R.~J.,  {Illingworth} G.~D.,  {Labb{\'e}} I.,   {Stefanon} M.,  2018, \mn@doi [\apj] {10.3847/1538-4357/aab03f}, \href {https://ui.adsabs.harvard.edu/abs/2018ApJ...855..105O} {855, 105}

\bibitem[\protect\citeauthoryear{{Oke} \& {Gunn}}{{Oke} \& {Gunn}}{1983}]{Oke1983}
{Oke} J.~B.,  {Gunn} J.~E.,  1983, \mn@doi [\apj] {10.1086/160817}, \href {https://ui.adsabs.harvard.edu/abs/1983ApJ...266..713O} {266, 713}

\bibitem[\protect\citeauthoryear{{Pallottini} et~al.,}{{Pallottini} et~al.}{2022}]{pallottini2022}
{Pallottini} A.,  et~al., 2022, \mn@doi [\mnras] {10.1093/mnras/stac1281}, \href {https://ui.adsabs.harvard.edu/abs/2022MNRAS.513.5621P} {513, 5621}

\bibitem[\protect\citeauthoryear{{Pannella} et~al.,}{{Pannella} et~al.}{2009}]{Pannella2009}
{Pannella} M.,  et~al., 2009, \mn@doi [\apjl] {10.1088/0004-637X/698/2/L116}, \href {https://ui.adsabs.harvard.edu/abs/2009ApJ...698L.116P} {698, L116}

\bibitem[\protect\citeauthoryear{{Pannella} et~al.,}{{Pannella} et~al.}{2015}]{Pannella2015}
{Pannella} M.,  et~al., 2015, \mn@doi [\apj] {10.1088/0004-637X/807/2/141}, \href {https://ui.adsabs.harvard.edu/abs/2015ApJ...807..141P} {807, 141}

\bibitem[\protect\citeauthoryear{{Pavesi} et~al.}{{Pavesi} et~al.}{2018}]{Pavesi_MF3D}
{Pavesi} R.,  et~al., 2018, \mn@doi [\apj] {10.3847/1538-4357/aacb79}, \href {https://ui.adsabs.harvard.edu/abs/2018ApJ...864...49P} {864, 49}

\bibitem[\protect\citeauthoryear{{Reddy}, {Steidel}, {Fadda}, {Yan}, {Pettini}, {Shapley}, {Erb}  \& {Adelberger}}{{Reddy} et~al.}{2006}]{Reddy2006}
{Reddy} N.~A.,  {Steidel} C.~C.,  {Fadda} D.,  {Yan} L.,  {Pettini} M.,  {Shapley} A.~E.,  {Erb} D.~K.,   {Adelberger} K.~L.,  2006, \mn@doi [\apj] {10.1086/503739}, \href {https://ui.adsabs.harvard.edu/abs/2006ApJ...644..792R} {644, 792}

\bibitem[\protect\citeauthoryear{{Sargsyan}, {Samsonyan}, {Lebouteiller}, {Weedman}, {Barry}, {Bernard-Salas}, {Houck}  \& {Spoon}}{{Sargsyan} et~al.}{2014}]{sargsyan2014}
{Sargsyan} L.,  {Samsonyan} A.,  {Lebouteiller} V.,  {Weedman} D.,  {Barry} D.,  {Bernard-Salas} J.,  {Houck} J.,   {Spoon} H.,  2014, \mn@doi [\apj] {10.1088/0004-637X/790/1/15}, \href {https://ui.adsabs.harvard.edu/abs/2014ApJ...790...15S} {790, 15}

\bibitem[\protect\citeauthoryear{{Schaerer} et~al.,}{{Schaerer} et~al.}{2020}]{Schaerer_20}
{Schaerer} D.,  et~al., 2020, \mn@doi [\aap] {10.1051/0004-6361/202037617}, \href {https://ui.adsabs.harvard.edu/abs/2020A&A...643A...3S} {643, A3}

\bibitem[\protect\citeauthoryear{{Schimek} et~al.,}{{Schimek} et~al.}{2023}]{Schimek23}
{Schimek} A.,  et~al., 2023, \mn@doi [arXiv e-prints] {10.48550/arXiv.2306.00583}, \href {https://ui.adsabs.harvard.edu/abs/2023arXiv230600583S} {p. arXiv:2306.00583}

\bibitem[\protect\citeauthoryear{{Schreiber}, {Elbaz}, {Pannella}, {Ciesla}, {Wang}  \& {Franco}}{{Schreiber} et~al.}{2018}]{schreiber18}
{Schreiber} C.,  {Elbaz} D.,  {Pannella} M.,  {Ciesla} L.,  {Wang} T.,   {Franco} M.,  2018, \mn@doi [\aap] {10.1051/0004-6361/201731506}, \href {https://ui.adsabs.harvard.edu/abs/2018A&A...609A..30S} {609, A30}

\bibitem[\protect\citeauthoryear{{Solomon}, {Downes}  \& {Radford}}{{Solomon} et~al.}{1992}]{solomon}
{Solomon} P.~M.,  {Downes} D.,   {Radford} S.~J.~E.,  1992, \mn@doi [\apjl] {10.1086/186569}, \href {https://ui.adsabs.harvard.edu/abs/1992ApJ...398L..29S} {398, L29}

\bibitem[\protect\citeauthoryear{{Sommovigo} et~al.,}{{Sommovigo} et~al.}{2022a}]{sommovigo22temp}
{Sommovigo} L.,  et~al., 2022a, \mn@doi [\mnras] {10.1093/mnras/stac302}, \href {https://ui.adsabs.harvard.edu/abs/2022MNRAS.513.3122S} {513, 3122}

\bibitem[\protect\citeauthoryear{{Sommovigo} et~al.,}{{Sommovigo} et~al.}{2022b}]{sommovigo2022}
{Sommovigo} L.,  et~al., 2022b, \mn@doi [\mnras] {10.1093/mnras/stac2997}, \href {https://ui.adsabs.harvard.edu/abs/2022MNRAS.517.5930S} {517, 5930}

\bibitem[\protect\citeauthoryear{{Speagle}, {Steinhardt}, {Capak}  \& {Silverman}}{{Speagle} et~al.}{2014}]{speagle14}
{Speagle} J.~S.,  {Steinhardt} C.~L.,  {Capak} P.~L.,   {Silverman} J.~D.,  2014, \mn@doi [\apjs] {10.1088/0067-0049/214/2/15}, \href {https://ui.adsabs.harvard.edu/abs/2014ApJS..214...15S} {214, 15}

\bibitem[\protect\citeauthoryear{{Stefanon} et~al.,}{{Stefanon} et~al.}{2019}]{stefanon2019}
{Stefanon} M.,  et~al., 2019, \mn@doi [\apj] {10.3847/1538-4357/ab3792}, \href {https://ui.adsabs.harvard.edu/abs/2019ApJ...883...99S} {883, 99}

\bibitem[\protect\citeauthoryear{{Talia}, {Cimatti}, {Giulietti}, {Zamorani}, {Bethermin}, {Faisst}, {Le F{\`e}vre}  \& {Smol{\c{c}}i{\'c}}}{{Talia} et~al.}{2021}]{talia2021}
{Talia} M.,  {Cimatti} A.,  {Giulietti} M.,  {Zamorani} G.,  {Bethermin} M.,  {Faisst} A.,  {Le F{\`e}vre} O.,   {Smol{\c{c}}i{\'c}} V.,  2021, \mn@doi [\apj] {10.3847/1538-4357/abd6e3}, \href {https://ui.adsabs.harvard.edu/abs/2021ApJ...909...23T} {909, 23}

\bibitem[\protect\citeauthoryear{{Traina} et~al.,}{{Traina} et~al.}{2023}]{Traina}
{Traina} A.,  et~al., 2023, \mn@doi [arXiv e-prints] {10.48550/arXiv.2309.15150}, \href {https://ui.adsabs.harvard.edu/abs/2023arXiv230915150T} {p. arXiv:2309.15150}

\bibitem[\protect\citeauthoryear{{Trakhtenbrot}, {Lira}, {Netzer}, {Cicone}, {Maiolino}  \& {Shemmer}}{{Trakhtenbrot} et~al.}{2017}]{trakhtenbrot17}
{Trakhtenbrot} B.,  {Lira} P.,  {Netzer} H.,  {Cicone} C.,  {Maiolino} R.,   {Shemmer} O.,  2017, \mn@doi [\apj] {10.3847/1538-4357/836/1/8}, \href {https://ui.adsabs.harvard.edu/abs/2017ApJ...836....8T} {836, 8}

\bibitem[\protect\citeauthoryear{{Ucci} et~al.,}{{Ucci} et~al.}{2023}]{Ucci23}
{Ucci} G.,  et~al., 2023, \mn@doi [\mnras] {10.1093/mnras/stac2654}, \href {https://ui.adsabs.harvard.edu/abs/2023MNRAS.518.3557U} {518, 3557}

\bibitem[\protect\citeauthoryear{{Venemans}, {Walter}, {Zschaechner}, {Decarli}, {De Rosa}, {Findlay}, {McMahon}  \& {Sutherland}}{{Venemans} et~al.}{2016}]{Venemans2016}
{Venemans} B.~P.,  {Walter} F.,  {Zschaechner} L.,  {Decarli} R.,  {De Rosa} G.,  {Findlay} J.~R.,  {McMahon} R.~G.,   {Sutherland} W.~J.,  2016, \mn@doi [\apj] {10.3847/0004-637X/816/1/37}, \href {https://ui.adsabs.harvard.edu/abs/2016ApJ...816...37V} {816, 37}

\bibitem[\protect\citeauthoryear{{Venemans} et~al.}{{Venemans} et~al.}{2018}]{venemans2018}
{Venemans} B.~P.,  et~al., 2018, \mn@doi [\apj] {10.3847/1538-4357/aadf35}, \href {https://ui.adsabs.harvard.edu/abs/2018ApJ...866..159V} {866, 159}

\bibitem[\protect\citeauthoryear{{Venemans} et~al.,}{{Venemans} et~al.}{2020}]{venemans_kilo}
{Venemans} B.~P.,  et~al., 2020, \mn@doi [\apj] {10.3847/1538-4357/abc563}, \href {https://ui.adsabs.harvard.edu/abs/2020ApJ...904..130V} {904, 130}

\bibitem[\protect\citeauthoryear{{Viero}, {Sun}, {Chung}, {Moncelsi}  \& {Condon}}{{Viero} et~al.}{2022}]{viero2022}
{Viero} M.~P.,  {Sun} G.,  {Chung} D.~T.,  {Moncelsi} L.,   {Condon} S.~S.,  2022, \mn@doi [\mnras] {10.1093/mnrasl/slac075}, \href {https://ui.adsabs.harvard.edu/abs/2022MNRAS.516L..30V} {516, L30}

\bibitem[\protect\citeauthoryear{{Wang} et~al.,}{{Wang} et~al.}{2019}]{Wang2019}
{Wang} T.,  et~al., 2019, \mn@doi [\nat] {10.1038/s41586-019-1452-4}, \href {https://ui.adsabs.harvard.edu/abs/2019Natur.572..211W} {572, 211}

\bibitem[\protect\citeauthoryear{{Wang} et~al.,}{{Wang} et~al.}{2023}]{Wang23}
{Wang} F.,  et~al., 2023, \mn@doi [\apjl] {10.3847/2041-8213/accd6f}, \href {https://ui.adsabs.harvard.edu/abs/2023ApJ...951L...4W} {951, L4}

\bibitem[\protect\citeauthoryear{{Whitaker} et~al.}{{Whitaker} et~al.}{2018}]{whitaker2018}
{Whitaker} K.~E.,  et~al., 2018. p. 328.01

\bibitem[\protect\citeauthoryear{{Wilkins} et~al.,}{{Wilkins} et~al.}{2023}]{Flares_wilkins}
{Wilkins} S.~M.,  et~al., 2023, \mn@doi [\mnras] {10.1093/mnras/stac3281}, \href {https://ui.adsabs.harvard.edu/abs/2023MNRAS.518.3935W} {518, 3935}

\bibitem[\protect\citeauthoryear{{Wolfire}, {McKee}, {Hollenbach}  \& {Tielens}}{{Wolfire} et~al.}{2003}]{wolfire03}
{Wolfire} M.~G.,  {McKee} C.~F.,  {Hollenbach} D.,   {Tielens} A.~G.~G.~M.,  2003, \mn@doi [\apj] {10.1086/368016}, \href {https://ui.adsabs.harvard.edu/abs/2003ApJ...587..278W} {587, 278}

\bibitem[\protect\citeauthoryear{{Wolfire}, {Vallini}  \& {Chevance}}{{Wolfire} et~al.}{2022}]{wolfire22}
{Wolfire} M.~G.,  {Vallini} L.,   {Chevance} M.,  2022, \mn@doi [\araa] {10.1146/annurev-astro-052920-010254}, \href {https://ui.adsabs.harvard.edu/abs/2022ARA&A..60..247W} {60, 247}

\bibitem[\protect\citeauthoryear{{Zana}, {Gallerani}, {Carniani}, {Vito}, {Ferrara}, {Lupi}, {Di Mascia}  \& {Barai}}{{Zana} et~al.}{2022}]{Zana2022}
{Zana} T.,  {Gallerani} S.,  {Carniani} S.,  {Vito} F.,  {Ferrara} A.,  {Lupi} A.,  {Di Mascia} F.,   {Barai} P.,  2022, \mn@doi [\mnras] {10.1093/mnras/stac978}, \href {https://ui.adsabs.harvard.edu/abs/2022MNRAS.513.2118Z} {513, 2118}

\bibitem[\protect\citeauthoryear{{Zavala} et~al.}{{Zavala} et~al.}{2021}]{Zavala2021}
{Zavala} J.~A.,  et~al., 2021, \mn@doi [\apj] {10.3847/1538-4357/abdb27}, \href {https://ui.adsabs.harvard.edu/abs/2021ApJ...909..165Z} {909, 165}

\bibitem[\protect\citeauthoryear{{da Cunha} et~al.,}{{da Cunha} et~al.}{2013}]{dacunha2013}
{da Cunha} E.,  et~al., 2013, \mn@doi [\apj] {10.1088/0004-637X/766/1/13}, \href {https://ui.adsabs.harvard.edu/abs/2013ApJ...766...13D} {766, 13}

\bibitem[\protect\citeauthoryear{{van der Vlugt}, {Hodge}, {Algera}, {Smail}, {Leslie}, {Radcliffe}, {Riechers}  \& {R{\"o}ttgering}}{{van der Vlugt} et~al.}{2022}]{Vlugt22}
{van der Vlugt} D.,  {Hodge} J.~A.,  {Algera} H.~S.~B.,  {Smail} I.,  {Leslie} S.~K.,  {Radcliffe} J.~F.,  {Riechers} D.~A.,   {R{\"o}ttgering} H.,  2022, \mn@doi [\apj] {10.3847/1538-4357/ac99db}, \href {https://ui.adsabs.harvard.edu/abs/2022ApJ...941...10V} {941, 10}

\bibitem[\protect\citeauthoryear{{van der Vlugt} et~al.,}{{van der Vlugt} et~al.}{2023}]{vlugt23}
{van der Vlugt} D.,  et~al., 2023, \mn@doi [\apj] {10.3847/1538-4357/acd549}, \href {https://ui.adsabs.harvard.edu/abs/2023ApJ...951..131V} {951, 131}

\makeatother
\end{thebibliography}

%%%%%%%%%%%%%%%%%%%%%%%%%%%%%%%%%%%%%%%%%%%%%%%%%%

%%%%%%%%%%%%%%%%% APPENDICES %%%%%%%%%%%%%%%%%%%%%
\newpage
\appendix

\section{Low-redshift contamination}
\label{contamination}

\begin{figure}
    \centering
    \includegraphics[width=\columnwidth]{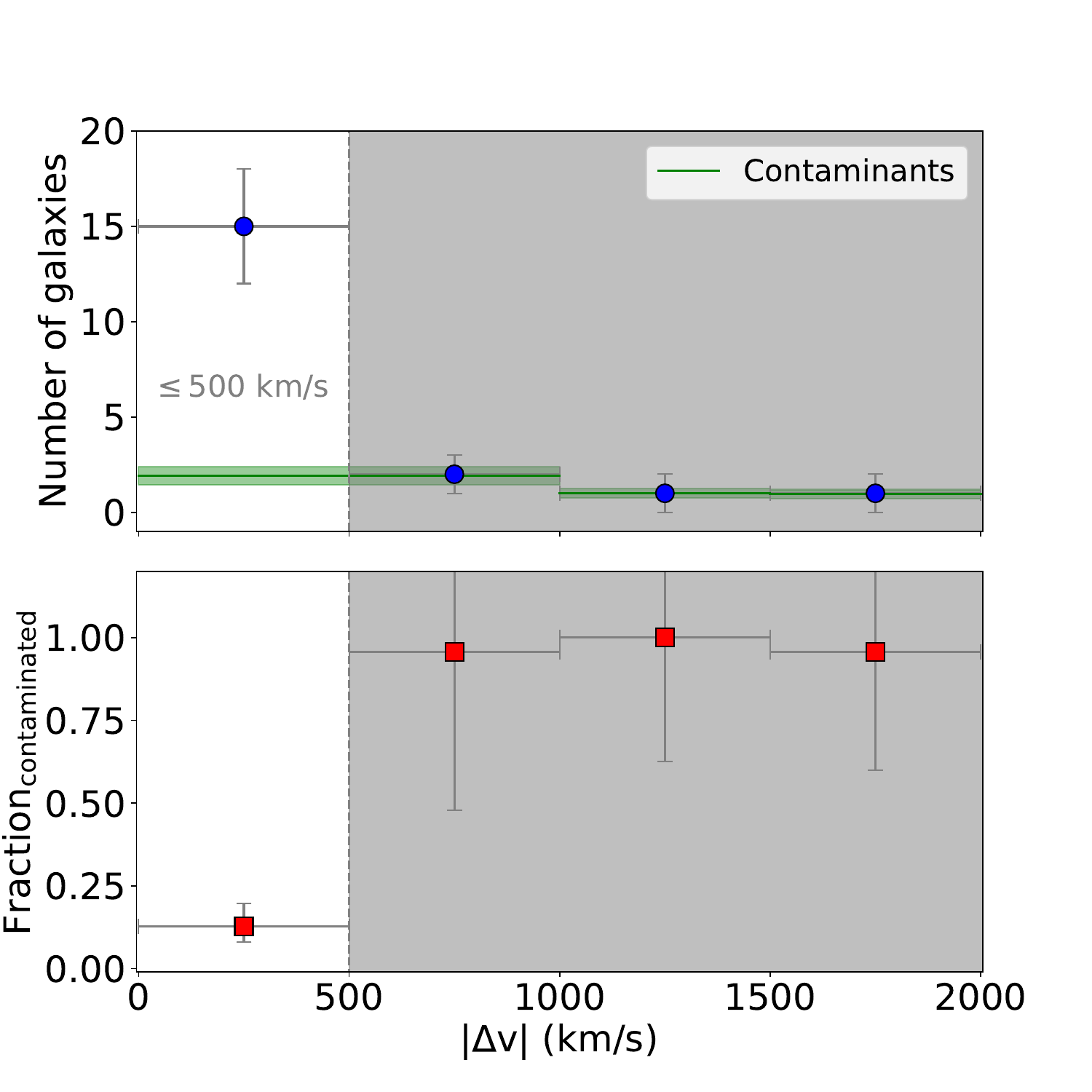}
    \caption{Top panel: the number of serendipitous lines identified by \citet{venemans_kilo} as a function of the absolute velocity difference ($\Delta \mathrm{v} = \frac{z_{\mathrm{comp}} - z_{\mathrm{target}}}{1 + z_{\mathrm{target}}} c $). In green the number of expected contaminants within one velocity bin is plotted. 
    Lower panel: the fraction of galaxies that are contaminants as function of the velocity difference. For galaxies with $\mathrm{\Delta v \leq 500\ km\ s^{-1}}$ the fraction of contaminants is the smallest ($\approx0.13$) and therefore the probability of them being high redshift sources is high. }
    \label{fig:contaminates}
\end{figure}

One potential uncertainty in this analysis is the nature of the serendipitous emission lines detected in the same data cubes as the primary $z \geq 4$ ALMA targets.  While the vast majority of these discovered lines very likely correspond to [CII], some of the discovered lines might originate from sources at lower redshift and in fact correspond to a CO transition.  To estimate the probability of this, we use 27 target quasar host galaxies and 27 serendipitously identified galaxies of \citet{venemans_kilo}. We focus on the velocity difference between the serendipitous lines and the quasar host galaxy as given in Eq. \ref{eq:veldiff}. To estimate the number of low redshift contaminants, we look at the number of galaxies with $\mathrm{\Delta v > 2000\ km\ s^{-1}}$, similarly to what is done in \citet{venemans_kilo}. Serendipitous lines with larger velocity differences are substantially less likely to be associated with the quasar host galaxy than lines with smaller velocity differences.  At large velocity differences, serendipitously identified sources very likely correspond to the same type of sources found in random search fields, i.e., CO transitions from lower redshift galaxies \citep[e.g.][]{GonzalezLopez2019_ASPECS}. There are 8 serendipitous lines with $\mathrm{\Delta v > 2000\ km\ s^{-1}}$, which we assume to be low redshift contaminants. The setup of the 27 quasars that are observed varied slightly from source to source, but on average two spectral windows overlapped in order to provide a coverage of 3.3 GHz around the quasar host galaxy, corresponding to 3600 km s$^{-1}$ at $z = 6$ . Two spectral windows are used at 15 GHz away in order to probe the continuum emission, so in total about 7200 km s$^{-1}$ is covered per observation. Using the amount of contaminants at $\mathrm{\Delta v > 2000\ km\ s^{-1}}$ as a proxy for the low redshift sources we can expect per GHz, we estimate there would be 6 lower-redshift contaminants lying within $\mathrm{\Delta v \leq 2000\ km\ s^{-1}}$ of the main target. In Fig. \ref{fig:contaminates} the upper panel shows the number of serendipitous lines as a function of the velocity difference. The green line shows the expected number of contaminates within one velocity bin. The lower panel shows the fraction of galaxies that are expected to be contaminants as a function of the velocity difference. The values plotted are the median values with 16th and 84th percentiles from bootstrapping 10000 times. We see that the fraction of contaminates is low ($\approx$ 0.13) for serendipitous lines with $\mathrm{\Delta v \leq 500\ km\ s^{-1}}$ and within that velocity limit we would expect to find about 2 contaminants for the sample of \citet{venemans_kilo}. At $\mathrm{\Delta v > 500\ km\ s^{-1}}$ the fraction increases significantly and would approach $\approx 1.0$. These numbers are an upper limit as some galaxies with $\mathrm{\Delta v > 2000\ km\ s^{-1}}$ could nevertheless be associated with the quasar host galaxy and be at $z >$ 4.  Choosing the velocity limit to be  $\mathrm{\Delta v \leq 500\ km\ s^{-1}}$ is therefore a safe limit to assume for the serendipitously identified galaxies to be at high redshift as the fraction of expected contaminants is low. 

\section{Neighbour candidates}
\begin{figure*}
    \centering
    \includegraphics[width=\textwidth,trim={0cm 6.5cm 0cm 0},clip]{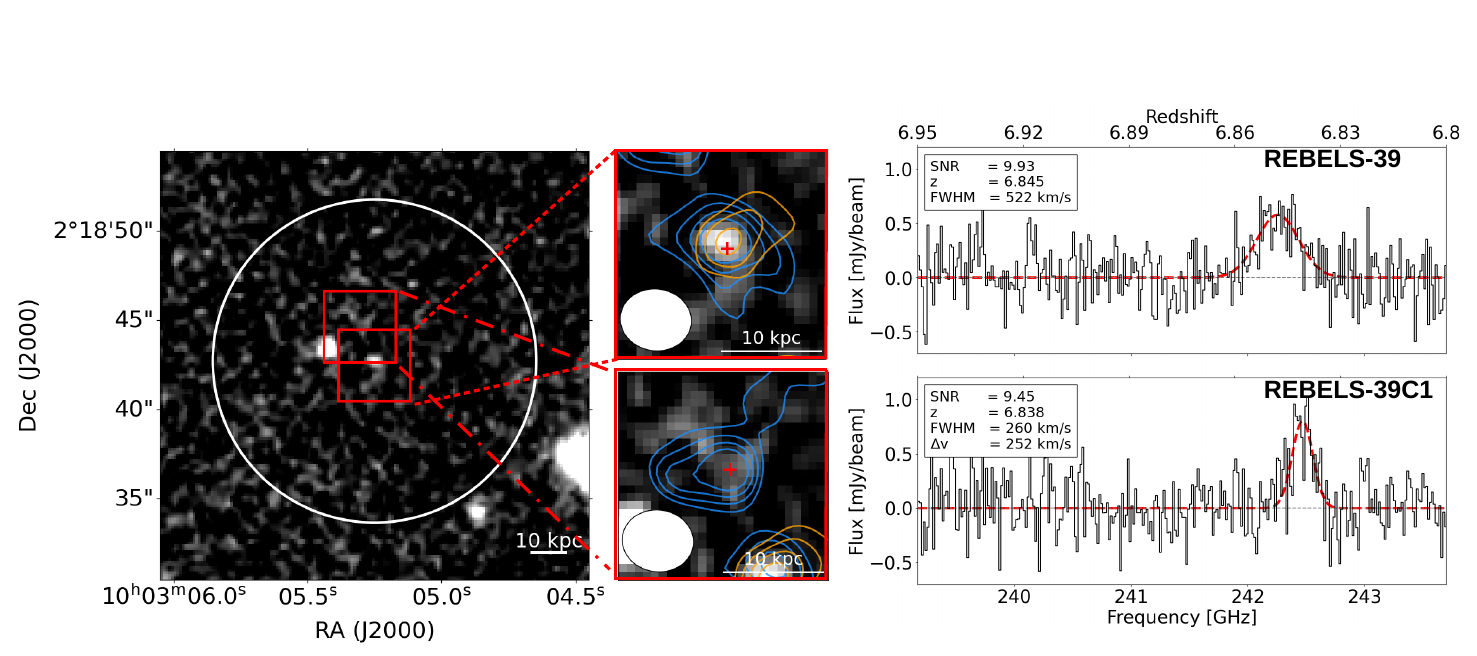}
    \caption{Similar to Figure~\ref{fig:J22} from the main text but for the REBELS target REBELS-39 and the background image is the J-filter from VISTA/VIRCAM.  Here we present a newly identified companion galaxy REBELS-39C1.}
    \label{fig:R39}
\end{figure*}

In this work we identify two neighbour candidates in archival ALMA data that had not been previously reported.  In \S\ref{neighbours} we show one of our [CII] neighbours candidates: J2228+0152C1. Here we show the second candidate J1208-0200C1 in Figure~\ref{fig:J22}. Similar to J2228+0152C1 J1208-0200C1 shows no detection in the available observations probing the rest-UV.   Moreover, in Figure~\ref{fig:R39} we show the companion galaxy to REBELS-39, which shows a companion galaxy with a $2\sigma$ detection in the rest-UV, in contrast to the dust-obscured galaxies from \citet{fudamoto_nature}.

\begin{figure*}
    \centering
    \includegraphics[width=\textwidth,trim={0cm 6.5cm 0cm 0},clip]{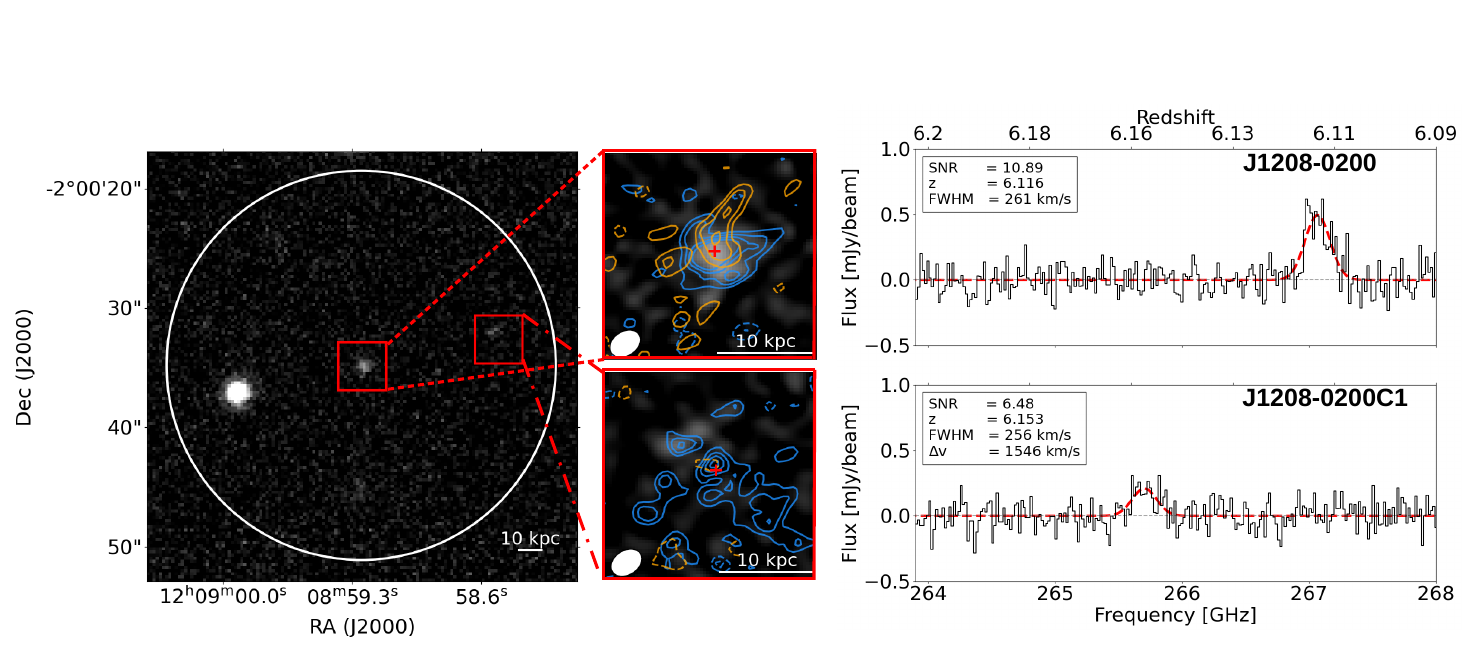}
    \caption{Similar to Figure~\ref{fig:J22} from the main text but for the QSO target J1208-0200.  Here we present a newly identified companion galaxy J1208-0200C1.}
    \label{fig:J12}
\end{figure*}

\section{SExtractor}
As explained in \S\ref{Rest-uv} we use SExtractor to measure the rest-UV fluxes of the sources imaged with HST and DECaLS. Unlike the ALPINE sources, we do not have sufficient data on these sources to use MOPHONGO. Table~\ref{tab:sextractor} shows the aperture fluxes and corresponding magnitudes of the \citet{venemans_kilo} quasars and neighbour galaxies. The aperture fluxes of the neighbour sources found in this study are presented in Table~\ref{tab:sextractor2}.

\begin{table}
\label{tab:sextractor}
\caption{The rest-UV aperture flux and magnitudes of the \citet{venemans_kilo} quasars and companion galaxies as measured by SExtractor. 
Whenever possible, various HST images probing the rest-UV light are stacked to increase the S/N.}
\centering
\begin{tabular}{lcc}
\hline
Source & Flux$_{\text{aper}}$ (nJy) \\ \hline
J0100+2802Q & $291383 \pm 143 $ \\
J0100+2802C1 & $236 \pm 143$  \\
J0305-3150Q & $17734 \pm 15$  \\
J0305-3150C1 & $16\pm 15$  \\
J0842+1218Q & $41163 \pm 14$  \\
J0842+1218C1 & $19\pm 14$  \\
J0842+1218C2 & $122 \pm 14$  \\
J1319+0950Q & $42189 \pm 23$  \\
J1319+0950C2 & $ 31 \pm 23$  \\
P231-20Q & $36992  \pm 13$ \\
P231-20C1 & $-0.5 \pm 13$  \\
J2054-0005Q & $20576 \pm 10$  \\
J2054-0005C1 & $7 \pm 10$  \\
J2100-1715Q & $16832 \pm 26$  \\
J2100-1715C1 & $-24\pm 26$ \\
\hline
\end{tabular}
\end{table}
\quad
\begin{table}
\caption{The rest-UV aperture flux and magnitudes of the two quasars and neighbours from the ALMA Science Archive as obtained through SExtractor. We used the observations of the DECaLS survey in the $z$-filter.}
\centering
\begin{tabular}{lcc}
% \label{}
\hline
Source & Flux$_{\text{aper}}$ (nJy)  \\ \hline

J1208-0200Q & $ 1358  \pm 80 $ \\
J1208-0200C1 & $171 \pm 80$  \\
J2228+0152Q & $784 \pm 84$  \\
J2228+0152C1 & $1 \pm 84$  \\ \hline
\label{tab:sextractor2}
\end{tabular}
\end{table}

\section{ALPINE companion galaxies}

\begin{figure*}
    \centering
    \includegraphics[width=\textwidth, trim={0 5cm 0 1cm}]{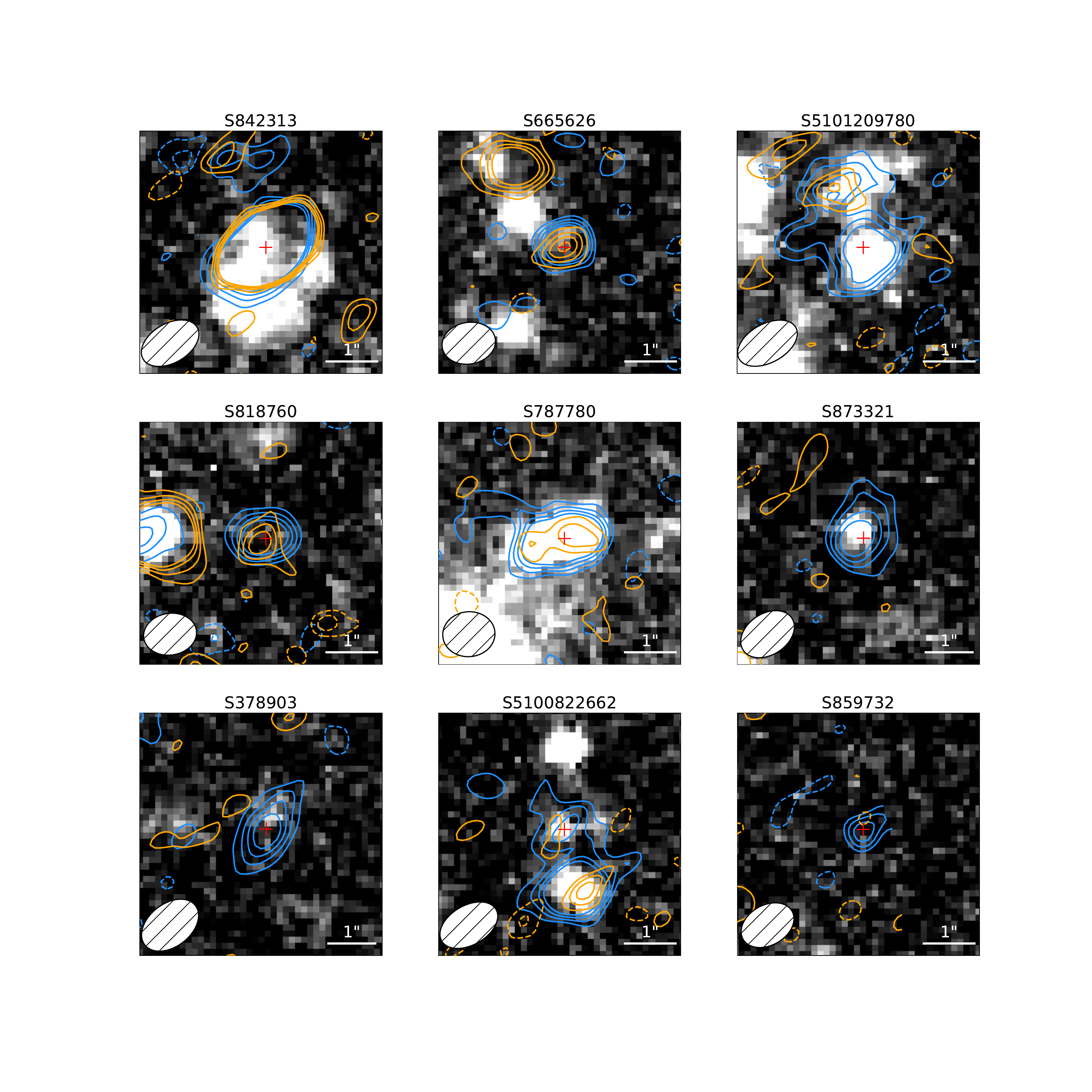}
    \caption{Companion sources identified in the neighborhood of target galaxies from ALPINE and following upon the analysis by \citet{Loiacona_alpine}.  The blue and orange contours show the $2\sigma$, $3\sigma$, $4\sigma$, $5 \sigma$ [CII] and the continuum emission, respectively. All background images correspond to rest-UV $z$-band light seen with Hyper Suprime-Cam. In the lower left corners are the beam sizes of the ALMA observations.}
    \label{fig:alpineserend}
\end{figure*}

In this work we make use of companion sources identified around [CII]-emitting galaxies from ALPINE in constraining the obscured fraction of SFR.  Given the lack of published coordinates for the companion sources identified by \citet{Loiacona_alpine}, we conducted our own search for these sources with MF3D.  The 9 sources we found are shown in Figure~\ref{fig:alpineserend}. The red crosses mark the positions of the peaks found by MF3D. In blue and orange we show the 2 to 5$\sigma$ contours of the [CII] and continuum emission, respectively.  We find no source with similar properties to S5100969402 with MF3D, but \citet{Loiacona_alpine} only report a fidelity of 0.51 for the source, so this is not surprising.  We have excluded S848185 from our sample given this source is positioned near the edge of the ALMA cube. S510327576 is also excluded given the large differences in apparent velocities for the putative [CII] line ($\mathrm{\Delta v > 2000 km\ s^{-1}}$), significantly decreasing our confidence that S51032756 actually corresponds to a $z >$ 4 galaxy (see Fig.~\ref{fig:contaminates} and Appendix~\ref{contamination}).  The positions and properties of the 9 serendipitously identified ALPINE galaxies used in this work are summarized in Table~\ref{tab:alpine}. MOPHONGO is used to measure the rest-UV fluxes as discussed in \S\ref{Rest-uv}.

\begin{table*}
\caption{Characteristics we find for the [CII]-emitting companion galaxies from ALPINE (and which were previously reported by \citealt{Loiacona_alpine}).}
\label{tab:alpine}
\begin{tabular}{lccccccccccccc}
\hline
Source & RA & Dec & $\nu_{c}$  & FWHM & $\mathrm{F_{[CII]}}$  & $\mathrm{S_{IR}}$ & $\mathrm{M_{1500}}$ \\
~ & (ICRS) & (ICRS) & (GHz) & (km s$^{-1}$) & (Jy km s$^{-1}$) & (mJy) & ~ \\ \hline
S842313 & 10:00:54.42 & +02:34:35.83 & $343.162 \pm 0.011$ & $845 \pm 48$ & $6.99 \pm 0.1$ & $8.517 \pm 0.092$ & $-21.63 \pm 0.17$ \\
S665626 & 10:01:13.83 &  +02:18:40.42 & $340.791 \pm 0.315$ & $277 \pm 22$ & $1.14 \pm 0.11$ & $0.353 \pm 0.074$ & $> - 20.59 $\\
S5101209780 & 10:01:33.45 & +02:22:08.31& $341.272 \pm 0.442$ & $388 \pm 23$ & $2.40 \pm 0.28$ & $0.065 \pm 0.063 $ & $-22.88 \pm 0.06$ \\
S818760 &10:01:54.69 & +02:32:31.30& $341.465 \pm 0.186$ & $164 \pm 17$ & $0.62 \pm 0.06$ & $0.331 \pm 0.052$ & $ > -20.88 $\\
S787780 & 09:59:56.85 &  +02:29:48.08 & $344.984 \pm 0.302$ & $263 \pm 20$ & $2.33 \pm 0.15$ & $0.365 \pm 0.057$ & $ -23.38 \pm 0.03$  \\
S873321 &10:00:03.22 & +02:37:37.46 & $308.763 \pm 0.264$ & $256 \pm 27$ & $2.47 \pm 0.32$ & $-0.007 \pm 0.114$ & $-21.22 \pm 0.23$ \\
S378903 & 10:01:11.03 &  +01:52:07.88 & $295.905 \pm 0.293$ & $297\pm 37$ & $0.64 \pm 0.02$ & $0.194 \pm 0.070$ & $> -20.60$ \\
S5100822662 & 09:58:57.94 & +02:04:52.80 & $344.280 \pm 0.179$ & $156 \pm 9 $ & $0.54 \pm 0.06$ & $0.153 \pm 0.072$ & $ -20.25 \pm 0.22$ \\
S859732 & 09:59:59.75 & +02:36:22.62 & $343.145 \pm 0.111$ & $97 \pm 13$ & $1.00 \pm 0.11$ & $-0.195 \pm 0.255 $ & $> -20.36$  \\ \hline
\end{tabular}
\end{table*}

\section{Archival ALMA Programs}
\label{sec:archive}
A summary of all ALMA projects where we have explicitly reduced the ALMA observations and done our own independent search for high-redshift companion galaxies is provided in Table~\ref{tab:all_projects}.

\begin{table*}
    \centering
    \caption{A summary of all the ALMA programs and targets analysed in our search for [CII]-emitting companion galaxies.$^a$}
    \label{tab:all_projects}
    \begin{tabular}[t]{|c l c|}
    \hline
        Project ID   &  ALMA source name & Angular resolution (") \\ \hline
    2012.1.00536.S	& MSDM\textunderscore 80+3 $\times$ & 0.60\\
	                & MSDM\textunderscore 71-5 $\times$ & 0.46\\ 
                	& MSDM\textunderscore 29.5\textunderscore 5 $\times$ & 0.45 \\ \hline
    2012.1.00676.S  & CFHQSJ0055+0146 * & 0.41\\
                    & CFHQSJ2229+1457 * & 0.62\\ \hline
    2012.1.00719.S 	& BDF-521 $\times$ & 0.39 \\	
	                & SDF-46975	$\times$ & 0.79 \\ \hline
    2012.1.00962.S	& Abell383\textunderscore z1 & 0.23\\ \hline
    2013.1.00815.S  & CLM1 & 0.40\\
                    & WMH13 & 0.88\\ \hline
    2013.1.01241.S	& A383 & 0.49\\
	                & MS0451 $\times$ & 0.88 \\ \hline
    2015.1.00091.S	& RXJ1347\textunderscore 1216 & 0.45\\ \hline
    2015.1.00122.S 	& CR7 & 0.26\\ \hline
    2015.1.00834.S  & WMH\textunderscore 5 & 0.21\\ \hline
    2015.1.00997.S	& SDSS\textunderscore J231038.88+185519.7 * &  0.75\\
    &  SDSS\textunderscore J205406.49-000514.8 * &  0.59\\
	                & SDSS\textunderscore J012958.51-003539.7  * & 0.31\\ 
                    & ULAS\textunderscore J131911.29+095051.4 * & 0.97\\ \hline
    2015.1.01096.S	& UDF-640-1417 & 0.69  \\  \hline
    2015.1.01105.S  & COSMOS13679 $\times$ & 0.74\\
                    & COSMOS24108 $\times$ & 0.67 \\
                    & NTTDF6345  $\times$ & 0.94 \\
                    & UDS16291 $\times$ & 0.72 \\
                    & UDS4812 $\times$ & 0.24\\ \hline
	2015.1.01136.S	& Abell383-iD & 0.22\\ \hline
    2015.1.01178.S	& Abell\textunderscore 611 $\times$ & 0.90\\ \hline

    \end{tabular}
    \begin{tabular}[t]{|c l c|}
    \hline
    Project ID   &  ALMA source name  & Angular resolution (")\\ \hline
    2015.1.01406.S	& A1689-zD1 & 0.23\\ \hline
    2016.1.01240.S  & 20521 & 1.12 \\
                    & 2203 $\times$ & 1.18 \\
                    & 2313 & 0.67 \\ \hline
    2016.1.01423.S	& J0859+0022 *  & 0.42\\
	                & J1152+0055 * & 0.42 \\
	                & J1202-0057 * & 0.60\\
	                & J2216-0016 * & 0.40 \\ \hline
    2017.1.00541.S	& J1208-0200 * & 0.41\\
	                & J2228+0152 * & 0.41\\ \hline
    2017.1.01451.S 	& VR7 & 0.42\\ 
    	& MASOSA & 0.43 \\ \hline
    2017.A.00026.S	& MACS1149-JD1 & 0.52 \\ \hline
    2018.1.00566.S 	& J0439 * & 0.29 \\ \hline
    2018.1.00570.S	& z7\textunderscore PAR2\textunderscore17941793$
    \times$ & 0.43 \\ \hline
    2018.1.00781.S	& system\textunderscore D1\textunderscore T1 $\times$ & 0.61\\ \hline
  
        2019.1.01003.S	& PJ308-SMG1 $\times$ & 0.64\\
	                & PJ308-SMG2 $\times$ & 0.64\\
	                & PJ308-SMG3 $\times$ & 0.64\\
	                & PJ308-SMG4 & 0.64\\
	                & PJ308-SMG6 $\times$ & 0.64\\
                	& PJ308-SMG7 $\times$ & 0.64\\
	                & PJ308-SMG8  $\times$& 0.64\\ \hline
	 2019.1.01025.S & J0252m0503 * & 0.24 \\ 
	 	         & J0525m2406 * & 0.24\\
	                & J0923p0753 * & 0.24\\
	                & J1007p2115 * & 0.28\\ \hline
	2019.1.01436.S	& PSO\textunderscore J083+11 * & 0.27\\ \hline
\end{tabular}
\begin{flushleft}
$^a$ Quasar host galaxies targets are marked with a *. [CII] non-detections of the target galaxies are marked with a $\times$. The angular resolution is defined as the beam of the observations along the minor axis. 34 of the 55 primary targets presented here are detected with [CII] and could be used to search for companion galaxies at similar redshift.
\end{flushleft}\end{table*}

% Don't change these lines
\bsp	% typesetting comment
\label{lastpage}
\end{document}